\documentclass[11pt]{cernrep}
\usepackage{graphicx,epsfig}
\newcommand{\bea}{\begin{eqnarray}}
\newcommand{\beq}{\begin{equation}}
\newcommand{\eea}{\end{eqnarray}}
\newcommand{\eeq}{\end{equation}}
\newcommand{\nn}{\nonumber}
\newcommand{\Frac}[2]{\frac{\displaystyle{#1}}{\displaystyle{#2}}}
\newcommand{\lsim}{\raise0.3ex\hbox{$\;<$\kern-0.75em\raise-1.1ex\hbox{$\sim\;$}}}
\newcommand{\gsim}{\raise0.3ex\hbox{$\;>$\kern-0.75em\raise-1.1ex\hbox{$\sim\;$}}}
\newcommand{\bmat}{\left( \begin{array}}
\newcommand{\emat}{\end{array} \right)}
\newcommand{\tr}{\mbox{Tr}}
\newcommand{\ler}{\stackrel{\scriptstyle <}{\scriptstyle\sim}}
\newcommand{\ger}{\stackrel{\scriptstyle >}{\scriptstyle\sim}}
\newcommand{\eq}[1]{Eq.~(\ref{#1})}
\newcommand{\GUT}{{\mbox{\scriptsize GUT}}}
\newcommand{\LL}{{\mbox{\scriptsize LL}}}
\newcommand{\LR}{{\mbox{\scriptsize LR}}}
\newcommand{\RL}{{\mbox{\scriptsize RL}}}
\newcommand{\RR}{{\mbox{\scriptsize RR}}}

\newcommand{\eff}{{\mbox{\scriptsize eff}}}
\newcommand{\elm}{{\mbox{\scriptsize em}}}
\newcommand{\soft}{{\mbox{\scriptsize soft}}}
\newcommand{\unity}{{\hbox{1\kern-.8mm l}}}

\begin{document}

\title{\LARGE Flavour Physics and Grand Unification}

\author{A. Masiero$^1$, S.K. Vempati$^{2,3}$ and O. Vives$^4$}
\institute{$^1$ Dip. di Fisica `G. Galilei', Univ. di Padova and 
INFN, Sezione di Padova, Via Marzolo 8, I-35131, Padua, Italy.
\\$^2$ Centre de Physique Theorique\footnote{~ Unit{\'e} mixte 
du CNRS et de l'EP, UMR 7644.} ,
Ecole Polytechnique-CPHT, 91128 Palaiseau Cedex, France.
\\$^3$ Centre for High Energy Physics, Indian Institute of Science,
Bangalore 560 012, India. 
\\$^4$ Departament de F\'{\i}sica Te\`orica and IFIC, Univ. de Val\`encia-CSIC, E-46100. Burjassot, Spain.}
\maketitle  
\begin{flushright}IFIC/07-72,\, FTUV/07-1121 \end{flushright}
\begin{abstract}
In spite of the enormous success of the Standard Model (SM), we have
strong reasons to expect the presence of new physics beyond the SM at
higher energies. The idea of the Grand Unification of all the known
interactions in nature is perhaps the main reason behind these
expectations. Low-energy Supersymmetry is closely linked with grand 
unification as a solution of the hierarchy problem associated with the
ratio $M_\GUT / M_Z$. In these lectures we will provide a general overview
of Grand Unification and Supersymmetry with special emphasis on their 
phenomenological consequences at low energies. We will analyse
the flavour and CP problems of Supersymmetry and try to identify 
in these associated low-energy observables possible indications of the 
existence of a Grand Unified theory at high energies. 
\end{abstract}
\tableofcontents

\section{INTRODUCTION}
\label{sec:intro}
The success of the standard model predictions is remarkably high
and, indeed, to some extent, even beyond what one would have
expected. As a matter of fact, a common view before LEP started
operating was that some new physics related to the electroweak
symmetry breaking should be present at the TeV scale.  In that case,
one could reasonably expect such new physics to show up when
precisions at the percent level on some electroweak observable could
be reached. As we know, on the contrary, even reaching sensitivities
better than the percent has not given rise to any firm indication of
departure from the SM predictions. To be fair, one has to recognise
that in the almost four decades of existence of the SM we have
witnessed a long series of ``temporary diseases'' of it, with effects
exhibiting discrepancies from the SM reaching even more than four
standard deviations. However, such diseases represented only ``colds''
of the SM, all following the same destiny: disappearance after some
time (few months, a year) leaving the SM absolutely unscathed and, if
possible, even stronger than before.  Also presently we do not lack
such possible ``diseases'' of the SM. The electroweak fit is not equally
good for all observables : for instance the forward-backward asymmetry
in the decay of $Z \to b \bar b$ ; some of the penguin $b \to s$
decays, the anomalous magnetic moment of the muon, etc. exhibit
discrepancies from the SM expectations. As important as all these
hints may be, undoubtedly we are far from any {\it firm} signal of
insufficiency of the SM.

All what we said above  can be summarised in a powerful
statement about the ``low-energy'' limit of any kind of new physics beyond
the SM: no matter which new physics may lie beyond the SM, it has to reproduce the SM with great accuracy when we consider its limit at energy scales of the order of the electroweak scale.

The fact that with the SM we have a knowledge of fundamental interactions
up to energies of $O(100)$ GeV should not be underestimated: it represents
a tremendous and astonishing success of our gauge theory approach in
particle physics and it is clear that it represents one of the greatest
achievements in a century of major conquests in physics. Having said that,
we are now confronting ourselves with an embarrassing question: if the SM  
is so extraordinarily good, does it make sense do go beyond it? The  
answer, in our view, is certainly positive. This ``yes'' is not only
motivated by what we could define ``philosophical'' reasons (for instance, the fact that we should not have a ``big desert'' with many orders of magnitude in energy scale without any new physics, etc.),  but there are
specific motivations pushing us beyond the SM. We will group them into  two
broad categories: theoretical and ``observational'' reasons.

\subsection{Theoretical reasons for new physics}

There are three questions which ``we'' consider fundamental and yet do not find any satisfactory answer within the SM: the 
flavor problem, the unification of the fundamental interactions and the gauge
hierarchy problem.  The reason why ``we'' is put in quotes is because it is debatable whether the three above issues  (or at least some of them) are really to be taken as questions that the SM should address, but fails to do. Let us first briefly go over them and then we'll comment about alternative views. 

\vskip .65cm
{\bf Flavor problem}. All the masses and mixings of fermions are just free
(unpredicted) parameters in the SM. To be sure, there is not even any
hint in the SM about the number and rationale of fermion families. Leaving
aside predictions for individual masses, not even any even rough relation 
among fermion masses within the same generation or among different
generations is present. Moreover, what really constitutes a {\it problem}, is the huge variety of fermion masses which is present. From the MeV region, where the electron mass sits, we move to the almost two hundred GeV of the top quark mass, i.e. fermion masses span at least five orders of magnitude, even letting aside the extreme smallness of the neutrino masses. If one has in mind the usual Higgs mechanism to 
give rise to fermion masses, it is puzzling to insert Yukawa couplings (which are free parameters of the theory) ranging from $O(1)$ to $O(10^{-6})$ or so without any justification whatsoever. Saying it concisely, we can state 
that a ``Flavor Theory'' is completely missing in the SM. To be fair, we'll see that even when we proceed to 
BSM new physics, the situation does not improve much in this respect. 
This important issue is thoroughly addressed at this school:  in Yossi Nir's  lectures \cite{Nir:2005js} 
you find an ample discussion of the flavor and CP aspects mainly within the SM, but with some
insights on some of its extensions. In these lectures we'll deal with the flavor issue in the context 
of supersymmetric and grand unified extensions of the SM.

\vskip .65cm
{\bf Unification of forces}. At the time of the Fermi theory we had two
couplings to describe the electromagnetic and the weak interactions (the
electric constant and the Fermi constant, respectively). In the SM we are 
trading off those two couplings with two new couplings, the gauge
couplings of  $SU(2)$ and $U(1)$. Moreover, the gauge coupling of the
strong interactions is very different from the other two. We cannot say   
that the SM represents a true unification of fundamental interactions,
even leaving aside the problem that gravity is not considered at all by
the model. Together with the flavor issue, the unification of fundamental interactions constitutes
the main focus of the present lectures. First, also respecting the chronological evolution, 
we'll consider  grand unified theories without an underlying supersymmetry, while then
we'll move to spontaneously broken supergravity theories with a unifying gauge symmetry 
encompassing electroweak and strong interactions. 

\vskip .65cm
{\bf Gauge hierarchy}. Fermion and vector boson masses are ``protected'' by
symmetries in the SM (i.e., their mass can arise only when we break certain   
symmetries). On the contrary the Higgs scalar mass does not enjoy such a
symmetry protection. We would expect such mass to naturally jump to some
higher scale where new physics sets in (this new energy scale could be  
some grand unification scale or the Planck mass, for instance). The only
way to keep the Higgs mass at the electroweak scale is to perform
incredibly accurate fine tunings of the parameters of the scalar sector. Moreover such fine tunings are unstable under radiative corrections, i.e. they should be repeated at any subsequent order in perturbation theory (this is the so-called ``technical'' aspect of the gauge hierarchy problem).  

We close this Section coming back to the question  about how fundamental the above problems actually are, a 
{\it caveat}  that we mentioned at the beginning of the Section. Do we really need a flavor theory, or 
can we simply consider that fermion masses as  fundamental parameters which just take the values that
we observe  in our Universe ? Analogously, for the gauge hierarchy, is it really something that we have to {\it explain}, or could we take the view that just the way our Universe is requires that the $W$ mass is 17 orders  of magnitude smaller than the Planck mass, i.e. taking $M_W$ as a fundamental input much in the same way we 
``accept'' a fundamental constant as incredibly small as it is? And, finally, why should all fundamental interactions unify, is it just an aesthetical criterion that ``we'' try to impose after the success of the electro-magnetic unification? 
The majority of particle physicists (including the authors of the present contribution) consider the above three issues as genuine problems that a {\it fundamental} theory should address. In this view, the SM could be considered only as a low-energy limit of such deeper theory. Obviously, the relevant question becomes: at which energy scale should such alleged new physics set in? Out of the above three issues, only that referring to the gauge hierarchy problem requires a modification of the SM physics at scales close to the electroweak scale, i.e. at the TeV scale. On the other hand, the absence of clear signals of new physics at LEP, in FCNC and CP violating processes, etc. has certainly contributed to cast doubts in some researchers about  the actual existence of a gauge hierarchy {\it problem} . Here we'll take the point of view that the electroweak symmetry breaking calls for new physics close to the electroweak scale itself and we'll explore its implications for FCNC and CP violation in particular.

\subsection{``Observational'' reasons for new physics}

We have already said that all the experimental particle physics results of
these last years have marked one success after the other of the SM. What
do we mean then by ``observational'' difficulties for the SM? It is
curious that such difficulties do not arise from observations within the
strict high energy particle physics domain, but rather they originate from 
astroparticle physics, in particular from possible
``clashes'' of the particle physics SM with the standard model of cosmology
(i.e., the Hot Big Bang) or the standard model of the Sun.

\vskip .65cm
{\bf Neutrino masses and mixings}.

The statement that non-vanishing  neutrino masses imply new physics beyond the SM is almost tautological. We built the SM in such a way that neutrinos {\it had} to be massless (linking such property to the V-A character of weak interactions), namely we avoided Dirac neutrino masses by banning  the presence of the right-handed neutrino from the fermionic spectrum, while Majorana masses for the left-handed neutrinos were avoided by limiting the Higgs spectrum to isospin doublets. Then we can say that a massive neutrino is a signal of new physics {\it ``by construction''}.  However, there is something deeper in the link 
massive neutrino -- new physics than just the obvious correlation we mentioned. Indeed, the easiest way to make neutrinos massive is the introduction of a right-handed neutrino which can combine with the left-handed one to give rise to a (Dirac) mass term through the VEV of the usual Higgs doublet. However, once such right-handed neutrino appears, one faces the question of its possible Majorana mass. Indeed, while a Majorana mass for the left-handed neutrino is forbidden by the 
electroweak gauge symmetry, no gauge symmetry is able to ban a Majorana  mass for the right-handed neutrino given that such particle is sterile with respect to the whole gauge group of the strong and electroweak symmetries. If we write a Majorana mass of the same order as an ordinary Dirac fermion mass we end up with unbearably heavy neutrinos. To keep neutrinos light we need to invoke a large Majorana mass for the right-handed neutrinos, i.e. we have to introduce a scale larger than the electroweak scale. At this scale the right-handed neutrinos should be no longer (gauge) sterile particles and, hence, we expect new physics to set in at such new scale.  Alternatively, we could avoid the introduction of right-handed neutrinos providing (left-handed) neutrino masses via the VEV of a new Higgs scalar transforming as the highest component of an $SU(2)_L$ triplet. Once again the extreme smallness of neutrino masses would force us to introduce a new scale; this time it would be a scale much lower than the electroweak scale 
(i.e., the VEV of the Higgs triplet has to be much smaller than that of the usual Higgs doublet) and, consequently, new physics at a new physical mass scale would emerge. 

Although, needless to say, neutrino masses and mixings play a role, and, indeed, a major one, 
in the vast realm of flavor physics, given the specificity of the subject, there is an entire set 
of independent lectures devoted to neutrino physics at this school  \cite{Smirnov:2005unp}.
In our lectures we'll have a chance to touch now and then aspects of neutrino physics related
to grand unification, although we recommend the readers more specifically interested in the
neutrino aspects to refer to the thorough discussion in Alexei Smirnov's lectures at this school.  
 But at
least a point should be emphasised here: together with the issue of dark
matter that we are going to present next, massive neutrinos witness that new
physics beyond the SM {\it is} present together with a new physical scale
different from that is linked to the SM electroweak physics. Obviously, new
physics can be (and probably is) associated to different scales; as we said
above, we think that the gauge hierarchy problem is strongly suggesting that
(some) new physics should be present close to the electroweak scale. It could
be that such new physics related to the electroweak scale is {\it not} that
which causes neutrino masses (just to provide an example, consider
supersymmetric versions of the seesaw mechanism: in such schemes, low-energy
SUSY would be related to the gauge hierarchy problem with a typical scale of
SUSY masses close to the electroweak scale, while the lightness of the
neutrino masses would  result from a large Majorana mass of the right-handed
neutrinos).

\vskip .85cm
{\bf Clashes of the SM of particle physics and cosmology}: dark matter,
baryogenesis and inflation.

Astroparticle physics represents a major road to access new physics BSM. This
important issue is amply covered by Pierre Binetruy's lectures at this school
\cite{Binetruy:2005unp}. Here we simply point out the three main ``clashes''
between the SM of particle physics and cosmology.

\vskip .65cm 
{\it Dark Matter}. There exists an impressive evidence that not
only most of the matter in the Universe is {\it dark}, i.e. it doesn't emit
radiation, but what is really crucial for a particle physicist is that (almost
all) such dark matter (DM) has to be provided by particles other than the
usual baryons. Combining the WMAP data on the cosmic microwave background
radiation (CMB) together with all the other evidences for DM on one side, and
the relevant bounds on the amount of baryons present in the Universe from Big
Bang nucleosynthesis and the CMB information on the other side, we obtain the
astonishing result that at something like 10 standard deviations DM has to be
of non-baryonic nature.  Since the SM does not provide any viable non-baryonic
DM candidate, we conclude that together with the evidence for neutrino masses
and oscillations, DM represents the most impressive observational evidence we
have so far for new physics beyond the standard model. Notice also that it has
been repeatedly shown that massive neutrinos cannot account for such
non-baryonic DM, hence implying that we need wilder new physics beyond the SM
rather than the obvious possibility of providing neutrinos a mass to have a
weakly interactive massive particle (WIMP) for DM candidate.  Thus, the
existence of a (large) amount of non-baryonic DM push us to introduce new
particles in addition to those of the SM.  \vskip .65cm {\it Baryogenesis}.
Given that we have strong evidence that the Universe is vastly
matter-antimatter asymmetric (i.e. no sizable amount of primordial antimatter
has survived), it is appealing to have a dynamical mechanism to give rise to
such large baryon-antibaryon asymmetry starting from a symmetric situation. In
the SM it is not possible to have such an efficient mechanism for
baryogenesis. In spite of the fact that at the quantum level sphaleronic
interactions violate baryon number in the SM, such violation cannot lead to
the observed large matter-antimatter asymmetry (both CP violation is too tiny
in the SM and also the present experimental lower bounds on the Higgs mass do
not allow for a conveniently strong electroweak phase transition). Hence a
dynamical baryogenesis calls for the presence of new particles and
interactions beyond the SM (successful mechanisms for baryogenesis in the
context of new physics beyond the SM are well known).

\vskip .65cm
{\it Inflation}. Several serious cosmological problems (flatness,
causality, age of the Universe, ...) are beautifully solved if the early
Universe underwent some period of exponential expansion (inflation). The
SM with its Higgs doublet does not succeed to originate such an
inflationary stage. Again some extensions of the SM, where in particular
new scalar fields are introduced, are able to produce a temporary
inflation of the early Universe.       

\vskip .8cm 
As we discussed for the case of theoretical reasons to go
beyond the SM, also for the above mentioned observational reasons one
has to wonder which scales might be preferred by the corresponding new
physics which is called for.  Obviously, neutrino masses, dark matter,
baryogenesis and inflation are likely to refer to {\it different}
kinds of new physics with some possible interesting correlations. Just
to provide an explicit example of what we mean, baryogenesis could
occur through leptogenesis linked to the decay of heavy right-handed
neutrinos in a see-saw context. At the same time, neutrino masses
could arise through the same see-saw mechanism, hence establishing a
potentially tantalising and fascinating link between neutrino masses
and the cosmic matter-antimatter asymmetry. The scale of such new
physics could be much higher than the electroweak scale.

On the other hand, the dark matter issue could be linked to a much
lower scale, maybe close enough to the electroweak scale. This is what
occurs in one of the most appealing proposals for cold dark matter,
namely the case of a WIMP in the mass range between tens to hundreds
of GeV. What really makes such a WIMP a ``lucky'' CDM candidate is that
there is an impressive quantitative ``coincidence'' between Big Bang
cosmological SM parameters (Hubble parameter, Planck mass, Universe
expansion rate, etc.) and particle physics parameters (weak
interactions, annihilation cross section, etc.) leading to a surviving
relic abundance of WIMPs just appropriate to provide an energy density
contribution in the right ball-park to reproduce the dark matter
energy density. A particularly interesting example of WIMP is
represented by the lightest SUSY particle (LSP) in SUSY extensions of
the SM with a discrete symmetry called R parity (see below more about
it). Once again, and in a completely independent way, we are led to
consider low-energy SUSY as a viable candidate for new physics
providing some answer to open questions in the SM.

As exciting as the above considerations on dark matter and unification
are in suggesting us the presence of new physics at the weak scale, we
should not forget that they are just {\it strong suggestions}, but
alternative solutions to both the unification and dark matter puzzles
could come from (two kinds of) new physics at scales much larger than
$M_W$.

\vskip .8cm
\subsection{The SM as an effective low-energy theory}

The above theoretical and ``observational'' arguments strongly motivate us
to go beyond the SM. On the other hand, the clear success of the SM in
reproducing all the known phenomenology up to energies of the order of the
electroweak scale is telling us that the SM has to be recovered as the
low-energy limit of such new physics.
 Indeed, it may even well be the case  that we have a ``tower''
of underlying theories which show up at different energy scales.

If we accept the above point of view we may try to find signals of new
physics considering the SM as a truncation to renormalisable operators of
an effective low-energy theory which respects the  $SU(3)\otimes SU(2) \otimes
U(1)$
symmetry  and whose fields are just those of the SM. The renormalisable
(i.e. of canonical dimension less or equal to four) operators giving rise
to the SM enjoy three crucial properties which have no reason to be shared
by generic operators of dimension larger than four. They are the
conservation (at any order in perturbation theory) of Baryon (B) and
Lepton (L) numbers and an adequate suppression of Flavour Changing Neutral
Current (FCNC) processes through the GIM mechanism.  

Now consider the new physics (directly above the SM in the
``tower'' of new physics theories) to have a typical energy scale $\Lambda$.
In the low-energy effective Lagrangian such scale appears with a positive
power only in the quadratic scalar term (scalar mass) and in the dimension
zero operator which can be considered a cosmological constant. Notice that
$\Lambda$ cannot appear in dimension three operators related to fermion
masses because chirality forbids direct fermion mass terms in the
Lagrangian. Then in all operators of dimension larger than four $\Lambda$
will show up in the denominator with powers increasing with the dimension
of the corresponding operator. 

The crucial question that all of us, theorists and experimentalists, ask
ourselves is: where is $\Lambda$? Namely is it close to the electroweak
scale (i.e. not much above $100$ GeV) or is $\Lambda$ of the order of the
grand unification scale or the Planck scale? B- and L-violating processes
and FCNC phenomena represent a potentially interesting clue to answer this
fundamental question.

Take $\Lambda$ to be close to the electroweak scale. Then we may expect
non-renormalisable operators with B, L and flavour violations not to be
largely suppressed by the presence of powers of $\Lambda$ in the
denominator. Actually this constitutes in general a formidable challenge
for any model builder who wants to envisage new physics close to $M_W$.
Theories with dynamical breaking of the electroweak symmetry
(technicolour) and low-energy supersymmetry constitute examples of new
physics with a ``small'' $\Lambda$. In these lectures we will only focus on a
particularly interesting ``ultra-violet completion'' of the Standard Model,
namely low energy supersymmetry (SUSY). Other possibilities are considered 
in other lectures.

Alternatively, given the above-mentioned potential danger of having a small
$\Lambda$, one may feel it safer to send $\Lambda$ to super-large values.
Apart from kind of ``philosophical'' objections related to the unprecedented  
gap of many orders of magnitude without any new physics, the above
discussion points out a typical problem of this approach. Since the
quadratic scalar terms have a coefficient in front scaling with
$\Lambda^2$ we expect all scalar masses to be of the order of the
super-large scale $\Lambda$. This is the gauge hierarchy problem and it
constitutes the main (if not only) reason to believe that SUSY should be a
low-energy symmetry.

Notice that the fact that SUSY should be a fundamental symmetry of Nature
(something of which we have little doubt given the ``beauty'' of this
symmetry) 
does not imply by any means that SUSY should be a low-energy symmetry,
namely that it should hold unbroken down to the electroweak scale.
SUSY may well be present in
Nature but be broken at some very large scale (Planck scale or string
compactification scale). In that case SUSY would be of no use in tackling
the gauge hierarchy problem and its phenomenological relevance would be
practically zero. On the other hand if we invoke SUSY to tame the growth
of the scalar mass terms with the scale $\Lambda$, then we are forced to
take the view that SUSY should hold as a good symmetry down to a scale
$\Lambda$ close to the electroweak scale. Then B, L and FCNC may be useful
for us to shed some light on the properties of the underlying theory from
which the low-energy SUSY Lagrangian resulted. Let us add that there is an
independent argument in favour of this view that SUSY should be a
low-energy symmetry. The presence of SUSY partners at low energy creates
the conditions to have a correct unification of the strong and electroweak
interactions. If they were at $M_{\rm Planck}$ and the SM were all the physics
up to super-large scales, the program of achieving such a unification
would largely fail, unless one complicates the non-SUSY GUT scheme with a
large number of Higgs representations and/or a breaking chain with
intermediate mass scales is invoked.

In the above discussion we  stressed that we are not only insisting on
the fact that SUSY should be present at some stage in Nature, but we are
asking for something much more ambitious: we are asking for SUSY to be a
low-energy symmetry, namely it should be broken at an energy scale as low
as the electroweak symmetry breaking scale. This fact can never be
overestimated. There are indeed several reasons pushing us to introduce
SUSY : it is the most general symmetry compatible with a local,
relativistic quantum field theory, it softens the degree of divergence of
the theory, it looks promising for a consistent quantum description of
gravity together with the other fundamental interactions. However, all
these reasons are not telling us where we should expect SUSY to be broken.
for that matter we could even envisage the maybe ``natural'' possibility
that SUSY is broken at the Planck scale. What is relevant for
phenomenology is that the gauge hierarchy problem and, to some extent, the
unification of the gauge couplings are actually forcing us to ask for SUSY
to be unbroken down to the electroweak scale, hence implying that the SUSY
copy of all the known particles, the so-called s--particles should have a
mass in the $100-1000$ GeV mass range. If Tevatron is not going
to see any SUSY particle, at least the advent of LHC will be decisive in
establishing whether low-energy SUSY actually exists or it is just a fruit
of our (ingenious) speculations. Although even after LHC, in case of a
negative result for the search of SUSY particles, we will not be able to
``mathematically'' exclude all the points of the SUSY parameter space, we
will certainly be able to very reasonably assess whether the low-energy
SUSY proposal makes sense or not. 

Before the  LHC (and maybe Tevatron) direct searches for SUSY
signals we should ask ourselves whether we can hope to have some indirect
manifestation of SUSY through virtual effects of the SUSY particles. 

We know that in the past virtual effects (i.e. effects due to the
exchange of yet unseen particles in the loops) were precious in leading us
to major discoveries, like the prediction of the existence of the charm
quark or the heaviness of the top quark long before its direct
experimental observation. Here we focus on the potentialities of SUSY
virtual effects in processes which are particularly suppressed (or
sometime even forbidden) in the SM ; the flavour changing neutral current
phenomena and the processes where CP violation is violated. 

However, the above role of the studies of FCNC and CP violation in
relation to the {\it discovery} of new physics should not make us
forget they are equally important for another crucial task: this is
the step going from {\it discovery} of new physics to its {\it
understanding}.  Much in the same way that discovering quarks, leptons
or electroweak gauge bosons (but without any information about quark
mixings and CP violation) would not allow us to {\it reconstruct} the
theory that we call the GWS Standard Model, in case LHC finds, say, a
squark or a gluino we would not be able to {\it reconstruct} the
correct SUSY theory. Flavour and CP physics would play a fundamental
role in helping us in such effort. In this sense, we can firmly state
that the study of FCNC and CP violating processes is {\it
complementary} to the direct searches of new physics at LHC.

\subsection{Flavor, CP and New Physics}
\label{sec:FCNC}

The generation of fermion masses and mixings (``flavour problem'') gives 
rise to a first and important distinction among theories of new physics 
beyond the electroweak standard model. 

One may conceive a 
kind of new physics which is completely ``flavour blind'', i.e. new 
interactions which have nothing to do with the flavour structure. To 
provide an example of such a situation, consider a scheme where flavour 
arises at a very large scale (for instance the Planck mass) while new 
physics is represented by a supersymmetric extension of the SM 
with supersymmetry broken at a much lower scale and with the SUSY 
breaking transmitted to the observable sector by flavour-blind gauge 
interactions. In this case one may think that the new physics does not 
cause any major change to the original flavour structure of the SM, 
namely that the pattern of fermion masses and mixings is compatible with 
the numerous and demanding tests of flavour changing neutral currents.

Alternatively, one can conceive a new physics which is entangled 
with the flavour problem. As an example consider a technicolour scheme 
where fermion masses and mixings arise through the exchange of new gauge 
bosons which mix together ordinary and technifermions. Here we expect 
(correctly enough) new physics to have potential problems in 
accommodating the usual fermion spectrum with the adequate suppression 
of FCNC. As another example of new physics which is not flavour blind, 
take a more conventional SUSY model which is derived from a 
spontaneously broken N=1 supergravity and where the SUSY breaking 
information is conveyed to the ordinary sector of the theory through 
gravitational interactions. In this case we may expect that the scale at 
which flavour arises and the scale of SUSY breaking are not so different 
and possibly the mechanism itself of SUSY breaking and transmission is 
flavour-dependent. Under these circumstances we may expect 
a potential flavour problem to arise, namely that SUSY contributions to 
FCNC processes are too large.

\vskip .8cm
\subsubsection{The Flavor Problem in SUSY}

The potentiality of probing SUSY in FCNC phenomena was readily realised
when 
the era of SUSY  phenomenology started in the early 80's \cite{Ellis:1981ts,Barbieri:1981gn,Duncan:1983wz,Gerard:1984bg,Gerard:1984pc,Langacker:1984ak,Gerard:1984vb}.
 In particular, the 
major implication that the scalar partners of quarks of the same electric 
charge but belonging to different generations had to share a remarkably high 
mass degeneracy was emphasised.

Throughout the large amount of work  in this last decade it became clearer 
and clearer that generically talking of the implications of low-energy SUSY on 
FCNC may be rather misleading. We have a minimal SUSY extension of the SM, the 
so-called Minimal Supersymmetric Standard Model (MSSM) \cite{Barbieri:1982eh,Chamseddine:1982jx,Nilles:1983ge,Haber:1984rc,Ross:1985ai,Haber:1993wf,Martin:1997ns}
where the FCNC 
contributions can be computed in terms of a very limited set of unknown 
new SUSY parameters. Remarkably enough, this minimal model succeeds to
pass all  
the set of FCNC tests unscathed. To be sure, it is possible to severely 
constrain the SUSY parameter space, for instance using $b 
\to s \gamma$, in a way which is complementary to what is achieved by direct 
SUSY searches at colliders.

However,  the MSSM is by no means equivalent to low-energy SUSY. A first 
sharp distinction concerns the mechanism of SUSY breaking and 
transmission to the observable sector which is chosen. As we mentioned 
above, in models with gauge-mediated SUSY breaking (GMSB models) 
\cite{Giudice:1998bp} it may be possible to avoid the 
FCNC threat ``ab initio'' (notice that this is not an automatic feature of 
this class of models, but it depends on the specific choice of the 
sector which transmits the SUSY breaking information, the so-called 
messenger sector). The other more ``canonical'' class of SUSY theories 
that was mentioned above has gravitational messengers and a very large 
scale at which SUSY breaking occurs. In this talk we will focus only on 
this class of gravity-mediated SUSY breaking models. Even sticking to 
this more limited choice we have a variety of options with very 
different implications for the flavour problem. 

First, there exists an interesting large class of SUSY realisations 
where the customary R-parity (which is invoked to suppress proton decay) 
is replaced by 
other discrete symmetries which allow either baryon or lepton violating terms 
in the superpotential. But, even sticking to the more orthodox view of 
imposing R-parity, we are still left with a large variety of extensions of the 
MSSM at low energy. The point is that low-energy SUSY ``feels'' the new physics 
at the super-large scale at which supergravity  (i.e., local supersymmetry) 
broke down. In this last couple of years we have witnessed an increasing 
interest in supergravity realisations without the so-called flavour 
universality of the terms which break SUSY explicitly. Another class of 
low-energy SUSY realisations which differ from the MSSM in the FCNC sector 
is obtained from SUSY-GUT's. The interactions involving super-heavy particles 
in the energy range between the GUT and the Planck scale bear important 
implications for the amount and kind of FCNC that we expect at low energy.

Even when R parity is imposed the FCNC challenge is not over. It is true
that in this case, analogously to what happens in the SM, no tree
level FCNC contributions arise. However, it is well-known that this is a
necessary but not sufficient condition to consider the FCNC problem
overcome. The loop contributions to FCNC in the SM exhibit the presence of
the GIM mechanism and we have to make sure that in the SUSY case with R
parity some analog of the GIM mechanism is active. 

To give a qualitative idea of what we mean by an effective super-GIM
mechanism, let us consider the following simplified situation where the
main features emerge clearly. Consider the SM box diagram responsible for
the $K^0 - \bar{K}^0$ mixing and take only two generations, i.e. only the up
and charm quarks run in the loop. In this case the GIM mechanism yields a
suppression factor of $O((m_c^2 - m_u^2)/M_W^2)$. If we replace the W
boson and the up quarks in the loop with their SUSY partners and we take,
for simplicity, all SUSY masses of the same order, we obtain a
super-GIM factor which looks like the GIM one with the masses of the
superparticles instead of those of the corresponding particles. The
problem is that the up  and charm squarks have masses 
which are much larger
than those of the corresponding quarks. Hence the super-GIM factor tends to
be of $O(1)$ instead of being $O(10^{-3})$ as it is in the SM case. To
obtain this small number we would need a high degeneracy between the mass of
the charm and up squarks. It is difficult to think that such a degeneracy
may be accidental. After all, since we invoked SUSY for a naturalness
problem (the gauge hierarchy issue), we should avoid invoking
a fine-tuning to solve its problems! Then one can turn to some symmetry
reason. For instance, just sticking to this simple example that we are
considering, one may think that the main bulk of the charm and up squark
masses is the same, i.e. the mechanism of SUSY breaking should have some
universality in providing the mass 
to these two squarks with the same
electric charge.  Flavour universality is by no means a prediction of
low-energy SUSY.
The absence of flavour universality of soft-breaking terms may result from
radiative effects at the GUT scale or from effective supergravities
derived 
from string theory. Indeed, from the point of view of effective
supergravity theories derived from superstrings it may appear more natural
not to have such flavor universality. To obtain it one has to invoke
particular circumstances, like, for instance, strong dilaton over moduli
dominance in the breaking of supersymmetry, something which is certainly
not expected on general ground.

 Another possibility one may envisage is that the masses
of the squarks are quite high, say above few TeV's. Then even if they are
not so degenerate in mass, the overall factor in front of the four-fermion
operator responsible for the kaon mixing becomes smaller and smaller (it
decreases quadratically with the mass of the squarks) and, consequently, one
can respect the observational result. We see from this simple example
that the issue of FCNC may be closely linked to the crucial problem of the
way we break SUSY.

We now turn to some general remarks about the worries and hopes that CP
violation arises in the SUSY context.

\vskip .8cm
\subsubsection{CP Violation in SUSY}

CP violation has major potentialities to exhibit manifestations of new physics
beyond the standard model.
Indeed, the reason behind this statement is at least twofold:
CP violation is a ``rare'' phenomenon and hence it constitutes an
ideal ground for NP to fight on equal footing with the (small) SM
contributions; generically any NP present in the neighbourhood of the
electroweak scale is characterised by the presence of new ``visible''
sources of CP violation in addition to the usual CKM phase of the SM.
A nice introduction to this subject by R.~N.~Mohapatra
can be found in the  book ``CP violation'', Jarlskog, C. (Ed.),
Singapore: World Scientific (1989) \cite{Mohapatra:1988ph}. 

Our choice of low energy SUSY for NP is due on one side to the usual
reasons related to the gauge hierarchy problem, gauge coupling
unification and the possibility of having an interesting cold dark
matter candidate and on the other hand to the fact that it provides
the only example of a completely defined extension of the SM where the
phenomenological implications can be fully detailed
\cite{Nilles:1983ge,Haber:1984rc,Ross:1985ai,Haber:1993wf,Martin:1997ns}. 
SUSY fully
respects the above statement about NP and new sources of CP violation:
indeed a generic SUSY extension of the SM provides numerous new CP
violating phases and in any case even going to the most restricted
SUSY model at least two new flavour conserving CP violating phases are
present. Moreover the relation of SUSY with the solution of the gauge
hierarchy problem entails that at least some SUSY particles should
have a mass close to the electroweak scale and hence the new SUSY CP
phases have a good chance to produce visible effects in the coming
experiments
\cite{Grossman:1997pa,Misiak:1997ei,Masiero:2001ep,Masiero:2002xj}.  
This sensitivity of CP violating phenomena to SUSY
contributions can be seen i) in a ``negative'' way : the ``SUSY CP
problem'' i.e. the fact that we have to constrain general SUSY
schemes to pass the demanding experimental CP tests and ii) in a
``positive'' way : indirect SUSY searches in CP violating processes
provide valuable information on the structure of SUSY viable
realisations.  Concerning this latter aspect, we emphasise that not
only the study of CP violation could give a first hint for the
presence of low energy SUSY before LHC, but, even after the possible
discovery of SUSY at LHC, the study of indirect SUSY signals in CP
violation will represent a complementary and very important source of
information for many SUSY spectrum features which LHC will never be
able to detail \cite{Grossman:1997pa,Misiak:1997ei,Masiero:2001ep,Masiero:2002xj}.

Given the mentioned potentiality of the relation between SUSY and CP
violation and obvious first question concerns the selection of the
most promising phenomena to provide such indirect SUSY hints. It is
interesting to notice that SUSY CP violation can manifest itself both
in flavour conserving and flavour violating processes. As for the
former class we think that the electric dipole moments (EDMs) of the
neutron, electron and atoms are the best place where SUSY phases, even
in the most restricted scenarios, can yield large departures from the
SM expectations. In the flavour changing class we think that the study
of CP violation in several B decay channels can constitute an important
test of the uniqueness of the SM CP violating source and of the
presence of the new SUSY phases. CP violation in kaon physics remains
of great interest and it will be important to explore rare decay
channels ($K_L \to \pi^0 \nu \bar \nu$ and $K_L \to \pi^0 e^+ e^-$ for
instance) which can provide complementary information on the presence
of different NP SUSY phases in other flavour sectors. Finally let us
remark that SUSY CP violation can play an important role in baryo-
and/or lepto-genesis.  In particular in the leptogenesis scenario the
SUSY CP violation phases can be related to new CP phases in the
neutrino sector with possible links between hadronic and leptonic CP
violations.

\section{GRAND UNIFICATION AND SUSY GUTS}

Unification of  all the known forces in nature into a universal
interaction describing all the processes on equal footing has been 
for a long time  and keeps being nowadays a major goal for particle physics. 
In a sense, we witness a first, extraordinary example of  a ``unified explanation'' 
of apparently different phenomena under a common fundamental interaction  
in Newton's ``Principia'', where the universality of gravitational law succeeds to 
link together the fall of a stone with the rotation of the Moon. But it is with
Maxwell's ``Treatise of Electromagnetism''  at the end of 
 the 19th century that two seemingly unlinked interactions,  electricity 
and magnetism, merge into the common description of electromagnetism. Another 
amazing step 
along this path was completed in the second half of the last century
when electromagnetic and weak interactions were unified in the electroweak 
interactions giving rise to the Standard Model. However, the
Standard Model is by no means satisfactory because it still involves three
different gauge groups with independent gauge couplings $SU(3)\times
SU(2) \times U(1)$. Strictly speaking, if we intend ``unification'' of fundamental interactions as a reduction 
of the fundamental coupling constants, no much gain was achieved in the SM 
with respect to the time when weak and electromagnetic interactions were associated to the
Fermi and electric couplings, respectively. Nevertheless, one should recognise
that, even though, $e$ and $G_F$ are traded with $g_2$ and $g_1$ of  
$SU(2) \times U(1)$, in the SM electromagnetic and weak forces are no longer 
two separate interactions, but they are closely entangled.

Another distressing feature of the Standard Model is its strange matter
content. There is no apparent reason why a family contains a doublet of 
quarks, a doublet of leptons, two singlets of quarks and a charged lepton 
singlet with quantum numbers,
\bea
Q~(3,2,\frac{1}{3}), ~~ u_R~(\bar 3, 1, \frac{4}{3}), ~~
d_R~(\bar 3, 1, -\frac{2}{3}),~~L~(1,2,-1),~~
e_R~(1,1,-2).
\eea

The $U(1)$ quantum numbers are specially disturbing. In principle any
charge is allowed for a $U(1)$ symmetry, but, in the SM, charges are
quantised in units of $1/3$. 

These three problems find an answer in Grand Unified Theories (GUTs).
The first theoretical attempt to solve these questions was the Pati-Salam
model, $SU(4)_{\rm C}\times SU(2)_{\rm L}\times SU(2)_{\rm R}$ \cite{Pati:1974yy}. The original
idea of this model was to consider quarks and leptons as different components
of the same representation, extending $SU(3)$ to include {\it leptons as the 
fourth colour}. In this way the matter multiplet would be
\bea
F^e_{\rm L,R} = \left[\matrix{u_r & u_y & u_b & \nu_e \cr d_r & d_y & d_b 
& e^-}
\right]_{\rm L,R}, 
\eea
with $F^e_{\rm L}$ and $F^e_{\rm R}$ transforming as $(4,2,1)$ and $(4,1,2)$,
respectively under the gauge group. Thus, this theory simplifies the matter 
content of the SM to only two representations containing 16 states, with the sixteenth component
which is missing in the SM fermion spectrum, carrying the quantum numbers of
 a right-handed neutrino. More importantly, it provides a 
very elegant answer to the problem of charge quantisation in the SM. Notice
that, while the eigenvalues of Abelian groups are continuous, those 
corresponding to non-Abelian group are discrete. Therefore, if we embed the 
hypercharge interaction of the SM in a non-Abelian group, the charge will 
necessarily be quantised. In this case the electric charge is given by
$Q_{\rm em} = T_{3 {\rm L}} + T_{3 {\rm R}} + 1/2 (B-L)$, where $SU(3)_{\rm C} 
\times U(1)_{B-L}$ is the subgroup contained in $SU(4)_{\rm C}$. Still, this 
group contains three independent gauge couplings and it does not really unify all 
the known interactions (even imposing a discrete symmetry interchanging the
two $SU(2)$ subgroups, we are left with two independent gauge couplings).
    
The Standard Model has four diagonal generators corresponding to $T_3$ and 
$T_8$ of $SU(3)$, $T_3$ of $SU(2)$ and the hypercharge generator Y, i.e.
it has rank four. If we want to unify all these interactions into a simple 
group it must have rank four at least. Indeed, to achieve a unification of the gauge couplings, we have to require the 
gauge group of such unified theory to be simple or the product of identical simple factors whose coupling
constants can be set equal by a discrete symmetry. 

There exist 9 simple or semi-simple groups of rank four. Imposing that the viable candidate contains
an SU(3) factor and that it possesses some {\it complex} representations (in order to
accommodate the chiral fermions), one is left with $SU(3)\times SU(3)$ and $SU(5)$. Since in the
former case the quarks u, d and s should be put in the same triplet representation, one would run into evident
problems with exceeding FCNC contributions in d-s transitions. Hence, we are left with $SU(5)$ as the only viable candidate 
of rank four for grand unification. 
The minimal $SU(5)$ model was  originally proposed by Georgi and Glashow \cite{Georgi:1974sy}.
 In this theory there is 
a single gauge coupling $\alpha_{\rm GUT}$ defined at the grand unification 
scale $M_{\rm GUT}$. The whole SM particle content is contained in two $SU(5)$ 
representations ${\bf \bar 5} = (\bar 3,1,-\frac{2}{3}) + (1,2,-1)$ and 
${\bf 10} =
(3,2,\frac{1}{3}) + (\bar 3,1,\frac{4}{3}) + (1,1,-2)$ under $SU(3)\times
SU(2) \times U(1)$. Once more the $U(1)_Y$ generator is a combination of the
diagonal generators of the $SU(5)$ and electric charge is also quantised in 
this model. The minimal $SU(5)$ will be described below.

A perhaps more complete unification is provided by the $SO(10)$ model \cite{Fritzsch:1974nn,Georgi:1979dq}. 
We have also a single gauge coupling and charge quantisation but in 
$SO(10)$ a single representation, the ${\bf 16}$ includes
both the ${\bf \bar 5}$ and ${\bf 10}$ plus a singlet corresponding to
a right handed neutrino.

\subsection{Gauge couplings and Unification }

\begin{figure}
\begin{center}
\includegraphics[width=11cm]{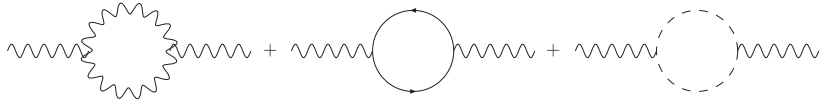}
\end{center}
\caption{One loop corrections to the gluon propagator.}
\label{fig:beta}
\end{figure}

A grand unified theory would require the equality of the three SM
gauge couplings to a single unified coupling
$g_1=g_2=g_3=g_\GUT$. However this requirement seems to be
phenomenologically unacceptable: the strong coupling $g_3$ is much
bigger than the electroweak couplings $g_2$ and $g_1$ that are also
different between themselves.  The key point in attempting a unification
of the coupling ``constants'' is the observation that they are, in
fact, not constant. The couplings  evolve with energy, they
``run''. The values of the renormalised couplings depend on the energy
scale at which they are measured through the renormalisation group equations
(RGEs). Georgi, Quinn and Weinberg \cite{Georgi:1974yf} realised that the equality of the gauge
couplings applies only at a high scale $M_\GUT$ where, possibly, but not necessarily, 
a new "grand unified" symmetry (like  $SU(5)$, for instance) sets in. 
The evolution of the couplings with energy is regulated by the equations of the 
renormalisation group (RGE): 

\bea
\label{gaugeRGE}
\Frac{d \alpha_i}{d \log \mu^2} = \beta_i \alpha_i^2 + O(\alpha_i^3),
\eea
where $\alpha_i = g_i^2/ (4 \pi)$ and $i=1,2,3$ refers to the $U(1)$, $SU(2)$
and $SU(3)$ gauge couplings. The coefficients $\beta_i$ receive contributions 
from vector-boson, fermion and scalar loops shown in figure \ref{fig:beta}.
These coefficients are obtained from the 1 loop renormalised gauge
couplings,
\bea
\label{betagen}
\beta_i= -\Frac{1}{4 \pi} \left[\frac{11}{3}~C_2(G_i) - \frac{2}{3} \sum_f
~T(R_f) -\frac{1}{3} \sum_S ~T(R_s)\right]
\eea 
with $C_2(G_i)=N$ the eigenvalue of Casimir operator of the group $SU(N)$,
and $T(R_f)=T(R_s)= 1/2$ for fermions and scalars in the fundamental 
representation. The sums is extended over all fermions and scalars in the 
representations $R_f$ and $R_s$. 

 For the particle content of the SM, the $\beta$ coefficients read:
\bea
\beta_3&=& -\Frac{1}{4 \pi} \left[\frac{11}{3}\cdot 3 - \frac{2}{3} \cdot 
4~n_g \cdot \frac{1}{2}\right] =  -\Frac{1}{4 \pi}\left[ 11 -4\right], \nn \\
\beta_2&=& -\Frac{1}{4 \pi} \left[\frac{11}{3}\cdot 2 - \frac{2}{3} \cdot 
4~n_g \cdot \frac{1}{2} -\frac{1}{3} \cdot n_H \cdot \frac{1}{2}\right] =
-\Frac{1}{4 \pi} \left[\frac{22}{3} - 4  -\frac{1}{6}\right], \nn \\
\beta_1&=& -\Frac{1}{4 \pi} \frac{3}{5}\left[- \frac{2}{3} \cdot 
\frac{10}{3}\cdot n_g -\frac{1}{3} \cdot n_H\cdot \frac{1}{2} \right]
= -\Frac{1}{4 \pi} \left[-4 -\frac{1}{10}\right],
\label{betai}
\eea 
with $n_g=3$ the number of generations and $n_H=1$ the number of Higgs 
doublets. Care must be taken in evaluating the $\beta_1$ coefficient of
$U(1)$ hypercharge.  Obviously its value depends on the normalisation one chooses for the hypercharge generator
(indeed, hypercharge is related to the electric charge Q and the isospin  $T_3$ by the relation
$Q = T_3 + aY$, with $a$ being the normalisation factor of the hypercharge generator Y). Asking for a unifying
gauge symmetry group G embedding the SM  to set in at the scale $M_\GUT$ at which the SM couplings obtain a common value
implies that {\it all} the SM generators are normalised in the same way. At this point the hypercharge normalisation is no longer arbitrary and the coefficient $\frac{3}{5}$ appearing in $\beta_1$ is readily explained (the interested reader is invited to explicitly 
derive this result , for instance considering one fermion family of the SM, computing $\tr(T)^2$ over such fermions and then imposing that $\tr(T)^2$=$\tr(Y)^2$).

\eq{gaugeRGE} can be easily integrated and for a superlarge scale $M$ we obtain 
\bea
\label{solalpha}
\Frac{1}{\alpha_i(Q^2)} = \Frac{1}{\alpha_i(M^2)} + \beta_i~ \log
\Frac{M^2}{Q^2} \nn \\ \alpha_i(Q^2) = {\alpha_i(M^2) \over \left(1
+\beta_i \alpha_i(M^2) \log \Frac{M^2}{Q^2}\right) } 
\eea 

This result is very encouraging because we can see that both
$\alpha_3$ and $\alpha_2$ decrease with increasing energies because
$\beta_3$ and $\beta_2$ are negative. Moreover $\alpha_3$ decreases
more rapidly because $|\beta_3|>|\beta_2|$. Finally $\beta_1$ is
positive and hence $\alpha_1$ increases. Now we can ask whether
starting from the measured values of the gauge couplings at low
energies and using the $\beta_i$ parameters of the SM in \eq{betai}
there is a scale, $M_\GUT$, where the three couplings meet. To do this
we have to solve the equations: 
\bea 
\Frac{1}{\alpha_3(\mu^2)} &=&
\Frac{1}{\alpha_\GUT} + \beta_3~ \log \Frac{M_\GUT^2}{\mu^2} \nn \\
\Frac{1}{\alpha_2(\mu^2)} &=& \Frac{\sin^2 \theta_{\rm
W}(\mu^2)}{\alpha_\elm(\mu^2)} = \Frac{1}{\alpha_\GUT} + \beta_2~ \log
\Frac{M_\GUT^2}{\mu^2} \nn \\ \Frac{1}{\alpha_1(\mu^2)} &=&
\frac{3}{5}~\Frac{\cos^2 \theta_{\rm W}(\mu^2)} {\alpha_\elm(\mu^2)}=
\Frac{1}{\alpha_\GUT} + \beta_1~ \log \Frac{M_\GUT^2}{\mu^2} 
\label{SMeqs}
\eea
Now, we can use $\alpha_3$ and $\alpha_\elm$ to determine $M_\GUT$
and then use the remaining equation to ``predict'' $\sin^2 \theta_{\rm
W}$: 
\bea 
\label{SMres}
\Frac{3}{5~\alpha_\elm(\mu^2)}-\Frac{8}{5~\alpha_3(\mu^2)} =
\Frac{67}{20 \pi} ~\log \Frac{M_\GUT^2}{\mu^2} \nn \\ \sin^2
\theta_{\rm W}(\mu^2) = \frac{3}{8}\left[1- \frac{109}{36 \pi}
\alpha_\elm \log\Frac{M_\GUT^2}{\mu^2}\right].  
\eea 
The result is astonishing (we say this without any exaggeration!):  
starting from the measured values of
$\alpha_3$ and $\alpha_\elm$ we find that all three gauge couplings
would unify at a scale $M_\GUT\simeq 2 \times 10^{15}$ GeV if $\sin^2
\theta_{\rm W}\simeq 0.21$. This is remarkably close to the experimental 
value of $\sin^2 \theta_{\rm W}$ and this constitutes a major triumph of the
grand unification idea and the strategy we adopted to implement it. 
Let us finally comment that in deriving these results we have used the 
so-called step approximation and one loop RGE equations. The step approximation 
consists in using the beta parameters of the SM (the three of them different) 
all the way from $M_W$ to $M_\GUT$ where the beta parameter would change 
with a step function to a
common beta parameter corresponding to $SU(5)$. In reality the 
$\beta$-function transitions from $\mu \ll M_\GUT$ to $\mu \gg M_\GUT$
are smooth ones that take into account the different threshold when new 
particles enter the RGE evolution. This effects can be included using mass dependent beta functions \cite{Antoniadis:1981gh}. Similarly, we have only 
used one loop 
RGE equations although two loop RGE equations are also available.
The inclusion of these additional refinements in our RGEs would not improve
substantially the agreement with the experimental results. 

\subsubsection{The minimal SU(5) model of Georgi and Glashow}
As has been discussed in the introduction, the SM gauge group has a rank four and
the simple groups which contain complex representations of rank four are just 
$SU(3) \times SU(3)$ and $SU(5)$. Georgi and Glashow have chosen the $SU(5)$ where
a single gauge coupling constant is manifestly incorporated. Further, the fermions of the 
Standard Model can be 
arranged in terms of the fundamental ${\bf \bar{5}}$ and the anti-symmetric 
${\bf 10}$ representation of the SU(5) \cite{Langacker:1980js}. 
The appropriate particle assignments
in these two representations are : 
\bea
{\bf \bar 5} = \left(\matrix{d^c\cr d^c \cr d^c \cr \nu_e \cr
e^-}\right)_L &{\bf 10} = \left(
\matrix{0  & u^c & u^c & u & d \cr
-u^c & 0 & u^c & u & d \cr
-u^c & -u^c & 0 & u & d \cr 
-u & -u &-u  & 0 & e^c \cr
-d &-d  &-d  & -e^c & 0} \right)_L,
\eea
where ${\bf \bar 5} = (\bar 3,1,-\frac{2}{3}) + (1,2,-1)$ and ${\bf 10} =
(3,2,\frac{1}{3}) + (\bar 3,1,\frac{4}{3}) + (1,1,-2)$ under $SU(3)\times
SU(2) \times U(1)$ (here, we consider Y normalised as $Q=T_3+Y$, for simplicity).
 It is easy to check that this combination of
the representations is anomaly free. The gauge theory of SU(5) contains 
24 gauge bosons. They are decomposed in terms of the standard model 
gauge group SU(3) $\times$ SU(2) $\times$ U(1) as : 
\beq
{\bf
24 = (8,1) + (1,3) + (1,1) + (3,2) + (\bar{3},2)}
\eeq 
The first component represents the gluon fields ($G$) mediating the colour, 
the second one corresponds to the Standard Model $SU(2)$ mediators ($W$) and 
the third component corresponds to the $U(1)$ mediator ($B$). The fourth and 
fifth components carry both colour as well as the $SU(2)$ indices and are
called the $X$ and $Y$ gauge bosons. Schematically, the gauge bosons can be represented 
in terms of the $5 \times 5$ matrix as: 
\beq
V= \bmat{ccccc} 
&&&X_1 & Y_1 \\
\multicolumn{3}{c}{\left[\Frac{(G-2B)}{\sqrt{30}}\right]^\alpha_\beta}&X_2&Y_2 \\
&&&X_3 &Y_3 \\
X_1&X_2 &X_3&\Frac{W^3}{\sqrt{2}} + \Frac{3B}{\sqrt{30}} & W^+ \\
Y_1&Y_2 &Y_3& W^-& -\Frac{W^3}{\sqrt{2}} + \Frac{3B}{\sqrt{30}}
\emat
\eeq
The particle spectrum is completed by the Higgs particles required to give 
masses to fermions as well as to break the GUT symmetry. To begin with, 
 let us study the fermion masses in the 
prototype $SU(5)$. Given that fermions are in ${\bf \overline{5}}$ and ${\bf 10}$
representations, after some simple algebra we conclude that the scalars that 
can form Yukawa couplings are
\bea
\label{1010}
{\bf 10 \times 10} &=&{\bf \overline{5} + \overline{45} + 50}\\
\label{105}
{\bf 10 \times \overline{5}}&=& {\bf 5 + 45 }
\eea 
From the above, we see that we need at least two Higgs representations 
transforming as the fundamental (${\bf 5}_H$) 
and the anti-fundamental (${\bf \overline{5}}_H$) to reproduce the fermion 
Yukawa couplings. The  corresponding Yukawa terms read: : 
\beq
\mathcal{L}^{\rm yuk}_{SU(5)} = h^u_{ij} {\bf 10_i 10_j \overline{5}}_H + h^d_{ij} 
{\bf 10_i \overline{5}_j 5}_H 
\eeq 
Though the Yukawa couplings written above are quite simple, they do not stand
the test of phenomenological constraints, as we will see later. 
 A Higgs in the adjoint representation 
can be used to break $SU(5)$ to the diagonal subgroup of the
Standard Model. Denoting the adjoint as 
$\Phi = \sum_{i=1}^{24} \lambda_i/\sqrt{2} \phi_i$ , where $\lambda_i$
are generators of the SU(5), the most general renormalisable scalar potential is 
\beq
\label{24break}
V(\Phi) = -{1 \over 2} m_1^2 \tr(\Phi^2) + {1 \over 4} a (\tr(\Phi^2))^2 
+ {1 \over 2} b \tr(\Phi^4) + {1 \over 3} c \tr(\Phi^3).
\eeq 
However, to simplify the potential, we impose a discrete symmetry 
($\Phi~ \leftrightarrow~-\Phi$) which sets $c$ to zero. The remaining
potential has the following minimum when $b >0$ and $a>-7/15~b$: 
\beq
<0|\Phi|0> = \bmat{ccccc} 
v & 0&0&0&0\\
0 & v&0&0&0\\
0 & 0&v&0&0\\
0 & 0&0& -3/2 v&0\\
0 & 0&0& 0&-3/2 v
\emat, 
\eeq
with $v$ determined by 
\beq
\label{m12}
m_1^2 = {15 \over 2} a v^2 + {7 \over 2} b v^2 
\eeq 
This completes the proto-GUT model containing all the required
features : gauge coupling unification, representations accomodating
all SM fermions, yukawa couplings for fermion masses, gauge symmetry
breaking down to SM gauge group. However as usual  in real life, 
things are a bit more complicated as we will see now. 

\subsubsection{Distinctive Features of GUTs and Problems 
in building a realistic Model}
\textit{(i)~Fermion Masses}\\
In the previous section, we have seen that in the typical
prototype SU(5) model, the fermions attain their masses
through a $5_H$ and $\bar{5}_H$ of Higgses. A simple consequence
of this approach is that there is an equality of $Y_d^T ~=~ Y_e$ 
at the GUT scale; which would mean equal charged lepton and down
quark masses at the $M_{\rm GUT}$ scale. Schematically, these are
given as:
\bea 
m_e (M_\GUT) &=& m_d (M_\GUT) \\
m_\mu (M_\GUT) &=& m_s (M_\GUT) \\
m_\tau (M_\GUT) &=& m_b (M_\GUT). 
\eea
We would have to verify these prediction by running the Yukawa
couplings from the SM to the GUT scale. Let us have a more closer 
look at these RGEs. For the bottom mass and the $\tau$ Yukawa 
these are given by \cite{Nanopoulos:1978hh,Nanopoulos:1981mv} : 
\bea
\label{rgeyukawa}
{d \log Y_b (\mu) \over d \log \mu} &=&  - 3~C_3^b \Frac{\alpha_3(\mu)}{4 \pi}
- 3~C_2^b \Frac{\alpha_2(\mu)}{4 \pi} - 3~C_1^b \Frac{\alpha_1(\mu)}{4 \pi}\\
{d \log Y_\tau (\mu) \over d \log \mu} &=&  - 3~C_3^\tau \Frac{\alpha_3(\mu)}{4 \pi}
- 3~C_2^\tau \Frac{\alpha_2(\mu)}{4 \pi} - 3~C_1^\tau \Frac{\alpha_1(\mu)}{4 \pi}
\eea
with $C_3^b=\frac{4}{3}$, $C_2^b=\frac{3}{4}$, $C_1^b=-\frac{1}{30}$,
$C_3^\tau=0$, $C_2^\tau=\frac{3}{4}$, $C_1^\tau=-\frac{3}{10}$.
Knowing the scale dependence of the gauge couplings, \eq{solalpha}, we can
integrate this equation, neglecting the effects of other Yukawa except
the top Yukawa in the RHS of the above equations. Taking the masses 
to be equal at $M_\GUT$, we obtain
\beq
{m_b (M_Z) \over m_\tau (M_Z)} \approx 
E_t^{-1/2} \left[ {\alpha_3 (M_Z) \over \alpha_3(M_\GUT)} \right]^ {-3 C_3 \over4 \pi \beta_3} \approx E_t^{-1/2} \left[ {\alpha_3 (M_Z) \over \alpha_3(M_\GUT)} \right]^ {4\over7} ,
\eeq
where $E_t = {\rm Exp}[{1 \over 2 \pi} \int_{M_Z}^{M_{\rm GUT}} Y_t(t)dt]$.
Taking these masses at the weak scale, we obtain a rough relation
\beq
{m_b (M_W) \over m_\tau (M_W)} \approx 3,
\eeq
which is quite in agreement with the experimental values. 
This can be considered as
one of the major predictions of the $SU(5)$ grand unification. However,
there is a caveat. If we extend similar analysis to the first two 
generations we end-up with relations :
\beq
{m_\mu \over m_e} = {m_s \over m_d},
\eeq
which don't hold water at weak scale. The question remains how can one
modify the \textit{bad} relations of the first two generations while
keeping the \textit{good} relation of the third generation intact.
Georgi and Jarlskog solved this puzzle \cite{Georgi:1979df} with a simple 
trick using an additional Higgs representation. As we have seen in 
Eq.~(\ref{105}), the ${\bf 10}$ and ${\bf \bar{5}}$ can couple to a 
${\bf 45}$ in addition to the ${\bf 5}$ representation.
The ${\bf 45}$ is a completely anti-symmetric representation and a texture can
be chosen such that the bad relations can be modified keeping the 
good relation intact.

\noindent
\textit{(ii)~Doublet-Triplet Splitting}\\
We have seen that in minimal
$SU(5)$ we need at least two Higgs representations transforming as
${\bf 5_H}$ and ${\bf \bar 5_H}$ to accommodate the fermion Yukawa
couplings. The ${\bf 5}$ representation of $SU(5)$ contains a $({\bf
3}, {\bf 1})$ and $({\bf 1}, {\bf 2})$ under $(SU(3)_C,SU(2)_{\rm
L})$. So, the ${\bf 5_H}$ and ${\bf \bar 5_H}$ representations contain
the required Higgs doublets that breaks the electroweak symmetry at low
energies, but they contain also colour triplets, extremely dangerous,
as we will see later,  because they mediate a fast proton decay if their mass
is much lower than the GUT scale. The doublet-triplet
splitting problem is then the question of how one can enforce the mass of the  Higgs
doublet to remain at the  electroweak scale, while the  Higgs triplet mass should jump 
to $M_\GUT$ \cite{Buras:1977yy}.

The $SU(5)$ symmetry is broken by the VEV of the Higgs $\Phi$ sitting in the adjoint representation,
as we saw in \eq{24break}. At the electroweak scale we need a second 
breaking step, $SU(3)_C \times SU(2)_{\rm L} \times U(1)_Y \to 
SU(3)_C \times U(1)_\elm$, which is obtained by the potential
\beq
\label{higgsweak}
V(H) = - \Frac{\mu^2}{2} {\bf 5_H}^\dagger {\bf 5_H} + \Frac{\lambda}{4} 
\left({\bf 5_H}^\dagger {\bf 5_H}\right)^2,
\eeq
with a VEV
\beq
\label{mu5}
\langle {\bf 5_H} \rangle =  \bmat{c} 0\\ 0\\ 0\\ 0\\ \frac{v_0}{\sqrt{2}} 
\emat, \qquad v_0^2 = \Frac{2 \mu^2}{\lambda}.
\eeq
However, the potential $V = V(\Phi) + V(H)$ does not give rise to a viable 
model. Clearly both the Higgs doublet and triplet fields remain with masses
at the $M_W$ scale which is catastrophic for proton decay.

This problem can find a solution if we consider also the following 
$\Phi$--${\bf 5_H}$ cross terms which are allowed by $SU(5)$
\bea
V(\Phi, H) = \alpha {\bf 5_H}^\dagger {\bf 5_H} \tr(\Phi^2) +
\beta {\bf 5_H}^\dagger \Phi^2 {\bf 5_H}.
\eea
Notice that even if one does not introduce the above mixed term at the tree level, one
expects it to arise at higher order given that the underlying $SU(5)$ symmetry does not prevent its appearance. 

Let's turn to the minimisation of the full potential $V = V(\Phi) + 
V(H) + V(\Phi, H)$. Now that $\Phi$ and ${\bf 5_H}$ are coupled, 
$\langle\Phi\rangle$ may also break $SU(2)_{\rm L}$ whilst $SU(3)_C$ must 
be rigorously unbroken. Therefore, we look for solutions with
$\langle\Phi\rangle = {\rm Diag.} \left( v,v,v,\left(-\frac{3}{2}-
\frac{\varepsilon}{2}\right) v ,\left(-\frac{3}{2}-
\frac{\varepsilon}{2}\right) v \right)$. In the absence of $\Phi$--${\bf 5_H}$ mixing,
i.e. $\alpha=\beta=0$, $\varepsilon$ must vanish. The solution with this 
properties has
\bea
\varepsilon = \Frac{3}{20}~\Frac{\beta v_0^2}{b v^2} + 
O\left(\Frac{v_0^4}{v^4}\right).
\eea
As $v\sim O\left(M_\GUT\right)$ and $v_0\sim O\left(M_W\right)$, we have
that the breaking of $SU(2)$ due to $\langle\Phi\rangle$ is much
smaller than that due to $\langle H \rangle$. Now, the expressions for $m_1^2$
(corresponding to \eq{m12}) and $\mu_5^2$ (corresponding to \eq{mu5}) 
are more complicated
\beq
\label{m12fin}
m_1^2 = {15 \over 2} a v^2 + {7 \over 2}{15} b v^2 + + \alpha v_0^2  + 
{9 \over 30} \beta v_0^2 
\eeq
and 
\beq
\label{mufin}
\mu^2 = {1 \over 2} \lambda v_0^2 + {15} \alpha v^2 + {9 \over 2} \beta v^2
- 3 \epsilon \beta v^2 . 
\eeq
We can see that \eq{m12fin} shows only a very small modification from
\eq{m12} being $v_0 \ll v$. What is very worrying is the result of
\eq{mufin}. Since the parameter in the Lagrangian $\mu \sim O(M_W)$,
i.e. $\mu \ll v$, the natural thing to happen would be that $v_0$
takes a value order $v$ to reduce the right-hand side of this equation
(remember that in this equation $v$ and $v_0$ are our unknowns). In
other words, without putting any particular constraint on $\alpha$ and
$\beta$, we would expect $v_0\sim O(v)$. However, this would completely
spoil the hierarchy between $M_W$ and $M_\GUT$. If we want to avoid
such a disaster, we have to fine-tune $\alpha$ and $\beta$ to one part
in $\left({v^2 \over v_0^2}\right) \sim 10^{24}$!!! Even more, such an
adjustment must be repeated at every order in perturbation theory, since
radiative correction will displace $\alpha$ and $\beta$ for more than
one part in $10^{24}$.  This is our first glimpse in the so-called
hierarchy problem.

\noindent
\textit{(iii).~Nucleon Decay }\\
As we saw in the previous section, perhaps the most prominent feature of 
GUT theories is the non-conservation of baryon (and lepton) number. In the 
minimal $SU(5)$ model this is due to the tree-level exchange of $X$ and $Y$
gauge bosons in the adjoint of $SU(5)$ with $({\bf 3},{\bf 2})$ quantum 
numbers  under $SU(3)\times SU(2)_{\rm L}$.  The couplings of these gauge
bosons to fermions are
\bea
{\cal L}_{X} = \sqrt{\frac{1}{2}} g X^a_{\mu \alpha} \left[\epsilon^{\alpha
    \beta\gamma}~ \overline u^c_\gamma \gamma^\mu q_{\beta a} + \epsilon^{a b}
\left(\overline q_{\alpha b} \gamma^\mu e^+ - \overline l_{b} \gamma^\mu
  d^c_\alpha \right) \right],
\label{Xcouplings}
\eea
where $\epsilon_{\alpha \beta \gamma}$ and $\epsilon_{a b}$ are the totally
antisymmetric tensors, $(\alpha, \beta, \gamma)$ are $SU(3)$ and $a,b$ are 
$SU(2)_{\rm L}$ indices. Thus the $SU(2)_{\rm L}$ doublets are
\bea
X_{\alpha a} = \left( X_\alpha, Y_\alpha\right), \quad 
q_{\alpha a} = \left( u_\alpha, u_\alpha\right), \quad 
l_{a} = \left( \nu_e, e\right).
\eea

\begin{figure}
\begin{center}
\includegraphics[width=8cm]{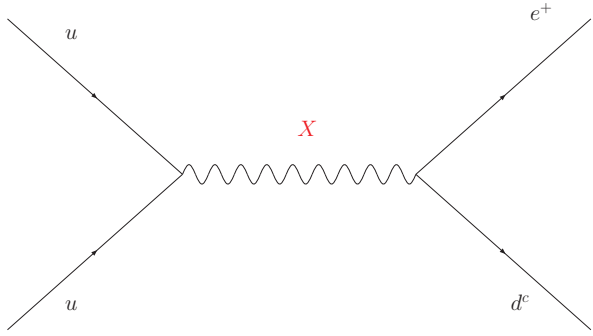}
\end{center}
\caption{Baryon number violating couplings of the $X$ boson.}
\label{dim6}
\end{figure}

We can see in \eq{Xcouplings} that the $(X,Y)$ bosons have two couplings to
fermions with different baryon numbers. They have a leptoquark coupling with
$B_1= -1/3$ and a diquark couplings with $B_2=2/3$. Therefore, through the
coupling of an $X$ boson we can change a $B=-1/3$ channel into a $B=2/3$
channel and a $\Delta B = 1$ process occurs at tree level as shown in Figure 
\ref{dim6}.
 If the mass of the 
$X$ boson, $M_X$ is large compared to the other masses, we can obtain the 
effective four-fermion interactions \cite{Weinberg:1979sa,Weinberg:1980bf,Wilczek:1979hc,Weldon:1980gi}
\bea
{\cal L}_{\Delta B = 1}^{\rm eff} = \Frac{g^2}{2 M_X^2}~\epsilon^{\alpha
    \beta\gamma} \epsilon^{a b} \left(\overline u^c_\gamma \gamma^\mu q_{\beta
      a} \right) \left(\overline d^c_\alpha \gamma_\mu l_{b} 
+ \overline e^+ \gamma_\mu q_{\alpha b}\right).
\eea
From this effective Lagrangian we can see that although baryon number is
violated, $(B-L)$ is still conserved, thus the decay $p \to e^+ \pi^0$ is
allowed but a decay $n \to e^- \pi^+$ is forbidden. From this effective
Lagrangian we can obtain the proton decay rate and we have $\Gamma_p \sim
10^{-3} m_p^5/M_X^4$ and therefore from the present bound on the proton
lifetime $\tau_p \geq 10^{33}$ yrs, we have that $M_X \geq 4 \times 10^{15}$ 
GeV. From this simple dimensional estimate of the proton decay lifetime, we can already see that 
the minimal non-supersymmetric $SU(5)$ can easily get into trouble because of matter stability.
Indeed, performing an accurate analysis of proton decay, even taking into account the relevant theoretical uncertainty
factors, like the evaluation of the hadronic matrix element,  one can safely conclude that
the minimal grand unified extension of the SM is ruled out because of the exceedingly high matter instability.
Analogously, the high precision achieved on electroweak observables (in particular 
thanks to LEP physics) allows us to further exclude the minimal SU(5) model: indeed, the low-energy
quantity one can {\it predict} solving the RGE's for the gauge coupling evolution (be it the electroweak angle $\theta_W$,
or the strong coupling $\alpha_s$) exhibits a large discrepancy with respect 
to its measured value. The precise $SU(5)$ prediction for $\sin^2 \theta_W$ is
\cite{marciano:1983sv}:
\bea
\sin^2 \theta_W (M_W) = 0.214 ^{+0.004}_{-0.003},
\eea
while the experimental value obtained from LEP data is:
\bea
\sin^2 \theta_W (M_W) = 0.23108\pm0.00005,
\eea
and both values only agree at 5 standard deviations.  

The fate  of the minimal $SU(5)$ should not induce the reader to conclude
``tout-court'' that non-supersymmetric grand unification
is killed by proton decay and $sin^2\theta_W$. Once one abandons the minimality criterion, for instance enlarging the Higgs spectrum or changing the grand unified gauge group, it is possible to rescue some GUT models. The price to pay for it 
is that we lose the simplicity and predictivity of minimal SU(5) ending up in more and more 
complicated grand unified realisations.

\subsection{Supersymmetric grand unification}

\subsubsection{The hierarchy problem and supersymmetry}
The Standard Model as a $SU(3)\otimes SU(2)\otimes U(1)$ gauge theory
with three generations of quarks and leptons and a Higgs doublet
provides an accurate description of all known experimental results.
However, as we have discussed, the SM cannot be the final theory, 
and instead we consider the SM as a low energy effective 
theory of some more fundamental theory at higher energies. Typically 
we have a Grand Unification (GUT) Scale around $10^{16}$ GeV where the
strong and electroweak interactions unify in a simple group like
$SU(5)$ or $SO(10)$ \cite{Pati:1974yy,Georgi:1974sy} and the Plank scale
of $10^{19}$ GeV where these 
gauge interactions unify with gravity. The presence of such different
scales in our theory gives rise to the so--called {\it hierarchy
problem} (see a nice discussion in \cite{Drees:1996ca}). This problem refers
to the difficulty to stabilise the large gap between the electroweak
scale and the GUT or Plank scales under radiative  corrections. Such
difficulty arises from a general property of the scalar fields in a gauge theory, namely 
 their  tendency of scalar  to get their
masses in the neighbourhood of the largest available energy  scale in the theory.
In the previous section, when dealing with the scalar potential of the minimal $SU(5)$ model, 
we have directly witnessed the existence of such problem. From such a particular example, let us 
move to more general considerations about what distinguishes the behaviour of scalar fields from that
of fermion and vector fields in gauge theories. 

\begin{figure}
\begin{center}
\includegraphics[width=8cm]{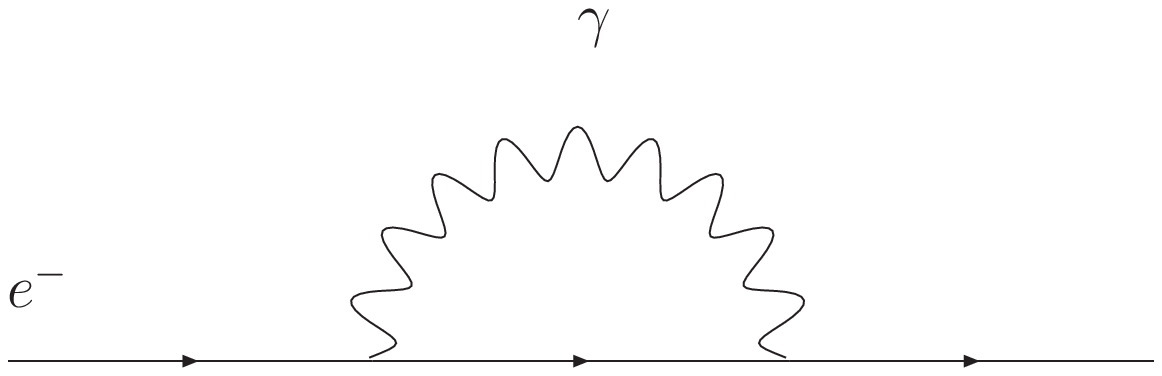}
\\[1cm]
\includegraphics[width=8.2cm]{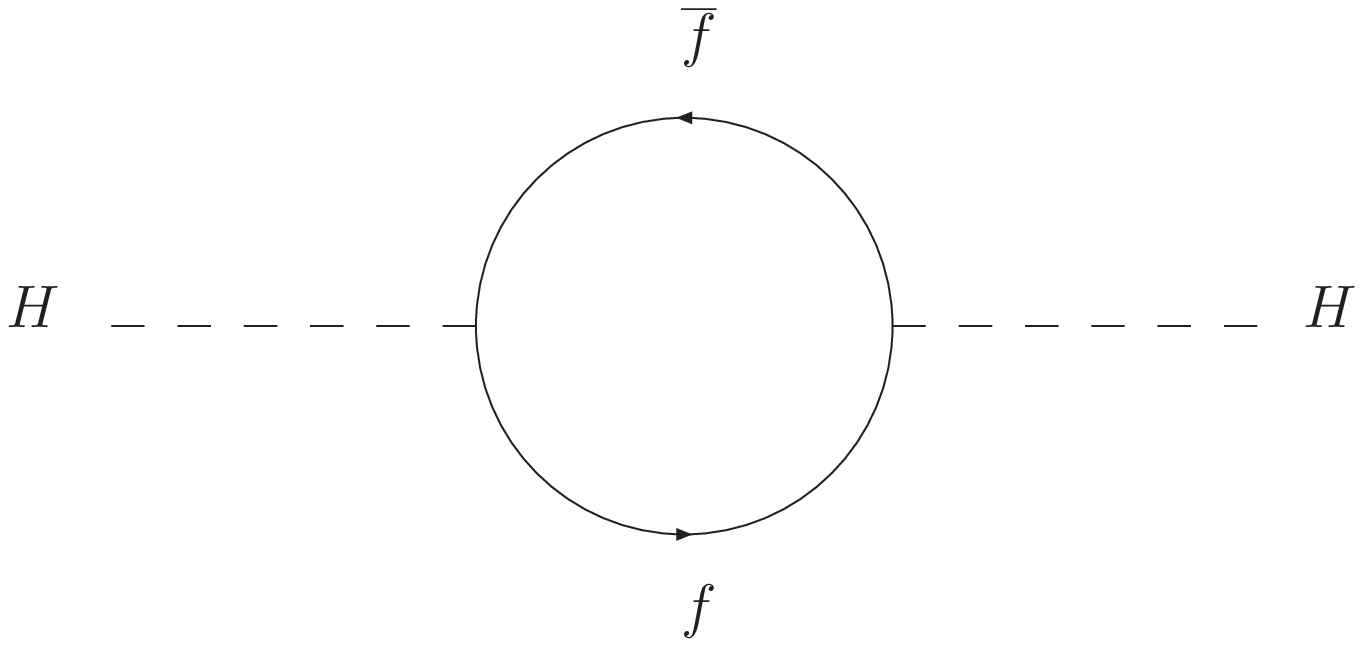}
\end{center}
\caption{One loop correction to fermion and scalar masses}
\label{FeynHe}
\end{figure}

To understand this problem let us compare the one loop corrections to
the electron mass and the Higgs mass. These one loop corrections are
given by the diagrams in Fig. \ref{FeynHe}.  
The self-energy contribution to the electron mass can be calculated
from this diagram to be, 
\bea
\delta m_e = 2 \Frac{\alpha_{em}}{\pi} m_e \log \Frac{\Lambda}{m_e}
\eea
and it is logarithmically divergent. Here we have
regulated the integral with an ultraviolet cutoff $\Lambda$. However, 
it is important to notice that this correction is proportional to the 
electron mass itself. This can be understood in terms of symmetry. In
the limit where $m_e \to 0$, our theory acquires a new chiral symmetry
where right-handed and left-handed electrons are decoupled. Were such a symmetry exact,
 the one loop corrections to the mass would
have to vanish. This chiral symmetry is only broken by the electron
mass itself and therefore any loop correction breaking this symmetry
must be proportional to $m_e$, the only source of chiral symmetry
breaking in the theory. This has important implications.  If we
replace the cutoff $\Lambda$ by the largest possible scale, the Planck
mass we get,     
\bea
\delta m_e = 2 \Frac{\alpha_{em}}{\pi} m_e \log \Frac{M_{Plank}}{m_e}
\simeq 0.24~ m_e,
\eea
which is only a small correction to the electron mass.

Analogously, for the gauge vector bosons there is the gauge symmetry itself which
constitutes the ``natural barrier'' preventing their masses to become arbitrarily
large. Indeed, if a vector boson V is associated to the generator of a certain 
symmetry G, as long as G is unbroken the vector V has to remain massless. Its mass
will be of the order of the scale at which the symmetry G is (spontaneously) broken. Hence,
once again, we have a symmetry protecting the mass of vector bosons. 

On the other hand, the situation is very different in the case of the
Higgs boson, 
\bea
\delta m_H^2 (f) = - 2 N_f \Frac{|\lambda_f|^2}{16 \pi^2} [
\Lambda^2 -2 m_f^2 \ln \Frac{\Lambda}{m_f} + \dots].
\label{Hfer}
\eea
But, in this case, the one loop contribution is quadratically
divergent !!. This is due to the fact that no symmetry protects the
scalar mass and in the limit $m_H^2 \to 0$ the symmetry of our model
is not increased. The combination $H H^\dagger$ is always neutral
under any symmetry independently of the charges of the field $H$.
So, the scalar mass should naturally be of the order of the largest
scale of the theory, as either at tree level or at loop level this
scale feeds into the scalar mass. 

So, if now we repeat the exercise we made with the electron mass and
replace the cutoff by the Plank mass, we obtain $\delta m_H^2\
\simeq\ 10^{30}\mbox{ GeV}^2$. In fact we could cancel these large
correction with a bare mass of the same order and opposite sign.
However, these two contributions should cancel with a precision of
one part in $10^{26}$ and even then we should worry about the two loop
contribution and so on. 
This is the so-called hierarchy problem and Supersymmetry constitutes so far
the most interesting answer to it (later on, we'll briefly comment on the existence of other
approaches tackling the hierarchy problem, although, in our view, not as effectively as 
low-energy supersymmetry does). 
 
As we have seen in the previous section, Supersymmetry associates a
fermion with every scalar in the theory with, in principle, identical
masses and gauge quantum numbers. Therefore, in a Supersymmetric
theory we would have a new contribution to the Higgs mass at one loop.
\begin{figure}
\begin{center}
\includegraphics[width=8.4cm]{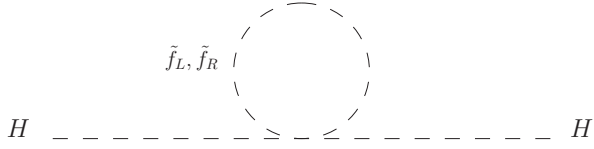}
\end{center}
\caption{Additional Supersymmetric contribution to the scalar mass.}
\label{newsusy}
\end{figure}
Now this graph gives a contribution to the Higgs mass as,
\bea
\delta m_H^2 (\tilde{f}) = - 2 N_{\tilde{f}}
\Frac{\lambda_{\tilde{f}}}{16 \pi^2} 
[\Lambda^2 -\ 2 m_{\tilde{f}}^2 \ln \Frac{\Lambda}{m_{\tilde{f}}}\ + \dots]
\label{Hsfer}
\eea
If we compare Eqs.~(\ref{Hfer}) and (\ref{Hsfer}) we see that
with $N_f = N_{\tilde{f}}$, $|\lambda_f|^2= - \lambda_{\tilde{f}}$ and 
$m_f= m_{\tilde{f}}$ we obtain a total correction $\delta m_H^2 (f)+
\delta m_H^2 (\tilde{f}) = 0$ !!.
This means we need a symmetry that associates a bosonic partner to
every fermion with equal mass and related couplings and this symmetry
is {\bf Supersymmetry}. 

Still, we have not found scalars exactly degenerate with the SM
fermions in our experiments. In fact, it would have been very easy to 
find a scalar partner of the electron if it existed. Thus,
Supersymmetry can not be an exact symmetry of nature, it must be
broken. Fortunately, we can break Supersymmetry while at the same time
preserving to an acceptable extent the Supersymmetric solution of the
hierarchy problem. To do that, we want to ensure the cancellation
of quadratic divergences and comparing \eq{Hfer} and \eq{Hsfer} we can
see that we must still require equal number of scalar and fermionic
degrees of freedom, $N_f = N_{\tilde{f}}$, and supersymmetric
dimensionless couplings $|\lambda_f|^2= - \lambda_{\tilde{f}}$.
Supersymmetry can be broken only in couplings with positive mass
dimension, as for instance the masses. This is called soft breaking
\cite{Girardello:1981wz}.
Now if we take $m_{\tilde{f}}^2 = m_{f}^2 + \delta^2$ we obtain a
correction to the Higgs mass,
\bea
\delta m_H^2 (f)+ \delta m_H^2 (\tilde{f}) \simeq 2 N_{f} 
\Frac{|\lambda_{f}|^2}{16 \pi^2}\ \delta^2\ \ln \Frac{\Lambda}{m_{\tilde{f}}} +
\dots 
\eea 
and this is only logarithmically divergent and proportional to the
mass difference between the fermion and its scalar partner. Still we
must require this correction to be smaller than the Higgs mass itself
(around the electroweak scale) implies that this mass difference, 
$\delta$, can not be too large, in fact $\delta \lsim 1
\mbox{ TeV}$. If Supersymmetry is the solution to the 
hierarchy problem it must be softly broken and the SUSY partners must
be roughly below 1 TeV. The rich SUSY phenomenology is thoroughly discussed in 
Marcela Carena and Carlos Wagner's lectures at this School 
\cite{Carena:2005unp}.

\subsection{Gauge coupling Unification in SUSY $SU(5)$}
The supersymmetric $SU(5)$ can be build analogously to the non-supersymmetric
$SU(5)$ with the ordinary fields replaced by superfields containing 
the SM field and its superpartner. 
What is relevant for our discussion on grand  unification at present  
is the effect of the
presence of new SUSY particles at 1 TeV in the evolution of the gauge
couplings. We saw in the previous section that the RGE equations in
the SM predict that the gauge couplings get very close at a large
scale $\simeq 2 \times 10^{15}$ GeV. Nevertheless this unification was
not perfect and, using the precise determination of the gauge
couplings at LEP we see that the SM couplings do not unify at
seven standard deviations. If we have new SUSY particles around 1 TeV, 
these RGE equations are modified. Using \eq{betagen}, it is straightforward to
obtain the new $\beta_i$ parameters in the MSSM. We have to take into 
account that for every gauge boson we have to add a fermion, called gaugino,
both in the adjoint representation. Therefore from gauge bosons and gauginos
we have 
\bea
\beta_i(V) = -\Frac{1}{4 \pi} \left[\frac{11}{3}~C_2(G_i) - \frac{2}{3} 
C_2(G_i)\right] = -\Frac{1}{4 \pi}~ 3~C_2(G_i).
\eea
While for every fermion we have a corresponding scalar partner in the same 
representation. Thus we have
\bea
\beta_i(F) = -\Frac{1}{4 \pi} \sum_F \left[- \frac{2}{3} 
~T(R_F) -\frac{1}{3} ~T(R_F)\right] = \Frac{1}{4 \pi}~ \sum_F T(R_F),
\eea
summed over all the chiral supermultiplet (fermion plus scalar) 
representations. Therefore the total $\beta_i$ coefficient in a 
supersymmetric model is 
\bea
\beta_i = -\Frac{1}{4 \pi} \left[3~C_2(G_i)  - \sum_F T(R_F)\right].
\eea
And for the MSSM
\bea
\beta_3&=& -\Frac{1}{4 \pi}\left[ 9 - 2~n_g \right] =-\Frac{3}{4 \pi} , \nn \\
\beta_2&=& -\Frac{1}{4 \pi} \left[6 - 2~n_g - \frac{1}{2}~n_H\right]= 
+\Frac{1}{4 \pi}, \nn \\
\beta_1&=& -\Frac{1}{4 \pi} \left[-\frac{10}{3}~n_g -\frac{1}{2}~n_H\right]=
+\Frac{11}{4 \pi} .
\label{betaiMSSM}
\eea 
From the comparison of \eq{betai} and \eq{betaiMSSM} we see that
the evolution of the gauge couplings is significantly modified.
We can easily calculate the grand unification scale and the ``predicted'' 
value of $\sin^2 \theta_{\rm W}$ as done in Eqs.~(\ref{SMeqs}) and 
(\ref{SMres}) and we obtain
\bea
\qquad M_\GUT^{\rm MSSM} = 1.5 \times 10^{16} \mbox{GeV}, &\qquad 
\sin^2 \theta_{\rm W} (M_Z) = 0.234,
\eea
which is remarkably close to the experimental vale $\sin^2 
\theta_{\rm W}^{\rm exp} (M_Z) = 0.23149 \pm 0.00017$. And we obtain easily
the grand unified coupling constant 
\begin{equation}
{5 \over 3} \alpha_1(M_\GUT) = \alpha_2(M_\GUT) = \alpha_3(M_\GUT) \approx {1 \over 24}
\end{equation}
In fact, the actual analysis, including
two loop RGEs and threshold effects predicts $\alpha_3(M_Z) = 0.129$ which
is slightly higher than the observed value (such discrepancy could be justified by the
presence of threshold effects when approaching the GUT scale in the running). 
 The couplings meet at the
value $M_X = 2 \times 10^{16}$ GeV \cite{Dimopoulos:1981yj,Marciano:1981un,Amaldi:1991zx}. The ``exact'' unification
of the gauge couplings within the MSSM may or may not be an accident. But
it provides enough reasons to consider supersymmetric standard models 
seriously  as it links supersymmetry and grand unification in an 
inseparable manner \cite{Mohapatra:1999vv}.  Let us know see how the 
other GUT features and problems which we have encountered earlier in
non-supersymmetric theories fare in the supersymmetric GUTs. 

\subsubsection{SUSY GUT predictions and problems}

\noindent
\textit{(i )~Doublet-Triplet Splitting}\\
As we saw in the non-supersymmetric case, a very accurate fine-tuning
in the parameters of the scalar potential was required to reproduce
the hierarchy between the electroweak and the GUT scale. In a
supersymmetric grand unified theory the problem is very similar. The
relevant terms in the superpotential are,
\beq
W = \alpha {\bf \overline 5_H} \Phi^2 {\bf 5_H} + \mu {\bf \overline 5_H} 
{\bf 5_H}.
\eeq
The breaking of $SU(5)$ in the $SU(3)\times SU(2)\times U(1)$ direction via
\beq
\langle \Phi \rangle = {2~m^\prime \over 3~\alpha}~ {\rm Diag.}~ 
\left(1,1,1,-\frac{3}{2},-\frac{3}{2}\right),
\eeq
leads to 
\beq
\label{spot23}
W = {\bf \overline 3_H} {\bf 3_H} \left( \mu + {2 \over 3} m^\prime \right) + 
{\bf \overline 2_H} {\bf 2_H} \left( \mu - m^\prime \right).
\eeq
Choosing $\mu = m^\prime$ (both them $\sim O(M_\GUT)$) renders the Higgs doublets massless. However,
although due to supersymmetry this equality is stable under radiative 
corrections, this extremely accurate adjustment is extremely unnatural.

There are several mechanisms in supersymmetric theories to render
doublet-triplet splitting natural. Here we will briefly discus the
``missing partner mechanism'' \cite{masiero:1982fe}. From \eq{spot23}
we see that if the direct mass term for the Higgses, $\mu$, was
absent the doublets would obtain super-heavy masses from the vacuum
expectation value of the adjoint Higgs $\Phi$. The strategy we will use 
to solve the doublet-triplet splitting problem is to introduce representations
that contain Higgs triplets but no doublets. We can choose the ${\bf 50}$ that
is decomposed under $SU(3) \times SU(2) $ as: 
\beq
{\bf
50 = (8,2) + (6,3) + (\bar 6,1) + (3,2) + (\bar{3},1) + (1,1) }
\eeq 
We need both the ${\bf 50}$ and ${\bf \bar 50}$ to get an anomaly-free 
model. In order to write mixing terms between ${\bf 5}$, ${\bf \bar 5}$ and
${\bf 50}$, ${\bf \bar 50}$ we need a field $\Sigma$ in the ${\bf 75}$ 
instead of the ${\bf 24}$ to break $SU(5)$. The relevant part of 
superpotential is then
\beq
\label{spotmiss}
W =  {M\over 2} \tr(\Sigma^2) + {a\over 3}~  \tr(\Sigma^3) + 
b~ {\bf 50}~ {\Sigma}~ {\bf 5_H} + c~ {\bf \bar 50}~ {\Sigma}~ {\bf \bar 5_H} 
+  \tilde M ~ {\bf \bar 50}~{\bf 50},
\eeq
where no mass term ${\bf \bar 5_H}~{\bf 5_H}$ is present. ${\Sigma}$ 
gets a VEV, $\langle {\Sigma}\rangle \sim \frac{M}{a}$, breaking $SU(5)$ to 
$SU(3)\times SU(2)\times U(1)$. The resulting $SU(3)\times SU(2)\times U(1)$
superpotential is,
\beq
\label{spotsu3}
W =  {\bf 50_3}~ {\frac{M~b}{a}}~ {H_3} + {\bf \bar 50_3}~ {\frac{M~c}{a}}~ 
{\bar H_3} +  \tilde M ~ {\bf \bar 50_3}~{\bf 50_3},
\eeq
with $H_3$ and ${\bf 50_3}$ the Higgs triplets in the ${\bf 5_H}$ and 
${\bf 50}$ representations respectively. Therefore the Higgs triplets get
a mass of the order of $M\sim\tilde M \sim M_\GUT$ and the Higgs doublets 
remain massless because there is no mass term for the doublets.
In this way we solve the doublet-triplet splitting problem without
unnatural fine-tuning of the parameters.

\noindent
\textit{(ii)~Proton Decay}\\
In the non-supersymmetric $SU(5)$ proton decay 
arises from four fermion operators, hence from operators of canonical dimension 6.
In addition to such dim=6 operators, in the supersymmetric case we encounter 
also dim=5 and even dim=4 operators leading to proton decay. 

\begin{figure}
\begin{center}
\includegraphics[width=8.4cm]{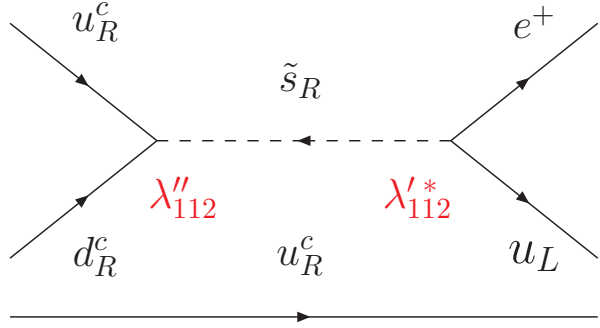}
\end{center}
\caption{Proton decay through R-parity violating couplings.}
\label{pdecay}
\end{figure}

Dimension 4 operators are not suppressed by any power of the GUT scale. 
In fact, these terms are gauge invariant and in principle are allowed to 
appear in the superpotential, 
\bea
W_{\Delta L =1} = \lambda^{ijk} L_i L_j e^c_{R k} + 
\lambda^{\prime\;ijk} L_i Q_j d^c_{R k} + \epsilon^i L_i H_2 \nonumber\\
W_{\Delta B =1} = \lambda^{\prime\prime\;ijk}\; u^c_{R i} d^c_{R j} d^c_{R k}
~~~~~~~~~~~~~~~
\label{RbreakW}
\eea
However, these terms violate baryon or lepton number by 1 unit. So,
these terms are very dangerous. Indeed, if $\lambda^{\prime}$ and 
$\lambda^{\prime\prime}$ are simultaneously present,  a very
fast proton decay arises through the diagram in Figure \ref{pdecay}.
Clearly, the major difference is that in the non-SUSY case the mediation of proton decay
occurs through the exchange of super-heavy (vector or scalar) bosons
whose masses are at the GUT scale. On the contrary, in Figure \ref{pdecay} the mediator
is a SUSY particle and, hence, at least if we insist in invoking low-energy SUSY to tackle
the hierarchy problem, its mass is at the electroweak scale instead of being at $M_\GUT$!
From the bounds to the decay $p^+ \to e^+ \pi^0$ we obtain 
$\lambda^{\prime\; *}_{112} \cdot \lambda^{\prime\prime}_{112} 
\leq 2 \times 10^{-27}$. Clearly this product is too small and it is
more natural to consider it as exactly zero. Other couplings from  
\eq{RbreakW} are not so stringently bounded but in general all of
them must be very small from phenomenological considerations (in particular, from 
FCNC constraints). 

One possibility is to introduce a new discrete symmetry, called
R-parity to forbid these terms. R-parity is defined as $R_P =
(-1)^{3B+L+2S}$ such that the SM particles and Higgs bosons have
$R_P = +1$ and all superpartners have $R_P = -1$. In the MSSM $R_P$ is
conserved and this has some interesting consequences. 
\begin{itemize}
\item $W_{\Delta L =1}$ and $W_{\Delta B =1}$ are absent in the MSSM.
\item The Lightest Supersymmetric Particle (LSP) is completely stable and
it provides a (cold) dark matter candidate.
\item Any sparticle produced in laboratory experiments decays into a final 
state with an odd number of LSP.
\item In colliders, Supersymmetric particles can only be produced (or destroyed) in pairs.
\end{itemize}

A second contribution to proton decay, already present in non-SUSY GUTs, 
comes from dimension 6 operators. The discussion is analogous to the analysis
in non-SUSY GUTS. Here we will only recall that a generic four-fermion
operator of the form $1/\Lambda^2~q~q~q~l$ results in a proton decay rate
of the order $\Gamma_p \sim 10^{-3}  m_p^5/\Lambda^4$. Given the
bound on the proton lifetime $\tau_p > 5 \times 10^{33}$ yrs, this constrains
the scale $\Lambda$ to be $\Lambda > 4 \times 10^{15}$ GeV. Therefore we can
see that with $\Lambda \simeq M_\GUT \simeq 2 \times 10^{16}$ GeV, dimension 6 
operators are still in agreement with the experimental bound.

\begin{figure}
\begin{center}
\includegraphics[width=8cm]{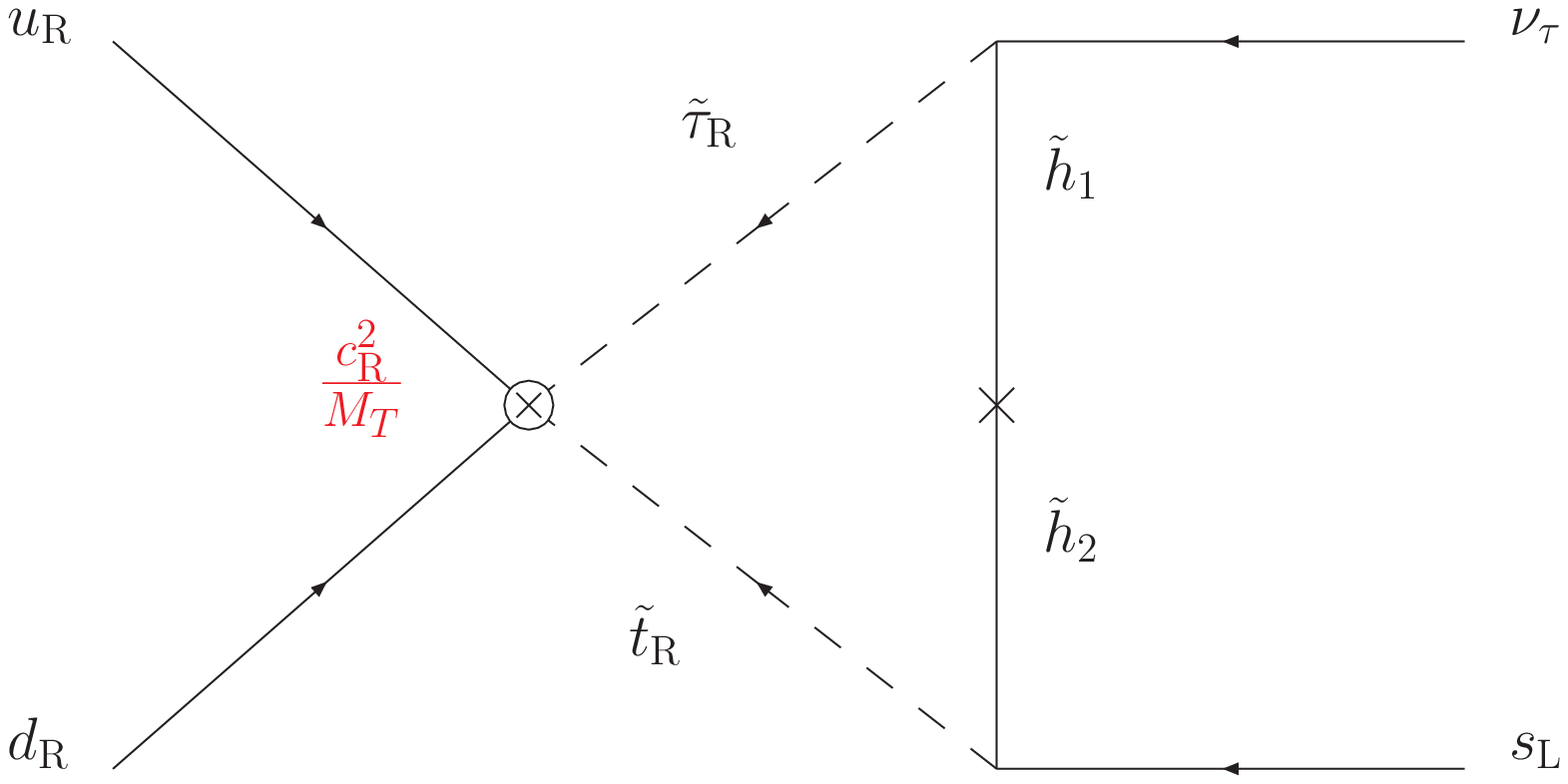}
\end{center}
\caption{Proton decay, $p \to K^+ \overline \nu_{\tau}$, through dimension 5
RRRR operator.}
\label{pdecay5}
\end{figure}

Dimension 5 operators are new in supersymmetric grand unified theories. They
are generated by the exchange of the coloured Higgs multiplet and are of the 
form 
\bea
W_5 = {c_L^{ijkl}\over M_T} (Q_k Q_l Q_i L_j) + {c_R^{ijkl}\over M_T} 
(u^c_i u^c_k d^c_j e^c_l),
\eea 
commonly called LLLL and RRRR operators respectively, with 
$M_T$ the mass of the coloured Higgs triplet. The coefficients $c_L^2$ and 
$c_R^2$ are model dependent factors depending on the Yukawa couplings. For
instance in Reference \cite{Arnowitt:1985iy,Nath:1985ub} they are
\bea
c_L^{ijkl} = \left(Y_D\right)_{ij} \left(V^T P Y_U V\right)_{kl}, \nn \\
c_R^{ijkl} = \left(P^*V^*Y_D\right)_{ij} \left(Y_U V\right)_{kl}, 
\eea
where $Y_D$ and $Y_U$ are diagonal Yukawa matrices, $V$ is the CKM mixing 
matrix and $P$ is a diagonal phase matrix.
The RRRR dimension 5 operator contributes to the decay $p \to K^+ \overline 
\nu_{\tau}$ through the diagram of Figure \ref{pdecay5}. The corresponding
amplitude is roughly given by
\bea
A_\tau(t_{\rm R}) \propto g^2 Y_d Y_t^2 Y_\tau V_{tb}^* V_{ud} V_{ts}
\Frac{\mu}{M_T m_{\tilde f}^2},
\eea
with $\mu$ the Higgs mass parameter in the superpotential and $m_{\tilde
  f}^2$ a typical squark or slepton mass. Notice that this amplitude is
proportional to $\tan^2 \beta$.
 
In fact these contributions from dimension 5 operators are extremely
dangerous. From the bound on the proton lifetime we have that, for $\tan \beta
= 2.5$ and $m_{\tilde f} \lsim 1$ TeV
\bea 
M_T \geq 6.5 \times 10^{16} \mbox{GeV},
\eea  
and this bound becomes more severe for larger values of $\tan \beta$ given
that the RRRR amplitude scales as $\tan^2 \beta/M_T$. On the other hand,
in minimal $SU(5)$, there is an upper bound on the Higgs triplet mass if we
require correct gauge coupling unification, $M_T \leq 2.5 \times 10^{16}$ GeV
at $90\%$ C.L.. This implies that the minimal SUSY SU(5) model would be 
excluded by proton decay if the sfermion masses are smaller than 1 TeV. 
Obviously, much in the same way that non-SUSY GUTs can be complicated enough 
to avoid the too fast proton decay present in minimal $SU(5)$, also in the SUSY case 
it is possible to avoid the mentioned problem in the minimal $SU(5)$ realisation by 
going to non-minimal SU(5) realisations or changing the gauge group altogether. 
How ``realistic" such non-minimal SUSY-GUTs are is  what we shortly discuss in the 
next subsection.

\subsubsection{``Realistic'' supersymmetric $SU(5)$ models}

Gauge coupling unification in supersymmetric grand unified theories 
is a big quantitative success. However, minimal $SU(5)$ models, face a 
series of other problems like proton decay or doublet-triplet splitting.
A sufficiently ``realistic'' model should be able to address and solve 
all these problems simultaneously \cite{Altarelli:2000fu}.  
The problems we would like this model to solve are: i) gauge coupling
unification with an acceptable value of $\alpha_s(M_Z)$ given $\alpha$
and $\sin^2 \theta_W$ at $M_Z$, ii) compatibility with the very
stringent bounds on proton decay and iii) natural doublet-triplet
splitting.

To solve the doublet-triplet problem we use the missing partner mechanism
presented above. The superpotential of this model will be that of \eq{spotmiss}
with the addition of the Yukawa couplings. Now the $SU(5)$ symmetry
is broken to the SM by a VEV of the representation ${\bf 75}$. This provides
a mass for the Higgs triplets while the doublets remain massless. Later
a $\mu$-term for the Higgs doublets of the order of the electroweak scale 
is generated through the Giudice-Masiero mechanism \cite{Giudice:1988yz}.  

Regarding gauge coupling unification, it is well known that in minimal
supersymmetric $SU(5)$ the central value of $\alpha_3(M_Z)$ required
by gauge coupling unification is too large: $\alpha_3(M_Z) \simeq
0.13$ to be compared with the experimental value $\alpha_3^{\rm
exp}(M_Z) \simeq 0.1187 \pm 0.002$. Using two loop RGE equations and 
taking into account the threshold effects we can write the corrected value of
$\alpha_3(M_Z)$ as
\bea
\label{corralpha}
\alpha_3(M_Z) &=& \Frac{\alpha_3^{(0)}(M_Z)}{1+\alpha_3^{(0)}(M_Z)~ \delta}\nn \\
\delta &=& k + \Frac{1}{2\pi} \log \Frac{M_{\rm SUSY}}{M_Z} -
\Frac{3}{5\pi} \log \Frac{M_T}{M_\GUT}, 
\eea 
with $\alpha_3^{(0)}$ the leading log value of this coupling equal to
the minimal $SU(5)$ value and $k$ contains the contribution from two
loop running, SUSY and GUT thresholds. $M_T$ is an effective mass defined as
\bea
m_T = \Frac{M_{T_1} M_{T_2}}{\tilde M},  
\eea
with $M_{T_1}$ and $M_{T_1}$ the two eigenvalues of the Higgs triplet
mass matrix and $\tilde M$ the mass of the ${\bf 50}$ in the
superpotential, \eq{spotsu3}.  The value of the parameter $k$ is
different in the minimal $SU(5)$ model and in the realistic model with
a ${\bf 75}$ breaking the $SU(5)$ symmetry:
\bea
\label{k values}
k^{\rm minimal} = -1.243 & k^{\rm realistic} = 0.614 \ .  \eea 

This difference is very important and improves substantially the comparison
of the prediction with the experimental value of $\alpha_3(M_Z)$.
In fact, for $k$ large and negative we need to take $M_{\rm SUSY}$ as large
as possible and $M_T$ as small as possible, but this runs into problems with 
proton decay. On the other hand if $k$ is positive and large, we can take 
$M_T > M_\GUT$. For instance, with $M_T = 6 \times 10^{17} {\rm GeV} \simeq
30 M_\GUT$ and $M_{\rm SUSY} = 0.25$ TeV we obtain $\alpha_3(M_Z) \simeq 0.116$
which is acceptable.
 
Regarding proton decay the main contribution comes again from
dimension five operators when the Higgs triplets are integrated out. 
Clearly these operators depend on $M_T$, but we have seen above that a large
$M_T$ is preferred in this model. Typical values would be
\bea
M_\GUT = 2.9 \times 10^{16} {\rm GeV},\qquad\tilde M 
= 2.0 \times 10^{16} {\rm GeV},\qquad~  \\
M_{T_1} = 1.2 \times 10^{17} {\rm GeV},~~ M_{T_2} = 1.0 
\times 10^{17} {\rm GeV},~~ M_{T} = 6 
\times 10^{17} {\rm GeV}.\nn
\eea
Notice that in this case the couplings of 
the triplets to the fermions is {\bf not} related to the fermion masses as the
Higgs triplets are now a mixing between the triplets in the ${\bf 5_H}$ and 
the triplets in the ${\bf 50}$. Therefore we have some unknown Yukawa coupling
$Y_{\bf\bar 50}$. Assuming a hierarchical structure in these couplings somewhat
analogous to the doublet Yukawa couplings \cite{Altarelli:2000fu} we would
obtain a proton decay rate in the range $8 \times 10^{31}$-- $3 \times 
10^{34}$ yrs for the channel $p \to K^+ \overline \nu$ and 
$2 \times 10^{32}$-- $8 \times 
10^{34}$ yrs for the channel $p \to \pi^+ \overline \nu$. The present bound 
at $90 \%$ C.L. on $\tau / {\rm BR}(p \to K^+ \overline \nu)$ is
$1.9 \times 10^{33}$ yrs. Thus we see that agreement with the stringent proton
decay bounds is possible in this model.

\subsection{Other GUT Models}
So, far we have discussed $SU(5)$, the prototype Grand Unified theory
in both supersymmetric and non-supersymmetric versions. Given that
supersymmetric Grand Unification ensures gauge coupling unification,
most models of Grand Unification which have been studied in recent
years have been supersymmetric. Other than the SUSY SU(5),
historically, one of the first unified models constructed was the
Pati-Salam model \cite{Pati:1974yy}. The gauge group was given by,
$SU(4)_c \times SU(2)_L \times SU(2)_R$. The fermion representations,
as explained above, require the presence of a right-handed neutrino.
Realistic models can be built incorporating bi-doublets of Higgs
giving rise to fermion masses and suitable representations for the
breaking of the gauge group.  However the Pati-Salam Model is not
truly a unified model in a strict sense. For this reason, one needs to
go for a larger group of which the Pati-Salam gauge group would be a
sub-group.  The simplest gauge group in this category is an orthogonal
group $SO(10)$ of rank 5.

\subsubsection{The seesaw mechanism}
There are several reasons to consider models beyond the simple $SU(5)$ 
gauge group we have considered here. One of the major reasons is 
the question of neutrino masses. This can be elegantly be solved 
through a mechanism which goes by the name \textit{seesaw} mechanism
\cite{Minkowski:1977sc,Yanagida:1979as,Gell-Mann:1980vs,Mohapatra:1979ia,Schechter:1980gr}. 
The seesaw mechanism requires an additional standard model singlet
fermion, which could be the right handed neutrino. Given that it is 
electrically neutral, this particle can have a Majorana mass (violating
lepton number by two units) in addition to the standard \textit{Dirac}
mass that couples it to the SM left handed neutrino. 
Representing the three left-handed fields by a column vector $\nu_L$ and
the three right handed fields by $\nu_R$, the Dirac mass terms are
given by
\beq
-{\mathcal L}^D = \bar{\nu}_L {\cal M}_D \nu_R + {\rm H.C.},
\eeq
where ${\cal M}^D$ represents the Dirac mass matrix. The Majorana
masses for the right handed neutrinos are given by
\beq
-{\mathcal L}^R = {1 \over 2} \bar{\nu}_R^c {\cal M}_R \nu_R + {\rm H.C.}.
\eeq
The total mass matrix is given as
\beq
-{\mathcal L}^{total} = {1 \over 2} \bar{\nu}_p {\cal M} \nu_p,
\eeq
where the column vector $\nu_p$ is
\beq
\nu_p = \left( \begin{array}{c} \nu_L \\ \nu_R^c \end{array} \right).
\eeq
And the matrix ${\cal M}$ is
\beq
{\cal M}= \left( \begin{array}{cc} 0 & {\cal M}_D^T 
\\ {\cal M}_D & {\cal M}_R \end{array} \right).
\eeq
Diagonalising the above matrix, one sees that the left handed neutrinos
attain Majorana masses of order,
\beq
{\cal M}^\nu = - {\cal M}_D^T~ {\cal M}_R^{-1} {\cal M}_D .
\eeq
This is called the seesaw mechanism. Choosing for example the
Dirac mass of the neutrinos to be typically of the order of
charged lepton masses or down quark masses, we see that
for a heavy right handed neutrino mass scale (${\cal M}_R \gg {\cal M}_D$), 
the left-handed neutrino masses are highly suppressed. In this way, the smallness of
neutrino masses can be explained naturally by the seesaw mechanism.
While the seesaw mechanism is elegant,
as mentioned in the Introduction, by construction we do not have right handed
neutrinos in the SM particle spectrum. The $SU(5)$ representations do not 
contain a right handed singlet particle either, as we have seen above.
However, in larger GUT groups like $SO(10)$ these additional particles are
naturally present. 

\subsubsection{SO(10)} 
The group theory of $SO(10)$ and its spinorial representations can be
simplified by using the $SU(N)$ basis for the $SO(2N)$ generators or
the tensorial approach. The spinorial representation of the $SO(10)$
is given by a 16-dimensional spinor, which could accommodate all the
SM model particles as well as the right handed neutrino. The question
of gauge coupling unification in $SO(10)$ is more complicated as now there
is a `natural' possibility of an intermediate scale. Rest assured it
can be achieved, though it depends on the $SO(10)$ breaking
mechanism  chosen. 

Let's now see how fermions attain their masses in this model. 
The product of two \textbf{16} matter representations can only couple to
\textbf{10}, \textbf{120} or \textbf{126} representations, which can be
formed by either a single Higgs field or a
non-renormalisable product of representations of several Higgs fields.
In either case, the  Yukawa matrices resulting from the couplings to
\textbf{10} and \textbf{126} are complex-symmetric, whereas they are
antisymmetric when the couplings are to the \textbf{120}.
Thus, the most general $SO(10)$ superpotential relevant to
fermion masses can be written as
\beq
W_{SO(10)} = h^{10}_{ij} {\bf 16_i}~ {\bf 16_j}~ {\bf 10} + h^{126}_{ij} {\bf 16_i}~ {\bf 16_j}~ {\bf 126}
+ h^{120}_{ij} {\bf 16_i}~{\bf 16_j}~ {\bf 120},
\eeq
where $i,j$ refer to the generation indices. In terms of the SM fields,
the Yukawa couplings relevant for fermion masses are given by \cite{Barbieri:1979ag,Langacker:1980js}
\footnote{Recently, SO(10) couplings have also been evaluated for various 
renormalisable and non-renormalisable couplings in \cite{Nath:2001uw}.}:
\bea
{\bf 16}\ {\bf 16}\ {\bf 10}\ &\supset&{\bf 5}\ ( u u^c + \nu \nu^c) + {\bf\bar 5}\
(d d^c + e e^c), \\
{\bf 16}\ {\bf 16}\ {\bf 126}\ &\supset&{\bf 1}\ \nu^c \nu^c + 15\ \nu \nu +
{\bf 5}\ ( u u^c -3~ \nu \nu^c) + {\bf\bar{45}}\ (d d^c -3~ e e^c), \nn\\
{\bf 16}\ {\bf 16}\ {\bf 120}\ &\supset&{\bf 5}\ \nu \nu^c +{\bf 45}\ u u^c +
{\bf\bar 5}\ ( d d^c + e e^c) +{\bf \bar{45}}\ (d d^c -3~ e e^c),\nn
\label{su5content}
\eea
where we have specified the corresponding $SU(5)$  Higgs representations
for each  of the couplings and all the fermions are left handed fields.
The resulting mass matrices can be written as
\bea
\label{upmats}
M^u &= & M^5_{10} + M^5_{126} + M^{45}_{120}, \\
\label{numats}
M^\nu_{LR} &= & M^5_{10} - 3~ M^5_{126} + M^{5}_{120}, \\
\label{downmats}
M^d &= & M^{\bar{5}}_{10} +  M^{\bar{45}}_{126} + M^{\bar{5}}_{120}
+ M^{\bar{45}}_{120}, \\
\label{clepmats}
M^e &= & M^{\bar{5}}_{10} - 3  M^{\bar{45}}_{126} + M^{\bar{5}}_{120}
- 3 M^{\bar{45}}_{120}, \\
\label{lneutmats}
M^\nu_{LL} &=& M^{15}_{126}, \\
M^\nu_{R} &=& M^{1}_{126}.
\eea
We can see here the relations between the different fermionic species. 
In particular, notice the relation between up-quarks and neutrino (Dirac)
mass matrices, Eqs.~(\ref{upmats}) and (\ref{numats}).

The breaking of $SO(10)$ to the Standard Model group on the other hand can
be quite complex compared to that of the $SU(5)$ model we have studied so far. 
In particular, the gauge group offers the possibility of the existence of
an intermediate scale where another ``gauge symmetry'', a subgroup of $SO(10)$,
can exist.  Some of the popular ones are summarised in the figure below:
\begin{figure}[ht]
\begin{center}
\includegraphics[scale=0.50]{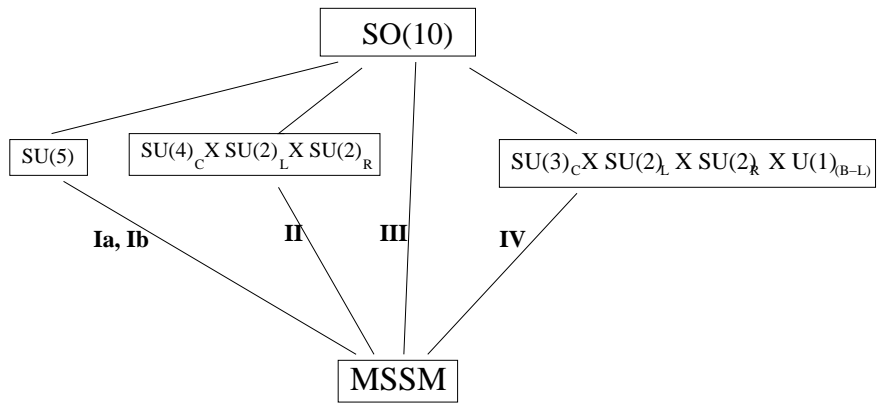}
\caption{The various breaking chains of $SO(10)$ are summarised in this
figure. }
\label{fig:breakingchains}
\end{center}
\end{figure}
Each of these breaking chains would have its own RG scaling
which can in principle lead to different results at the weak scale even though
the initial conditions at the $SO(10)$ scale are the same.
Different Higgs representations are used for the breaking in each of these
cases. In the recent years, attempts have been made to construct complete
(renormalisable) models of $SO(10)$ where it could be possible to have
precision studies of $SO(10)$ models. This studies do
give a good handle on the predictions on proton life time, gauge
coupling unification, and, to some extent, fermion masses. 
However, for processes involving the supersymmetric spectra 
like flavour changing neutral currents, etc, the situation is more
model dependent.

Apart from $SU(5)$ and $SO(10)$, there are other GUT models in the literature
based on gauge groups $E_6$, $SU(6)$ etc, which we have not touched
in this set of lectures. 

\section{FLAVOUR AND CP VIOLATION IN SUSY}
\label{sec:constraints}
Before entering into the issues of flavour in supersymmetric GUTs, it 
is instructive to study the issue of flavour within SUSY Standard Models.
The simplest Supersymmetric version of the Standard Model that we can
build is the so-called Minimal Supersymmetric Standard Model (MSSM). 
 Clearly, this model
must include all the SM interactions and particle spectrum together
with their Supersymmetric partners.  This means that to every quark
and lepton in the SM we add a scalar Supersymmetric partner, called
``squark'' or ``slepton'' respectively, with identical gauge quantum
numbers and, in principle, identical mass, forming a ``chiral
supermultiplet''. In the same way, to the SM Higgs or more exactly to
the Higgses in a 2 Higgs doublet version of the SM\footnote{As it is
well-known, Supersymmetry requires two different Higgs doublets to
give mass to fermions of weak isospin $+1/2$ and $-1/2$
\cite{Nilles:1983ge,Haber:1984rc,Ross:1985ai,Haber:1993wf,Martin:1997ns}} 
we add fermionic
partners called ``higgsinos'' with the same quantum numbers and masses
in another ``chiral Supermultiplet''.  Then every gauge boson is also
joined by a gaugino (``gluino'', ``wino'', ``bino''...) with spin
$1/2$ in the adjoint representation in a ``vector supermultiplet''
(for a complete formulation of Supersymmetric theories in superfield
notation see Ref. \cite{Wess:1992cp}).

The gauge interactions in our MSSM are completely fixed by the gauge
quantum numbers of the different particles in the usual way.  However,
we still need the Yukawa interactions of the Standard Model that give
masses to the fermions once we break the electroweak symmetry. These
interactions are included in the MSSM Superpotential, which is a gauge
invariant analytic function of the MSSM superfields (i.e. a function 
of fields $\phi_i$ but not of complex conjugate fields $\phi_i^*$) with 
dimensions of
mass cube. If we include all possible terms invariant under the gauge
symmetry then it turns out that some of these terms violate either
baryon or lepton number. As we have seen in the previous section, this 
endangers proton stability; hence one
usually imposes a discrete symmetry called R-parity under which the
ordinary particles are even while their SUSY partners are odd
\cite{Farrar:1978xj} \footnote{Since these terms violate either lepton or
baryon number, it is also possible to forbid only lepton number or
baryon number violation to ensure proton stability \cite{Dreiner:1997uz}}. 
The MSSM Superpotential (using standard notation) is then,
\bea
W= Y_d^{i j} Q_i H_1 d^c_{R j} + Y_e^{i j} L_i H_1 e^c_{R j} + Y_u^{i
j} Q_i H_2 u^c_{R j} + \mu H_1 H_2,
\label{WMSSM}
\eea 
and this gives rise to the interactions,
\bea
{\mathcal L}_W= \left|\Frac{\partial W}{\partial \phi_i}\right|^2 +
\psi_i \psi_j \Frac{\partial^2 W}{\partial \phi_i \partial \phi_j},
\label{LW}
\eea
with $\phi_i$ any scalar in the MSSM and $\psi_i$ its corresponding fermionic
partner. 

Still, we know that Supersymmetry is not an exact symmetry in nature
and it must be broken. If Supersymmetry is the solution to the
hierarchy problem, the breaking of Supersymmetry must be soft,
i.e. should not reintroduce the quadratic divergences which are
forbidden in the SUSY invariant case, and the scale of SUSY breaking
must be close to the electroweak scale. The most general set of
possible Soft SUSY breaking terms (SBT) \cite{Girardello:1981wz} under these
conditions are,
\begin{enumerate}
\item Gaugino masses\\[.2cm]
${\mathcal L}_\soft^{(1)} = \Frac{1}{2} \left(M_1\ \tilde B \tilde B + M_2\ \tilde W 
\tilde W + M_3\ \tilde g \tilde g\right) +$ h.c.
\item Scalar masses\\[.2cm]
${\mathcal L}_\soft^{(2)} = (M_{\tilde{Q}}^2)_{i j} \tilde Q_i \tilde Q_j^* +
(M_{\tilde{u}}^2)_{i j} \tilde u_{R i}^c \tilde u_{R j}^{c *} +
(M_{\tilde{d}}^2)_{i j} \tilde d_{R i}^c \tilde d_{R j}^{c *} +
(M_{\tilde{L}}^2)_{i j} \tilde L_i \tilde L_j^* +$\\
$~~~~~~~~~~~(M_{\tilde{e}}^2)_{i j} \tilde e_{R i}^c \tilde e_{R j}^{c *} + 
(m_{H_1}^2) H_{1} H_{1}^* +
(m_{H_2}^2) H_{2}  H_{2}^*$
\item Trilinear couplings and B--term\\[.2cm]
${\mathcal L}_\soft^{(3)} =  (Y^A_d)^{i j} \tilde Q_i H_1 \tilde d_{R j} + 
 (Y^A_e)^{i j} \tilde L_i H_1 \tilde e^c_{R j} + 
 (Y^A_u)^{i j}\tilde Q_i H_2 \tilde u^c_{R j} + B \mu H_1 H_2$
\end{enumerate}
where, $M_{\tilde{Q}}^2$, $M_{\tilde{u}}^2$, $M_{\tilde{d}}^2$, 
$M_{\tilde{L}}^2$ and
$M_{\tilde{e}}^2$ are hermitian $3 \times 3$ matrices in flavour space,
while $(Y^A_d)$, $(Y^A_u)$ and $(Y^A_u)$ are complex $3 \times 3$
matrices and $M_1$, $M_2$, $M_3$ denote the Majorana gaugino masses for the 
$U(1)$, $SU(2)$, $SU(3)$  gauge symmetries respectively.

This completes the definition of the MSSM. However, these conditions 
include a huge variety of models with very different phenomenology 
specially in the flavour and CP violation sectors. 

It is instructive to identify all the observable parameters in a general 
MSSM \cite{Haber:1997if}. Here we
distinguish the flavour independent sector which includes the gauge and Higgs
sectors and the flavour sector involving the three generations of chiral
multiplets containing the SM fermions and their Supersymmetric partners. 

In the flavour independent sector, we have three real gauge couplings,
$g_i$, and three complex gaugino masses, $M_i$. In the Higgs sector,
also flavour independent, we have a complex $\mu$ parameter in the 
superpotential, a complex
$B\mu$ soft term and two real squared soft masses $m_{H_1}^2$ and
$m_{H_2}^2$. However, not all the phases in these parameters are
physical \cite{Dimopoulos:1995kn}. In the limit of $\mu=B\mu=0$, vanishing gaugino masses and
zero trilinear couplings, $Y^A$, (we will discuss trilinear terms in the
flavour dependent sector), our theory has two global $U(1)$
symmetries: $U(1)_{R}$ and $U(1)_{PQ}$. This implies
that we can use these two global symmetries to remove two of the
phases of these parameters. For instance, we can choose a real $B\mu$
and a real gluino mass $M_3$. Then, in the flavour independent sector,
we have 10 real parameters ($g_i$, $|M_i|$, $|\mu|$, $B\mu$,
$m_{H_1}^2$ and $m_{H_2}^2$) and 3 phases ($\mbox{arg}(\mu)$,
$\mbox{arg}(M_1)$ and $\mbox{arg}(M_2))$.
 
Next, we have to analyse the flavour dependent sector. As a starting point,
let us not take into account the non-zero neutrino masses. Then, in the superpotential
we have the up quark, down quark and charged lepton Yukawa couplings,
$Y_u$, $Y_d$ and $Y_e$, that are complex $3 \times 3$ matrices. In the
soft breaking sector we have 5 hermitian mass squared matrices,
$M_{\tilde Q}^2$, $M_{\tilde U}^2$, $M_{\tilde D}^2$, $M_{\tilde L}^2$
and $M_{\tilde E}^2$ and three complex trilinear matrices, $Y^A_u$, $Y^A_d$ 
and $Y^A_e$. This
implies we have $6 \times 9$ moduli and $6 \times 9 $ phases from the
6 complex matrices ($Y_u$, $Y_d$, $Y_e$, $Y^A_u$, $Y^A_d$ and $Y^A_e$) and 
$5 \times 6$ moduli and $5 \times 3$ phases
from the 5 hermitian matrices. Therefore, in the flavour sector we
have 84 moduli and 69 phases. However, it is well-known that not all
these parameters are observable. In the absence of these flavour
matrices the theory has a global $U(3)_{Q_L}\otimes U(3)_{u_R}\otimes
U(3)_{d_R} \otimes U(3)_{L_L}\otimes U(3)_{e_R}$ flavour symmetry
under exchange of the different particles of the three generations. The
number of observable parameters is easily determined using the method
in Ref.~\cite{Santamaria:1993ah} as,
\bea
\label{numbpar}       
N = N_{fl} - N_{G} - N_{G^\prime}, 
\eea 
where $N_{fl}$ is the number of parameters in the flavour
matrices. $N_{G}$ is the number of parameters of the group of
invariance of the theory in the absence of the flavour matrices
$G=U(3)_{Q_L}\otimes U(3)_{u_R}\otimes U(3)_{d_R} \otimes
U(3)_{L_L}\otimes U(3)_{e_R}$. Finally $N_{G^\prime}$ is the number of
parameters of the group $G^\prime$, the subgroup of $G$ still unbroken 
by the flavour matrices. In this case, $G^\prime$ corresponds
to two $U(1)$ symmetries, baryon number conservation and lepton number
conservation and therefore $N_{G^\prime}=2$.  Furthermore \eq{numbpar}
can be applied separately to phases and moduli. In this way, and
taking into account that a $U(N)$ matrix contains $n(n-1)/2$ moduli
and $n(n+1)/2$ phases, it is straightforward to obtain that we have,
$N_{ph}=69 - 5 \times 6 + 2 = 41$ phases and $N_{mod} = 84 - 5 \times
3 = 69$ moduli in the flavour sector.  This amounts to a total of 123
parameters in the model\footnote{Notice that we did not include the
parameter $\theta_{QCD}$ which was also present in the 124
parameters MSSM of H. Haber \cite{Haber:1997if}.}, out of which 44 are CP
violating phases!!
As we know, in the SM, there is only one observable CP violating phase, 
the CKM phase, and therefore we have here 43 new phases, 40 in the 
flavour sector and three in the flavour independent sector.

Clearly, to explore completely the flavour and CP violating phenomena in 
a generic
MSSM is a formidable task as we have to determine a huge number of
unknown parameters \cite{Botella:2004ks}. However, this parameter counting 
corresponds to a
completely general MSSM at the electroweak scale but the number of
parameters is largely reduced in most of the theory motivated models
defined at high energies. In these models most of the parameters at
$M_W$ are fixed as a function of a handful of parameters at the scale
of the transmission of SUSY breaking, for instance $M_{Pl}$ in the case 
of supergravity mediation, and
therefore there are relations among the parameters at $M_W$.  
So, our task will be to
determine as many as possible of the CP violating and flavour
parameters at $M_W$ to look for possible relations among them that
will allow us to explore the physics of SUSY and CP breaking at high energies.

The so-called Constrained MSSM (CMSSM), or Sugra-MSSM, (for an early 
version of these models see, \cite{Barbieri:1982eh,Chamseddine:1982jx}) 
is the simplest
version we can build of the MSSM. For instance a realisation of this
model is obtained in string models with dilaton dominated SUSY
breaking \cite{Kaplunovsky:1993rd,Brignole:1993dj}. Here all the SBT
are universal.  The soft
masses are all proportional to the identity matrix and the trilinear
couplings are directly proportional to the corresponding Yukawa
matrix.  Moreover the gaugino masses are all unified at the high
scale. So, we have at $M_\GUT$, 
\bea 
M^2_{\tilde Q}~=~ M^2_{\tilde
U}~=~ M^2_{\tilde D}~=~ M^2_{\tilde L} ~=~ M^2_{\tilde E}~=~
m_0^2~~\unity, \nn \\ 
Y^A_u = A_0~ Y_u,\qquad Y^A_d = A_0~ Y_d, \qquad Y^A_e = A_0~ Y_e,\nn \\ 
m_{H_1}^2~ =~ M_{H_2}^2~ =~ m_0^2\qquad M_3~=~M_2~=~M_1~ =~ M_{1/2} 
\eea 
In this way the number of parameters is strongly reduced. If we repeat
the counting of parameters in this case we have only 27 complex
parameters in the Yukawa matrices, out of which only 12 moduli and 1
phase are observable. In the soft breaking sector we have only a real
mass square, $m_0^2$, and a complex trilinear term, $A_0$. We have a
single unified gauge coupling, $g_U$, and a complex universal gaugino
mass $M_{1/2}$ in the gauge sector. Finally in the Higgs sector there
are two complex parameters $\mu$ and $B\mu$.  Again two of these
phases can be reabsorbed through the $U(1)_R$ and $U(1)_{PQ}$
symmetries. Therefore, we have only 21 parameters, 18 moduli and 3
phases. In fact, 14 of these parameters are already known in the
Standard Model and we are left with only 7 unknown parameters from
SUSY: ($m_0^2$, $|M_{1/2}|$, $|\mu|$, $\mbox{arg}(\mu)$, $|A_0|$,
$\mbox{arg}(A_0)$ and $|B|$). If we require radiative electroweak
symmetry breaking \cite{Ibanez:1992rk} we get an additional constraint
which is used to relate $|B|$ to $M_W$.  In the literature it is also
customary to exchange $|\mu|$ by $\tan \beta= v_2/v_1$, the ratio of
the two Higgs vacuum expectation values, so that the set of parameters
usually considered in the MSSM with radiative symmetry breaking is
($m_0^2$, $|M_{1/2}|$, $\tan \beta$, $|A_0|$, $\mbox{arg}(A_0)$ and
$\mbox{arg}(\mu)$).  Regarding CP violation, we see that even in the
simplest MSSM version we have two new CP violating phases, which we
have chosen to be $\varphi_\mu\equiv\mbox{arg}(\mu)$ and $\varphi_A
\equiv\mbox{arg}(A_0)$.  These phases will have a very strong effect
on CP violating observables, mainly the Electric Dipole Moments (EDMs)
of the electron and the neutron as we will show in the next
section. We must remember that a generic MSSM will always include at
least these two phases and therefore the constraints from EDMs are
always applicable in any MSSM.

All these flavour parameters and phases are encoded at the electroweak scale 
in the different mass matrices of sfermions and gauginos/higgsinos.
For instance, after breaking the $SU(2)_L$ symmetry, the superpartners 
of $W^\pm$ and $H^\pm$ have the same unbroken quantum number and thus can mix
through a matrix, 
\bea
-\Frac{1}{2} \left(\matrix{\tilde W^- & \tilde H_1^-} \right) \; 
\left(\matrix{M_2  & \sqrt{2} M_W \sin \beta \cr \sqrt{2} M_W \cos 
\beta & \mu} \right) \left(\matrix{\tilde W^+ \cr \tilde H_2^+
} \right),
\eea
This non-symmetric (non-hermitian) matrix is diagonalised with two 
unitary matrices, $U^* \cdot M_{\chi^+} \cdot 
V^\dagger = \mbox{Diag.}(m_{\chi_1^+},m_{\chi_2^+})$.

In the same way, once we break the electroweak symmetry, neutral
higgsinos and neutral gauginos mix. In the basis $(\tilde B~\tilde W^0\tilde H_1^0\tilde H_2^0)$, the mass matrix is,
\bea 
\left(\matrix{M_1&0 & - M_Z c \beta s \theta_W& M_Z s \beta s \theta_W
\cr0& M_2&  M_Z c \beta c \theta_W& M_Z s \beta c \theta_W \cr
- M_Z c \beta s \theta_W& M_Z c \beta c \theta_W &0& -\mu\cr
M_Z s \beta s \theta_W& -M_Z s \beta c \theta_W & -\mu&0} \right),
\eea
with $c \beta (s \beta)$ and $c \theta_W (s \theta_W)$, $\cos (\sin) \beta$
and $\cos (\sin) \theta_W$ respectively.
This is diagonalised by a unitary matrix $N$,
\bea
N^* \cdot M_{\tilde N}
\cdot N^\dagger = \mbox{Diag.} 
(m_{\chi_1^0}, m_{\chi_2^0}, m_{\chi_3^0}, m_{\chi_4^0})
\eea

Finally, the different sfermions, as $\tilde f_L$ and $\tilde f_R$, mix after
EW breaking. In fact they can also mix with fermions of different
generations and in general we have a $6 \times 6$ mixing matrix.
\bea
&M_{\tilde f}^2 \; = \; 
\left(\matrix{m_{\tilde f_\LL}^2 & m_{\tilde f_\LR}^2 \cr {m_{\tilde f_\LR}^{2\,\dagger}} &
m_{\tilde f_\RR}^2 } \right) ~~~~m_{\tilde f_\LR}^2\; = \; 
\big(  Y^A_f\cdot ^{v_2}_{v_1} - m_f \mu ^{\tan \beta}_{\cot \beta}\big) \;\; \mbox{for}\;
f=^{e,d}_u \nonumber \\[.2cm]
&m_{\tilde f_\LL}^2 =  M^2_{\tilde f_{L}} + m^2_f + M_Z^2 \cos 2 \beta
(I_3 + \sin^2 \theta_W Q_\elm) \nonumber \\
&m_{\tilde f_\RR}^2 =  M^2_{\tilde f_{R}} + m^2_f + M_Z^2 \cos 2 \beta
\sin^2 \theta_W Q_\elm  
\eea
These hermitian sfermion mass matrices are diagonalised by a unitary rotation,
\\$R_{\tilde f} \cdot M_{\tilde f} \cdot R_{\tilde f}^\dagger = 
\mbox{Diag.}(m_{\tilde f_1},m_{\tilde f_2},\dots,m_{\tilde f_6})$.

Therefore all the new SUSY phases are kept in these gaugino and sfermion
mixing matrices. However, it is not necessary to know the
full mass matrices to estimate the CP violation effects.  We have some
powerful tools as the Mass Insertion (MI) approximation
\cite{Hall:1985dx,Gabbiani:1988rb,Hagelin:1992tc,Gabbiani:1996hi}  
to analyse FCNCs and
CP violation. In this approximation, we use flavour diagonal gaugino
vertices and the flavour changing is encoded in non-diagonal sfermion
propagators. These propagators are then expanded assuming
that the flavour changing parts are much smaller than the flavour
diagonal ones. In this way we can isolate the relevant elements of the
sfermion mass matrix for a given flavour changing process and it is not
necessary to analyse the full $6\times 6$ sfermion mass matrix.
Using this method, the experimental limits lead to upper bounds on the 
parameters (or combinations of) $\delta_{ij}^f \equiv 
\Delta^f_{ij}/m_{\tilde{f}}^2$, known as mass insertions;
where $\Delta^f_{ij}$ is the flavour-violating off-diagonal entry
appearing in the $f = (u,d,l)$ sfermion mass matrices and
$m_{\tilde{f}}^2$ is the average sfermion mass. In addition, the
mass-insertions are further sub-divided into LL/LR/RL/RR types,
labelled by the chirality of the corresponding SM fermions. In
the following sections we will use both the full mass matrix diagonalisation
and this MI formalism to analyse flavour
changing  and CP violation processes. Now, we will start by studying the
EDM calculations and constraints which are common to all
Supersymmetric models.

\subsection{Electric Dipole Moments in the MSSM}
\label{sec:EDM}

The large SUSY contributions to the electric dipole moments of the
electron and the neutron are the main source of the so-called
``Supersymmetric CP problem''. This ``problem'' is present in any MSSM
due to the presence of the flavour independent phases $\varphi_\mu$
and $\varphi_A$. Basically Supersymmetry gives rise to contributions
to the EDMs at 1 loop order with no suppression associated to flavour
as these phases are flavour diagonal \cite{Ellis:1982tk,Buchmuller:1982ye,Polchinski:1983zd,Franco:1983xm,Dugan:1984qf,Fischler:1992ha}. Taking into account 
these facts,
this contribution can be expected to be much larger than the SM
contribution which appears only at three loops and is further
suppressed by CKM angles and fermion masses. In fact the SM
contribution to the neutron EDM is expected to be of the order
of $10^{-32}$ e cm, while the present experimental bounds are $d_n
\leq 6.3 \times 10^{-26}$ e cm (90\% C.L.) \cite{Harris:1999jx} and 
$d_e \leq 1.6 \times 10^{-27}$ e cm (90\% C.L.) \cite{Regan:2002ta}.  
As we will show here the Supersymmetric 1
loop contributions to the EDM for SUSY masses below several TeV 
can easily exceed the present experimental bounds. Therefore,
these experiments impose very stringent bounds on $\varphi_\mu$ and
$\varphi_A$.

\begin{figure}
\begin{center}
\includegraphics[width=8.6cm]{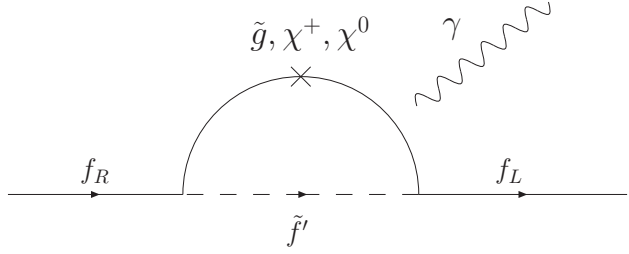}
\caption{1 loop contributing to a fermion EDM}
\label{fig:EDM}
\end{center}
\end{figure}

The typical diagram giving rise to a fermion EDM is shown in Figure 
\ref{fig:EDM}.
In the case of a quark EDM, the dominant contribution typically corresponds to
the diagram with internal gluino and squark states. Here all
the phases appear only in the squark mass matrix. If we neglect
intergenerational mixing (that can be expected to be small), we have a
$2 \times 2$ squark mass matrix. For instance the down squark mass
matrix, $M_{\tilde d}$ in the basis $(\tilde d_L, \tilde d_R)$ is,
\bea
\left(\matrix{
m_{\tilde d_L}^2 + m_d^2 - ( \frac{1}{2} - \frac{1}{3}~ s^2 \theta_W) 
~c 2 \beta M_Z^2 &  Y_d^{A\,*}~ v~ c \beta - m_d~ \mu~ tg \beta \cr
 Y_d^{A}~ v~ c \beta - m_d~ \mu^* ~tg \beta & m_{\tilde d_R}^2 + m_d^2 
- \frac{1}{3} ~s^2 \theta_W ~c 2 \beta M_Z^2}\right)
\eea
with $Y_d^{A} \simeq A_0 Y_d$ except 1 loop correction in the RGE
evolution from $M_\GUT$ to $M_W$. Therefore we have both $\varphi_\mu$
and $\varphi_A$ in the left-right squark mixing and these phases
appear then in the down squark mixing matrix, $R^d$, $R^{\tilde d}
M_{\tilde d} R^{\tilde d\,\dagger} = \mbox{Diag.}(m_{\tilde d_1},
m_{\tilde d_2})$. In terms of this mixing matrix the 1 loop gluino 
contribution to the EDM of the down quark is (in a similar way we
 would obtain the gluino contribution to the up quark EDM),
\bea
d^{d}_{\tilde g}= \frac{2 \alpha_s e}{9 \pi} \sum_{k=1}^2 
\mbox{Im}[R^{\tilde d}_{k 2 } R^{\tilde d\,*}_{k 1}] \frac{1}{m_{\tilde g}}
B(\frac{m_{\tilde g}^2}{m_{\tilde d_k}^2})
\label{EDMglu}
\eea
with,
\bea
B(r) = \frac{r}{2(1-r)^2}\left(1+r +\frac{2 r \log r}{1-r}\right)
\eea
It is interesting to obtain the corresponding formula in terms of the
$\left(\delta_{11}^f\right)_\LR$ mass insertion. To do this we
observe that given a $n \times n$ hermitian matrix $A = A^0 + A^1$
diagonalised by $U\cdot A\cdot U^\dagger = \mbox{Diag}(a_1,\dots, a_n)$, 
with $A^0 = \mbox{Diag} (a_1^0,\dots, a_n^0)$ and $A^1$ completely 
off-diagonal, we have at first order in $A^1$ 
\cite{Buras:1997ij,Hisano:1998fj},
\bea
\label{seriesherm}
U_{k i}^* f(a_k) U_{k j} \simeq \delta_{ij} f(a^0_i) + A^1_{ij} 
\Frac{f(a^0_i)-f(a^0_j)}{a^0_i-a^0_j}
\eea
Therefore, for small off-diagonal entries $A^1$ and taking into 
account that for approximately degenerate
squarks we can replace the finite differences by the derivative of the
function, $B^\prime(x)$, \eq{EDMglu} is converted into,
\bea
d^{d}_{\tilde g}&\simeq&\frac{2 \alpha_s e}{9 \pi}   
\frac{m_{\tilde g}}{m_{\tilde d}^2} B^\prime(\frac{m_{\tilde g}^2}
{m_{\tilde d}^2})~\mbox{Im}\left[\frac{Y_d^{A\,*}~ v \cos \beta - 
m_d~ \mu \tan \beta}{m_{\tilde d}^2}\right] \nn \\
&\equiv&\frac{2 \alpha_s e}{9 \pi}   
\frac{m_{\tilde g}}{m_{\tilde d}^2} B^\prime(\frac{m_{\tilde g}^2}
{m_{\tilde d}^2})~\mbox{Im}\left[\left(\delta_{11}^d\right)_\LR\right]
\label{EDMgluMI}
\eea
with $m_{\tilde d}^2$ the average down squark mass. From this equation it is 
straightforward to obtain a simple numerical estimate. Taking $m_{\tilde g} =
m_{\tilde d} = 500$ GeV, $Y^A_d = A_0 Y_d$ and $\mu\simeq A_0\simeq 500$ GeV,
we have,
\bea
d^{d}_{\tilde g}&\simeq& 2.8 \times 10^{-20}~~ \mbox{Im}\left[\left(
\delta_{11}^d\right)_\LR\right] \mbox{ e cm}\\ 
&=& 2.8 \times 10^{-25}~~
\left(\sin \varphi_A - \tan \beta \sin \varphi_\mu \right) \mbox{ e cm}
\nn
\eea
where we used $\alpha_s = 0.12$ and $m_d = 5$ MeV.  Comparing with the
experimental bound on the neutron EDM and using, for simplicity the
quark model relation $d_n = \frac{1}{3}( 4 d_d - d_u)$, we see
immediately that $\varphi_A$ and  $(\tan \beta~ \varphi_\mu) \leq 0.16$.
This is a simple aspect of the ``supersymmetric CP problem''.
As we will see the constraints from the electron EDM give rise to 
even stronger bounds on these phases.

\begin{figure}
\begin{center}
\includegraphics[width=8.6cm]{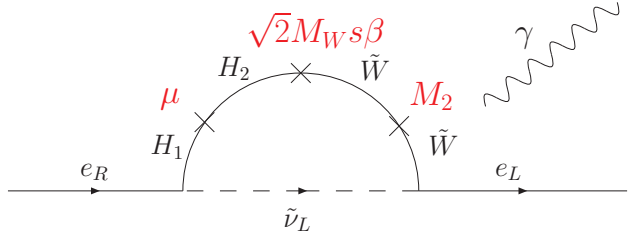}
\caption{1 loop chargino contribution to the electron EDM at leading
  order in chargino mass insertions.}
\label{fig:eEDM}
\end{center}
\end{figure}

In addition, we have also contributions from chargino and
neutralino loops which are usually subdominant in the quark EDMs but
are the leading contribution in the electron EDM.  A simple example is 
the chargino contribution to the electron EDM.
The corresponding diagram is shown in Figure \ref{fig:EDM} with the
chargino and sneutrino in the internal lines,
\bea
d^{e}_{\chi^+}= - \frac{\alpha e}{4 \pi \sin^2 \theta_W}
~ \frac{m_e}{\sqrt{2} M_W \cos \beta}\sum_{j=1}^2 
\mbox{Im}[ U_{j2} V_{j1}] 
\frac{m_{\chi^+_j}}
{m_{\tilde \nu _e}^2} A(\frac{m_{\chi_j^+}^2}
{m_{\tilde \nu_e}^2})
\eea
with,
\bea
A(r) = \frac{1}{2(1-r)^2}\left(3-r +\frac{2 \log r}{1-r}\right)
\eea
It is also useful to use a technique similar to \eq{seriesherm} to
expand the chargino mass matrix. In this case we have to be careful
because the chargino mass matrix is not hermitian. However due to the
necessary chirality flip in the chargino line we know that the EDM is a
function of odd powers of $M_{\chi^+}$ \cite{Clavelli:2000ua}, 
\bea
\sum_{j=1}^2  U_{j2} V_{j1} m_{\chi^+_j} A(m_{\chi_j^+}^2)
= \sum_{j,k,l=1}^2 U_{lk} m_{\chi^+_l} V_{l1} ~ U_{j2}
A(m_{\chi^+_j}^2) U^*_{jk}.
\eea
where we have simply introduced an identity $\delta_{l j} = 
\sum_k  U_{lk} U^*_{jk}$.
Now, assuming $M_W \ll M_2, \mu$, we can use \eq{seriesherm} to
develop the loop function A(x) as a function of the hermitian matrix
$M_{\chi^+} M_{\chi^+}^\dagger$ and we get,
\bea 
d^{e}_{\chi^+}&\simeq& \frac{- \alpha~ e~ m_e}{4 \pi
\sin^2 \theta_W}~
\frac{\mbox{Im}\left[\sum_{k}\left(M_{\chi^+}M_{\chi^+}^\dagger\right)_{2
k} \left(M_{\chi^+}\right)_{k 1}\right]}{\sqrt{2} M_W \cos \beta~ m_{\tilde \nu_e}^2} 
~\Frac{A(r_1) - A(r_2)}{ m_{\chi_1^+}^2-
m_{\chi_2^+}^2}\nn\\
& =& \frac{- \alpha~ e~ m_e \tan \beta}{4 \pi \sin^2
\theta_W}~ \frac{\mbox{Im}[M_2~ \mu]}{m_{\tilde
\nu_e}^2}~\Frac{A(r_1)-A(r_2)}{ m_{\chi_1^+}^2-
m_{\chi_2^+}^2} 
\eea
with $r_i = m_{\chi_i^+}^2/m_{\tilde \nu_e}^2$.
This structure with three chargino MIs is shown in
figure \ref{fig:eEDM}.
Here we can see that only $\varphi_\mu$ enters in the chargino contribution.
In fact arg$(M_2~ \mu)$ is the rephasing invariant expression of the 
observable phase that we usually call $\varphi_\mu$. Again we can make 
a rough estimate with 
$\mu\simeq M_2\simeq m_{\tilde\nu}\simeq 200$ GeV (taking the derivative of 
$A(r)$),
\bea
d^{e}_{\chi^+}\simeq 1.5 \times 10^{-25}~~
\tan \beta \sin \varphi_\mu \mbox{ e cm}.
\eea
Now, comparing with the experimental bound on the electron EDM, we
obtain a much stronger bound, $(\tan \beta~ \varphi_\mu) \leq 0.01$.
These two examples give a clear idea of the strength of the 
``SUSY CP problem''.

As we have seen in these examples typically the bound on $\varphi_\mu$ is
stronger than the bound on $\varphi_A$. There are several reason for this, 
as we can see $\varphi_\mu$ enters the down-type sfermion mass matrix together
with $\tan \beta$ while $\varphi_A$ is not enhanced by this factor. 
Furthermore, $\varphi_\mu$ appears also in the chargino and neutralino mass
matrices. This difference is increased if we consider the bounds
on the original parameters at $M_\GUT$. The $\mu$ phase is unchanged in the
RGE evolution, but $\varphi_A = \mbox{arg}(M_{1/2} A_0)$ (where $M_{1/2}$ 
is the gaugino mass) is reduced due to large gaugino contributions to the
trilinear couplings in the running from $M_\GUT$ to $M_{W}$.  
The bounds we typically find in the literature\cite{Abel:2001vy,Barger:2001nu}
are,
\bea
\label{boundsphi}
\varphi_\mu \leq 10^{-2}-10^{-3}, ~~~~~~~~\varphi_A \leq 10^{-1}-10^{-2}.
\eea

Nevertheless, a full computation should take into account all the
different contributions to the electron and neutron EDM. In the case
of the electron, we have both chargino and neutralino contributions at
1 loop. For the neutron EDM, we have to include also the gluino
contribution, the quark chromoelectric dipole moments and the
dimension six gluonic operator \cite{Weinberg:1989dx,Ibrahim:1997nc,Ibrahim:1997gj,Chang:1998uc,Pospelov:2000bw,Abel:2001vy}. When all these
contributions are taken into account our estimates above may not be
accurate enough and the bound can be loosened.

In fact, there can be regions on the parameter space where different
contributions to the neutron or electron EDM have opposite signs and similar
size. Thus the complete result for these EDM can be smaller than the
individual contributions. In this way, it is possible to reduce the stringent
constraints on these phases and  $\varphi_A={\cal{O}}(1)$ and
$\varphi_\mu={\cal{O}}(0.1)$ can be still allowed
\cite{Ibrahim:1998je,Brhlik:1998zn,Bartl:1999bc,Brhlik:1999ub,Brhlik:1999pw,Ibrahim:1999af,Bartl:2001wc}.
However, when all the EDM constraints, namely electron, neutron and also
mercury atom EDM, are considered simultaneously the cancellation regions
practically disappear and the bounds in \eq{boundsphi} remain basically
valid \cite{Abel:2001mc,Lebedev:2004va}.

\subsection{Flavour changing neutral currents in the MSSM}
\label{sec:flavour}

In the previous section we have analysed the effects of the 
``flavour independent'' SUSY phases, $\varphi_\mu$ and $\varphi_A$, on the
EDMs of the electron and the neutron. However, we have seen that a generic
MSSM contains many other observable phases and flavour changing parameters. 
This huge number of new parameters in the SUSY soft breaking sector can 
easily generate dangerous contributions in FCNC and flavour changing CP 
violation processes.

Given the large number of unknown parameters involved in FC processes, 
it is particularly helpful to make use of the Mass Insertion formalism. 
The mass insertions are
defined in the so-called Super CKM (SCKM) basis. This is the basis
where the Yukawa couplings for the down or up quarks are diagonal and
we keep the neutral gaugino couplings flavour diagonal. In this basis
squark mass matrices are not diagonal and therefore the flavour
changing is exhibited by the non-diagonality of the sfermion
propagators. Denoting by $\Delta^f_{ij}$ the flavour-violating
off-diagonal entry appearing in the $f =
(u_L,d_L,u_R,d_R,u_\LR,d_\LR)$ sfermion mass matrices, the sfermion
propagators are expanded as a series in terms of
$\left(\delta^f\right)_{ij} = \Delta^f_{ij}/m_{\tilde f}$, which are
known as mass insertions (MI). Clearly the goodness of this approximation 
depends on the smallness of the expansion parameter $\delta^f_{ij}$. As we 
will see, indeed the phenomenological constraints require these parameters to
be small and it is usually enough to keep the first terms in this expansion.   
The use of the MI approximation
presents the major advantage that it is not necessary to know and
diagonalise the full squark mass matrix to perform an analysis of FCNC
in a given MSSM. It is enough to know the single entry contributing to
a given process and in this way it is easy to isolate the relevant
phases.

In terms of the MI, and taking all diagonal elements approximately equal to 
$m_{\tilde d}^2$, the down squark mass matrix is,
\bea
\Frac{M_{\tilde d}^2}{ m_{\tilde d}^2} \simeq  
\left(\matrix{1 & \left(\delta^d_{12}\right)_\LL&
  \left(\delta^d_{13}\right)_\LL  &\left(\delta^d_{11}\right)_\LR &
  \left(\delta^d_{12}\right)_\LR &\left(\delta^d_{13}\right)_\LR \cr
 \left(\delta^d_{12}\right)_\LL^*& 1 &
  \left(\delta^d_{23}\right)_\LL  &\left(\delta^d_{21}\right)_\LR &
  \left(\delta^d_{22}\right)_\LR &\left(\delta^d_{23}\right)_\LR\cr
\left(\delta^d_{13}\right)_\LL^*& \left(\delta^d_{23}\right)_\LL^* &1
 &\left(\delta^d_{31}\right)_\LR &
  \left(\delta^d_{32}\right)_\LR &\left(\delta^d_{33}\right)_\LR\cr
\left(\delta^d_{11}\right)_\LR^*&
  \left(\delta^d_{21}\right)_\LR^*  &\left(\delta^d_{31}\right)_\LR^* &
  1&\left(\delta^d_{12}\right)_\RR &\left(\delta^d_{13}\right)_\RR\cr 
\left(\delta^d_{12}\right)_\LR^*&
  \left(\delta^d_{22}\right)_\LR^*  &\left(\delta^d_{32}\right)_\LR^* &
  \left(\delta^d_{12}\right)_\RR^* &1&\left(\delta^d_{23}\right)_\RR\cr 
\left(\delta^d_{13}\right)_\LR^*&
  \left(\delta^d_{23}\right)_\LR^*  &\left(\delta^d_{33}\right)_\LR^* &
  \left(\delta^d_{13}\right)_\RR^* &\left(\delta^d_{23}\right)_\RR^*&1 
\cr
}\right), 
\eea
with all the off-diagonal elements complex which means we have 15 new moduli 
and 15 phases. The
same would be true for the up squark mass matrix, although the
$\left(\delta^u_{ij}\right)_\LL$ would be related to
$\left(\delta^d_{ij}\right)_\LL$  by a CKM rotation. Therefore there would be
a total of 27 moduli and  27 phases in the squark sector.

An illustrative example of the usage of the MI formalism is provided
by the SUSY contribution to $K$--$\bar K$ \cite{Hall:1985dx,Gabbiani:1988rb,Hagelin:1992tc,Gabbiani:1996hi} mixing.
\begin{figure}
\begin{center}
\includegraphics[width=8.6cm]{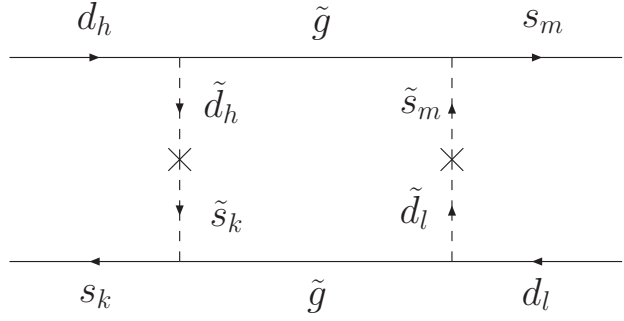}
\caption{1 loop contribution to $K$--$\bar K$ mixing}
\label{fig:Epsk}
\end{center}
\end{figure}
 The relevant diagram
at leading order in the MI approximation is shown in
Fig. \ref{fig:Epsk}. Here the MI are treated as new vertices in our
theory. We have to compute the contribution to the Wilson coefficients
of the different four--fermion operators in the $\Delta S=2$ effective
Hamiltonian \cite{Gabbiani:1996hi,Ciuchini:1998ix}. For example the Wilson coefficient
associated with the operator, $Q_1= \bar d^\alpha_L
\gamma^\mu s^\alpha_L ~ d^\beta_L \gamma_\mu s^\beta_L$, would be,
\bea
C_1 = -\frac{\alpha_s^2}{216 m_{\tilde q}^2}\left( 24 x f_6(x) + 66 
\tilde f_6(x) \right) \left(\delta^d_{12} \right)_\LL^2
\eea
with $x=m_{\tilde g}^2/m_{\tilde q}^2$ and the functions $f_6(x)$ and $\tilde f_6(x)$ given by,
\bea
\label{functions}
f_6(x) = \frac{6(1+3x)\log x + x^3 -9 x^2 - 9 x + 17}{6 (x-1)^5} \nn\\
\tilde f_6(x) = \frac{6x(1+x)\log x - x^3 -9 x^2 + 9 x + 1}{3 (x-1)^5}. 
\eea 
It is straightforward to understand the different factors in this
formula: we have four flavour diagonal gluino vertices providing a factor 
$g_s^4$ and the two MI which supply the necessary flavour transition.
The remainder corresponds only to the loop functions.
A full computation of the whole set of Wilson coefficients can be
found in Refs. \cite{Gabbiani:1996hi,Ciuchini:1998ix}.

The complete leading order expression for $K^0$--$\bar K^0$ mixing, using 
the Vacuum Insertion Approximation (VIA) for the matrix elements of the 
different operators, is \cite{Gabbiani:1996hi},
\bea
\label{kkmix}
\langle K^0 | {\mathcal H}_\eff^{\Delta S=2} | \bar{K}^0\rangle  &=&
 -\frac{\alpha_s^2}{216 m_{\tilde q}^2}~\frac{1}{3} m_K f_K^2 
\left\{ \right.\\
&&\left. \left(\left(\delta^d_{12} \right)_\LL^2 + \left(\delta^d_{12} 
\right)_\RR^2 \right)\left( 24 x f_6(x) + 66 
\tilde f_6(x) \right) \right. \nn \\ \
&+&\left. \left(\delta^d_{12} \right)_\LL \left(\delta^d_{12} \right)_\RR 
\left[ \left( 84 \left(\frac{m_K}{m_s+m_d}\right)^2 + 72 \right) x f_6(x) 
 \right. \right.  \nn \\ 
&&\qquad\qquad\qquad +\left.\left. \left( -24 \left(\frac{m_K}{m_s+m_d}
\right)^2 + 36 \right)
\tilde f_6(x) \right]  \right. \nn \\ 
&+&\left.\left(\left(\delta^d_{12} \right)_\LR^2 + \left(\delta^d_{12} 
\right)_\RL^2 
\right)\left( -132 \left(\frac{m_K}{m_s+m_d}\right)^2 \right) x f_6(x) \right. 
\nn\\
&+&\left. \left(\delta^d_{12} \right)_\LR \left(\delta^d_{12} \right)_\RL
\left[ -144 \left(\frac{m_K}{m_s+m_d}\right)^2 - 84 \right] \tilde f_6(x)
\right\} \nn
\eea
The neutral kaon mass difference and the mixing CP violating
parameter, $\varepsilon_K,$ are given by,
\bea
\Delta M_K = 2 \Re \langle K^0 | {\mathcal H}_\eff^{\Delta S=2} 
| \bar{K}^0\rangle \nn \\
\varepsilon_K = \frac{1}{\sqrt{2}\Delta M_K} \Im \langle K^0 | 
{\mathcal H}_\eff^{\Delta S=2} | \bar{K}^0\rangle 
\label{defepsk}
\eea
To obtain a model independent bound on the different MI, we assume
that each time only one of these MI is different from zero
neglecting accidental cancellations between different MIs. Moreover, it
is customary to consider only the gluino contributions leaving aside
other SUSY contributions as chargino, charged Higgs or neutralino. In
fact, in the presence of sizable MI, the gluino contribution provides
typically a large part of the full SUSY contribution.  Barring
sizable accidental cancellations between the SM and SUSY
contributions a conservative limit on the $\delta$s is obtained by
requiring the SUSY contribution by itself not to exceed the
experimental value of the observable under consideration.

The different MI bounds for the $\left(\delta^d_{12}\right)_a$
($a=$ LL,RR,LR) are presented in Table 1. As can be seen
explicitly in \eq{kkmix} gluino contributions are completely
symmetrical under the interchange $L\leftrightarrow R$ and therefore
the limits on $\left(\delta^d_{12}\right)_\RR$ are equal to those on
$\left(\delta^d_{12}\right)_\LL$ and the limits on
$\left(\delta^d_{12}\right)_\RL$ to those on
$\left(\delta^d_{12}\right)_\LR$. In this table we present the bounds
at tree level in the four fermion effective Hamiltonian (TREE), 
i.e. using directly \eq{kkmix} without any further QCD corrections and we
compare them with bounds obtained using the NLO QCD evolution with lattice
B parameters in the matrix elements \cite{Ciuchini:1998ix}. As we can see, although QCD
corrections may change the bounds even a factor 2, the tree level estimates
remain valid as order of magnitude bounds.
\begin{table}[t]
\begin{center}
{\tabcolsep10pt
\begin{tabular}{||c|c|c|c|c||}  \hline \hline
& \multicolumn{2}{c|}{$\sqrt{\vert\Re(\delta^{d}_{12})_\LL^2\vert}$ }
& \multicolumn{2}{c|}{$\sqrt{\vert\Im(\delta^{d}_{12})_\LL^2\vert}$ } \\
\hline
$x$ &TREE & NLO & TREE & NLO \\

0.3& $1.4\times 10^{-2}$ 

& $2.2\times 10^{-2}$

& $1.8\times 10^{-3}$ 

& $2.9\times 10^{-3}$
\\
1.0& $3.0\times 10^{-2}$ 

& $4.6\times 10^{-2}$

& $3.9\times 10^{-3}$ 

& $6.1\times 10^{-3}$
\\
4.0& $7.0\times 10^{-2}$ 

& $1.1\times 10^{-1}$

& $9.2\times 10^{-3}$ 

& $1.4\times 10^{-2}$ \\
\hline
& \multicolumn{2}{c|}{$\sqrt{\vert\Re(\delta^{d}_{12})_\LL(\delta^{d}_{12})_\RR\vert}$}
& \multicolumn{2}{c|}{$\sqrt{\vert\Im(\delta^{d}_{12})_\LL(\delta^{d}_{12})_\RR\vert}$}\\
\hline

$x$ & TREE & NLO & TREE & NLO\\

0.3& $1.8\times 10^{-3}$

& $8.6\times 10^{-4}$

& $2.3\times 10^{-4}$ 

& $1.1\times 10^{-4}$
\\
1.0& $2.0\times 10^{-3}$

& $9.6\times 10^{-4}$

& $2.6\times 10^{-4}$ 

& $1.3\times 10^{-4}$
\\
4.0& $2.8\times 10^{-3}$

& $1.3\times 10^{-3}$

& $3.7\times 10^{-4}$ 

& $1.8\times 10^{-4}$
\\
\hline 
& \multicolumn{2}{c|}{$\sqrt{\vert\Re(\delta^{d}_{12})_\LR^2\vert}$ }
& \multicolumn{2}{c|}{$\sqrt{\vert\Im(\delta^{d}_{12})_\LR^2\vert}$}\\
\hline

$x$ & TREE & NLO & TREE & NLO\\ 

0.3& $3.1\times 10^{-3}$ 

& $2.6\times 10^{-3}$ 

& $4.1\times 10^{-4}$ 

& $3.4\times 10^{-4}$ 
\\ 
1.0& $3.4\times 10^{-3}$ 

& $2.8\times 10^{-3}$

& $4.6\times 10^{-4}$ 

& $3.7\times 10^{-4}$ 
\\ 
4.0& $4.9\times 10^{-3}$ 

& $3.9\times 10^{-3}$ 

& $6.5\times 10^{-4}$ 

& $5.2\times 10^{-4}$  \\
\hline  \hline
 \end{tabular}}
\caption{Maximum allowed values for   $\vert\Re\left(\delta^d_{12}\right)_{AB}
\vert$ and $\vert\Im\left(\delta^d_{12}\right)_{AB}\vert$, with $A,B=(L,R)$
  for an average squark mass $m_{\tilde{q}}=500$ GeV and for different values
 of $x=m_{\tilde{g}}^2/m_{\tilde{q}}^2$. The bounds are given at tree level
in the effective Hamiltonian and at NLO in QCD corrections as explained 
in the text. For different values of $m_{\tilde q}$ the bounds scale roughly 
as $m_{\tilde q}/ 500$ GeV.}
\end{center}
\end{table}
The main conclusion we can draw from this table is that MI bounds in $s\to d$
transitions are very tight and this is specially true on the imaginary
parts.   This poses a very stringent constraint in most attempts to build 
a viable MSSM or any realistic supersymmetric flavour model
\cite{Ross:2004qn,Babu:2004dp,Babu:2005yr}. Conversely we can say that $s\to d$
transitions are very sensitive to the presence of relatively small
SUSY contributions and a deviation from SM predictions here could
provide the first indirect sign of SUSY \cite{Masiero:2000rg,Masiero:2000ni}.

CP violating supersymmetric contributions can also be very interesting in the
B system \cite{Bertolini:1990if,Ciuchini:1997zp}. Similarly to the previous 
case, we can 
build the $\Delta B =2$ effective Hamiltonian to obtain the
bounds from $B_d$--$\bar B_d$ mixing. A full calculation is presented in 
Ref. \cite{Becirevic:2001jj}, in Table 2 we present the results. 
\begin{table}[t]
\begin{center}
{\tabcolsep10pt
\begin{tabular}{@{}||c|c|c|c|c||}  \hline \hline
& \multicolumn{2}{c|}{$\vert\Re(\delta^{d}_{13})_\LL\vert$ }
& \multicolumn{2}{c|}{$\vert\Re(\delta^{d}_{13})_{\mbox{LL=RR}}\vert$}\\
\hline
$x$ &TREE & NLO & TREE & NLO \\

0.25& $4.9\times 10^{-2}$

& $6.2\times 10^{-2}$

& $3.1\times 10^{-2}$ 

& $1.9\times 10^{-2}$
\\
1.0& $1.1\times 10^{-1}$ 

& $1.4\times 10^{-1}$

& $3.4\times 10^{-2}$ 

& $2.1\times 10^{-2}$
\\
4.0& $6.0\times 10^{-1}$ 

& $7.0\times 10^{-1}$

& $4.7\times 10^{-2}$ 

& $2.8\times 10^{-2}$ \\
\hline
& \multicolumn{2}{c|}{$\vert\Im(\delta^{d}_{13})_\LL\vert$ }
& \multicolumn{2}{c|}{$\vert\Im(\delta^{d}_{13})_{\mbox{LL=RR}}\vert$}\\
\hline

$x$ & TREE & NLO & TREE & NLO\\

0.25& $1.1\times 10^{-1}$

& $1.3\times 10^{-1}$

& $1.3\times 10^{-2}$ 

& $8.0\times 10^{-3}$
\\
1.0& $2.6\times 10^{-1}$ 

& $3.0\times 10^{-1}$

& $1.5\times 10^{-2}$ 

& $9.0\times 10^{-3}$
\\
4.0& $2.6\times 10^{-1}$ 

& $3.4\times 10^{-1}$

& $2.0\times 10^{-2}$ 

& $1.2\times 10^{-2}$
\\
\hline 
& \multicolumn{2}{c|}{$\vert\Re(\delta^{d}_{13})_\LR\vert$ }
& \multicolumn{2}{c|}{$\vert\Re(\delta^{d}_{13})_{\mbox{LR=RL}}\vert$}\\
\hline

$x$ & TREE & NLO & TREE & NLO\\ 

0.25& $3.4\times 10^{-2}$ 

& $3.0\times 10^{-2}$ 

& $3.8\times 10^{-2}$ 

& $2.6\times 10^{-2}$ 
\\ 
1.0& $3.9\times 10^{-2}$ 

& $3.3\times 10^{-2}$

& $8.3\times 10^{-2}$

& $5.2\times 10^{-2}$ 
\\ 
4.0& $5.3\times 10^{-2}$

& $4.5\times 10^{-2}$ 

& $1.2\times 10^{-1}$ 

& $-$  \\
\hline
& \multicolumn{2}{c|}{$\vert\Im(\delta^{d}_{13})_\LR\vert$ }
& \multicolumn{2}{c|}{$\vert\Im(\delta^{d}_{13})_{\mbox{LR=RL}}\vert$}\\
\hline

$x$ & TREE & NLO & TREE & NLO\\

0.25& $7.6\times 10^{-2}$ 

& $6.6\times 10^{-2}$

& $1.5\times 10^{-2}$ 

& $9.0\times 10^{-3}$
\\
1.0& $8.7\times 10^{-2}$ 

& $7.4\times 10^{-2}$

& $3.6\times 10^{-2}$ 

& $2.3\times 10^{-2}$
\\
4.0& $1.2\times 10^{-1}$ 

& $1.0\times 10^{-1}$

& $2.7\times 10^{-1}$ 

& $-$
\\
\hline  \hline
 \end{tabular}}
\caption{ Maximum allowed values for   $\vert\Re\left(\delta^d_{13}
 \right)_{AB}\vert$ and $\vert\Im\left(\delta^d_{13}\right)_{AB}\vert$, 
with $A,B=(L,R)$ for an average squark mass $m_{\tilde{q}}=500$ GeV and 
different values of $x=m_{\tilde{g}}^2/m_{\tilde{q}}^2$. with NLO evolution and lattice $B$ parameters, denoted by NLO. The
missing entries correspond to cases in which no constraint was found for
$\vert\left(\delta^d_{ij}\right)_{AB}\vert < 0.9$.}
\end{center}
\label{tab:MIBd}
\end{table}
As we can see here, the constraints in the $B_d$ system are less stringent than
in the $K$ sector specially in the imaginary parts of the MI which come from
$\varepsilon_K$ and $\sin 2 \beta$ \cite{Aubert:2002ic,Abe:2001xe,Abe:2003yu,Affolder:1999gg}. 
At first sight this may be surprising as
it is well-known that CP violation is more prominent in the B system. 
To understand this difference we analyse more closely these two observables.

Let us assume that the imaginary part of $K^0$--$\bar K^0$ and 
$B_d^0$--$\bar B_d^0$ is entirely provided by SUSY from a single
$\left(\delta^d_{ij}\right)_\LL$ MI, while the real part is mostly given by SM
loops.
The Standard Model contribution to $K^0$--$\bar K^0$ mixing is given by,
\bea
\label{SMmix}
\langle K^0 | {\mathcal H}_\eff^{\Delta S=2} | \bar{K}^0\rangle = 
- \frac{\alpha_\elm^2}{8 M_W^2 \sin^4 \theta_W} ~\frac{m_{c}^2}{M_{W}^2} 
\frac{f_K^2 m_K}{3} (V_{cs} V_{cd}^*)^2
\eea
Replacing this expression and \eq{kkmix} in \eq{defepsk} we have,
\bea
\label{Kestim}
 \varepsilon_K^{\mbox{\tiny SUSY}} =
\frac{\left.\mbox{Im}~ M_{12}\right|_{\mbox{\tiny SUSY}}}{\left.\sqrt{2}
~\Delta M_K\right|_{\mbox{\tiny SM}}}
\simeq
\frac{\alpha_s^2 \sin^2 \theta_W}{\alpha_\elm^2}~
\frac{ M_W^4}{M_{\mbox{\tiny SUSY}}^2~m_c^2}~\frac{\mbox{Im} \left\{
(\delta^d_{12})_\LL^2\right\}}{\left(V_{cd} V_{cs}^*\right)^2} \nn \\
~\frac{8 (24
x  f_6(x) + 66 \tilde f_6(x))}{216 \sqrt{2}}
\simeq  12.5 \times 84 \times \frac{\mbox{Im} \left\{
(\delta^d_{12})_\LL^2\right\}}{0.05}\times 0.026, \nn\\ 
\varepsilon_K^{\mbox{\tiny SUSY}} \leq 2.3 \times 10^{-3} \Rightarrow 
\sqrt{\mbox{Im} \left\{(\delta^d_{12})_\LL^2\right\}} \leq 2.0 \times 10^{-3}
\eea
where we used $x=1$ and $M_{\mbox{\tiny SUSY}} = 500$ GeV.
In the same way, we can obtain an estimate of the MI bound from the $B^0$ 
CP asymmetries. The gluino and SM contributions to $B^0$--$\bar B^0$ mixing 
are analogous to \eq{kkmix} and \eq{SMmix} respectively changing 
$f_K^2 m_K \to f_B^2 m_B$, $m_s \to m_b$, 
$m_c\to m_t$ and $(V_{cs} V_{cd}^*)$ by $(V_{tb} V_{td}^*)$. Then we have,
\bea
\label{Bestim}
\left. a_{J/\psi} \right|_{\mbox{\tiny SUSY}} =
\frac{\left.\mbox{Im}~ M_{12}\right|_{\mbox{\tiny SUSY}}}{\left|
M_{12}\right|_{\mbox{\tiny SM}}} \simeq
\frac{\alpha_s^2\sin^2\theta_W}{\alpha_\elm^2}~\frac{M_W^4}{M_{\mbox{\tiny
SUSY}}^2~m_t^2}~\frac{\mbox{Im} \left\{
(\delta^d_{13})_\LL^2\right\}}{\left(V_{tb} V_{td}^*\right)^2} \nn \\
~\frac{8 (24
x  f_6(x) + 66 \tilde f_6(x))}{216}
\simeq  12.5 \times 0.005 \times \frac{\mbox{Im} \left\{
(\delta^d_{13})_\LL^2\right\}}{(0.008)^2} \times 0.037, \nn\\ 
\left. a_{J/\psi} \right|_{\mbox{\tiny SUSY}} \leq 0.74 
\Rightarrow \sqrt{\mbox{Im} \left\{
(\delta^d_{13})_\LL^2\right\}} \leq 0.14
\eea
From here we see that, although there is a difference due to masses
and mixings, $m_c^2 (V_{cs} V_{cd}^*)^2$ versus $m_t^2 (V_{tb}
V_{td}^*)^2$, the main reason for the difference in the MI bounds is
the experimental sensitivity to CP violation observables. In the kaon
system we can measure imaginary contributions to $K$--$\bar K$ mixings
three orders of magnitude smaller than the real part while in the B
system we can only distinguish imaginary contributions if they are of
the same order as the mass difference. It is clear that we need much
larger MI in the B system that in the K system to have observable
effects \cite{Masiero:2000rg}. On the other hand, as we will show in 
the next section, in
realistic flavour models we expect larger MI in b transitions that in
s transitions. Whether the B--system or K--system is more sensitive to
SUSY will finally depend on the particular model considered. 

Similarly, $b \to s$ transitions can be very interesting in SUSY models
\cite{Lunghi:2001af,Goto:2002xt,Chang:2002mq,Khalil:2002fm,Harnik:2002vs,Ciuchini:2002uv,Baek:2003kb,Agashe:2003rj,Kane:2003zi,Mishima:2003ta,Endo:2004xt,Endo:2004dc}. 
In fact, the only phenomenological constraints in this sector come from the 
$b \to s \gamma$ process. As we can see in Table 3, the bounds are stringent 
only for the $(\delta^d_{23})_{\LR}$ while they are very weak for 
$(\delta^d_{23})_{\LL,\RR}$. A
large $(\delta^d_{23})_{\LL,\RR,\LR}$ could have observable effects in several 
decays like $B \to \Phi K_S$ that can still differ from the SM predictions
\cite{Aubert:2002nx,Abe:2003yt}.
\begin{table}[t]
 \begin{center}
{\tabcolsep10pt
 \begin{tabular}{||c|c|c||}  \hline \hline
  & & \\
  $x$ &   ${\left|\left(\delta^{d}_{23}  \right)_\LL
\right|} $   &  
 ${\left|\left(\delta^{d}_{23}\right)  _\LR\right| }$   
\\
  & & \\\hline
 $
   0.3
 $ &
 $
4.4
 $ & $
1.3\times 10^{-2}
 $ \\
 $
   1.0
 $ &
 $
8.2
 $ & $
1.6\times 10^{-2}
 $ \\
 $
   4.0
 $ &
 $
26
 $ & $
3.2\times 10^{-2}
 $ \\ \hline \hline
 \end{tabular}}
\caption{Limits on $|\left(\delta_{13}^{d}\right)|$, from the $b\to s \gamma$
decay, for an average squark
 mass $m_{\tilde{q}}=500\mbox{GeV}$ and for different values of
 $x=m_{\tilde{g}}^2/m_{\tilde{q}}^2$. For different values of
 $m_{\tilde{q}}$, the limits can be obtained multiplying the ones in
 the table by $\left(m_{\tilde{q}}(\mbox{GeV})/500\right)^2$.}
 \end{center}
\label{tab:bsg}
 \end{table}

Another interesting CP violating process in SUSY is 
$\varepsilon^\prime/\varepsilon$
\cite{Gabrielli:1994ff,Masiero:1999ub,Eyal:1999gk,Barbieri:1999ax,Babu:1999xf,Khalil:2000ci,Buras:2000qz}. 
We present the corresponding MI bounds from $\varepsilon^{\prime}/\varepsilon
< 2.7 \times 10^{-3}$  \cite{Barr:1993rx,Alavi-Harati:1999xp} in Table 4.
\begin{table}[t]
 \begin{center}
{\tabcolsep10pt
 \begin{tabular}{||c|c|c||}  \hline \hline
  & & \\
  $x$ &   ${\left|\Im \left(\delta^{d}_{12}  \right)_\LL
\right|} $   &  
 ${\left|\Im \left(\delta^{d}_{12}\right)  _\LR\right| }$   
\\
  & & \\\hline
 $
   0.3
 $ &
 $
1.0\times 10^{-1}
 $ & $
1.1\times 10^{-5}
 $ \\
 $
   1.0
 $ &
 $
4.8\times 10^{-1}
 $ & $
2.0\times 10^{-5}
 $ \\
 $
   4.0
 $ &
 $
2.6\times 10^{-1}
 $ & $
6.3\times 10^{-5}
 $ \\ \hline \hline
 \end{tabular}}
\caption{Limits from $\varepsilon^{\prime}/\varepsilon < 2.7 \times
 10^{-3}$  
on $\Im\left(\delta_{12}^{d}\right)$, for an average squark
 mass $m_{\tilde{q}}=500\mbox{GeV}$ and for different values of
 $x=m_{\tilde{g}}^2/m_{\tilde{q}}^2$. For different values of
 $m_{\tilde{q}}$, the limits can be obtained multiplying the ones in
 the table by $\left(m_{\tilde{q}}(\mbox{GeV})/500\right)^2$.}
 \end{center}
\label{tab:eps'}
 \end{table}
This observable is more sensitive to chirality changing MI due to the
dominance of the gluonic and electroweak penguin operators. The bounds
on $\Im \left(\delta^{d}_{12}\right) _\LR$ look really tight and in
fact these are the strongest bounds attainable on this MI.  However,
it is important to remember that these off-diagonal LR mass insertions
come from the trilinear soft breaking terms which in realistic models
are always proportional to fermion masses. Thus this MI typically
contains a suppression $m_s/M_{\mbox{\tiny SUSY}} \simeq 2 \times
10^{-4}$ for $M_{\mbox{\tiny SUSY}} = 500$ GeV. So, if we consider
this ``intrinsic'' suppression the bounds are less impressive.

In summary, these MI bounds show the 
present sensitivity of CP violation experiments to the presence of new 
phases and flavour structures in the SUSY soft breaking terms. 
An important lesson we can draw from the stringent bounds in the
tables is that, in fact we already posses a crucial information on the
enormous (123-dimensional) parameter space of a generic MSSM: most of
this parameter space is already now excluded by flavour physics, and
indeed the ``realistic'' MSSM realisation should not depart too
strongly from the CMSSM, at least barring significant accidental
cancellations.

 \subsection{Mass Insertion bounds from leptonic processes}

 \label{sec:leptons}
 In this section, we study the constraints on slepton mass matrices in
 low energy SUSY imposed by several LFV transitions, namely
 $l_i\rightarrow l_j \gamma$, $l_i\rightarrow l_jl_kl_k$ and
 $\mu$--$e$ transitions in nuclei \cite{Paradisi:2005fk}. The present and 
projected bounds on 
 these processes are summarized in Table \ref{tab:exp}.  
\begingroup
 \begin{table}
 \begin{center}
 \begin{tabular*}{0.8\textwidth}{@{\extracolsep{\fill}}||c|c|c||}
 \hline\hline
Process & Present Bounds & Expected Future Bounds  \\[0.2pt] 
 \hline
 BR($\mu \to e\,\gamma$) & $1.2~ \times~ 10^{-11}$ &
 $\mathcal{O}(10^{-13} - 10^{-14})$ \\
 BR($\mu \to e\,e\,e$) & $1.1~ \times~ 10^{-12}$ &
 $\mathcal{O}(10^{-13} - 10^{-14})$ \\
 BR($\mu \to e$ in Nuclei (Ti)) & $1.1~ \times~ 10^{-12}$ &
$\mathcal{O}(10^{-18})$ \\
 BR($\tau \to e\,\gamma$) & $1.1~ \times~ 10^{-7}$ &
 $\mathcal{O}(10^{-8}) $ \\
 BR($\tau \to e\,e\,e$) & $2.7~ \times~ 10^{-7}$ &
 $\mathcal{O}(10^{-8}) $ \\
 BR($\tau \to e\,\mu\,\mu$) & $2.~ \times~ 10^{-7}$ &
 $\mathcal{O}(10^{-8}) $ \\
 BR($\tau \to \mu\,\gamma$) & $6.8~ \times~ 10^{-8}$ &
 $\mathcal{O}(10^{-8}) $ \\
 BR($\tau \to \mu\, \mu\, \mu$) & $2~ \times~ 10^{-7}$ &
 $\mathcal{O}(10^{-8}) $ \\
 BR($\tau \to \mu\, e\,e$) & $2.4~ \times~ 10^{-7}$ &
 $\mathcal{O}(10^{-8}) $ \\
\hline\hline
 \end{tabular*}
 \end{center}
 \caption{Present and Upcoming experimental limits on various leptonic 
processes at 90\% C.L.}
 \label{tab:exp}
 \end{table}
 \endgroup
 These processes are mediated by chargino and neutralino loops and
 therefore they depend on all the parameters entering chargino and
 neutralino mass matrices.  In order to constrain the leptonic MIs
 $\delta^{ij}$, we will first obtain the spectrum at the weak scale
 for our SU(5) GUT theory as has been mentioned in detail in section
 \ref{sec:constraints}.  Furthermore, we take all the flavor
 off-diagonal entries in the slepton mass matrices equal to zero
 except for the entry corresponding to the MI we want to bound.  To
 calculate the branching ratios of the different processes, we work in
 the mass eigenstates basis through a full diagonalization of the
 slepton mass matrix.  So, imposing that the contribution of each
 flavor off-diagonal entry to the rates of the above processes does
 not exceed (in absolute value) the experimental bounds, we obtain the
 limits on the $\delta^{ij}$'s, barring accidental cancellations.

 The process that sets the most stringent bounds is the
 $l_{i}\rightarrow l_{j}\gamma$ decay, whose amplitude has the form
 \bea
 T=m_{l_i}\epsilon^{\lambda}\overline{u}_j(p-q)[iq^\nu\sigma_{\lambda\nu}
 (A_{L}P_{L}+A_{R}P_{R})]u_i(p)\,, \eea where $p$ and $q$ are momenta
 of the leptons $l_k$ and of the photon respectively, $P_{R,L}=
 \frac{1}{2}(1 \pm \gamma_5)$ and $A_{L,R}$ are the two possible
 amplitudes entering the process. The lepton mass factor $m_{l_i}$ is
 associated to the chirality flip present in this transition.  In a
 supersymmetric framework, we can implement the chirality flip in
 three ways: in the external fermion line (as in the SM with massive
 neutrinos), at the vertex through a higgsino Yukawa coupling or in
 the internal gaugino line together with a chirality change in the
 sfermion line.  The branching ratio of $l_{i}\rightarrow l_{j}\gamma$
 can be written as
 \bea
 \frac{BR(l_{i}\rightarrow  l_{j}\gamma)}{BR(l_{i}\rightarrow 
l_{j}\nu_i\bar{\nu_j})} = 
 \frac{48\pi^{3}\alpha}{G_{F}^{2}}(|A_L^{ij}|^2+|A_R^{ij}|^2)\,,
  \nonumber
 \eea
 with the SUSY contribution to each amplitude given by the sum of two terms
$A_{L,R}=A_{L,R}^{n}+A_{L,R}^{c}$. Here $A_{L,R}^{n}$ and $A_{L,R}^{c}$ denote the
contributions from the neutralino and chargino loops respectively.

 \begingroup
 \begin{table}
 \begin{center}
 \begin{tabular*}{0.6\textwidth}{@{\extracolsep{\fill}}||c|c|c|c||}
\hline\hline 
Type of $\delta^l_{12}$ &$\mu \to e\, \gamma$ & $\mu \to e\,e\,e$ & $\mu \to e$
conversion in $Ti$ 
 \\[0.2pt] 
 \hline
 LL & $6 \times 10^{-4}$ & $2\times 10^{-3}$ & $2\times 10^{-3}$ \\
 RR & - & $0.09$ & - \\
 LR/RL & $1 \times 10^{-5}$ & $3.5 \times 10^{-5}$ & $3.5 \times 10^{-5}$ \\
\hline\hline
 \end{tabular*}
 \end{center}
 \caption{Bounds on leptonic $\delta^l_{12}$ from various $\mu \to e$ 
 processes. The bounds are obtained by making a scan of $m_0$ and $M_{1/2}$ 
 over the ranges $m_{0}<380$\,\rm{GeV} and $M_{1/2}<$160\,\rm{GeV}
 and varying $\tan\beta$ within $5<\tan\beta<15$.
 The bounds are rather insensitive to the sign of the $\mu$ mass term.
}
 \label{tab:mue}
 \end{table}
 \endgroup
 Even though all our numerical results presented in Tables 
\ref{tab:mue}--\ref{tab:taumu} 
 are obtained performing an exact diagonalization of sfermion and gaugino
 mass matrices, it is more convenient for the discussion to use the
 expressions for the $l_i \rightarrow l_j \gamma$ amplitudes in the MI
 approximation.
 In particular, we treat both the slepton mass matrix
 and the chargino and neutralino mass matrix off-diagonal elements as
 mass insertions.\footnote{This approximation is well justified and
 reproduces the results of the full computation very accurately in a
 large region of the parameter space \cite{Paradisi:2005fk}.}
 In this approximation, we have the following expressions
 \bea
 \label{MIamplL}
 A^{ij}_{L}&=&\Frac{\alpha_{2}}{4\pi} \frac{
   \left(\delta^l_{ij}
   \right)_{\LL}}{m_{\tilde l}^{2}}
 ~\Bigg[~f_{1n}(a_2)\!+\!f_{1c}(a_2)\!+\!
 \frac{\mu M_{2}\tan\beta}{(M_{2}^2\!-\!\mu^2)}
 \bigg(f_{2n}(a_2,b)\!+\!f_{2c}(a_2,b)\bigg)\\\nonumber
 &&\qquad\quad+ \tan^2\theta_{W}\,
 \bigg(f_{1n}(a_1)+\mu M_{1}\tan\beta\bigg(\frac{f_{3n}(a_1)}{m_{\tilde l}^{2}}+
 \frac{f_{2n}(a_1,b)}{(\mu^2\!-\!M_{1}^2)}\bigg)\!\bigg)\Bigg]\\
 &+& \Frac{\alpha_{1}}{4\pi}~
 \frac{\left(\delta^l_{ij}
   \right)_{\RL}}{m_{\tilde
l}^{2}}~\left(\frac{M_1}{m_{l_i}}\right)~2~f_{2n}(a_1)\,,\nonumber
 \eea
 \bea
 \label{MIamplR}
 A^{ij}_R=\frac{\alpha_{1}}{4\pi}&\!\!\Bigg(\!\!&\frac{ \left(\delta^l_{ij}
   \right)_{\RR}}{m_{\tilde
l}^{2}}
 \left[4f_{1n}(a_1)+\mu M_{1}\tan\beta\left(\frac{f_{3n}(a_1)}{m_{\tilde l}^{2}}-
 \frac{2f_{2n}(a_1,b)}{(\mu^2\!-\!M_{1}^2)}\right)\right]\\
 &+&
 \frac{ \left(\delta^l_{ij}
   \right)_{\LR}}{m_{\tilde l}^{2}}~\left(\frac{M_1}{m_{l_i}}\right)~
 2~f_{2n}(a_1) \Bigg)\,, \nn
 \eea
where $\theta_W$ is the weak mixing angle, $a_{1,2}=M^{2}_{1,2}/\tilde{m}^{2}$,
$b=\mu^2/m_{\tilde l}^{2}$ and $f_{i(c,n)}(x,y)=f_{i(c,n)}(x)-f_{i(c,n)}(y)$.
The loop functions $f_i$ are given as
\begin{eqnarray}
f_{1n}(x) &=& (-17x^3+9x^2+9x-1+6x^2(x+3)\ln x)/(24(1-x)^5)\nonumber, \\
f_{2n}(x) &=& (-5x^2+4x+1+2x(x+2)\ln x)/(4(1-x)^4), \nonumber \\
f_{3n}(x) &=& (1+9x-9x^2-x^3+6x(x+1)\ln x)/(3(1-x)^5), \nonumber \\
f_{1c}(x) &=&  (-x^3-9x^2+9x+1+6x(x+1)\ln x)/(6(1-x)^5), \nonumber \\
f_{2c}(x) &=& (-x^2-4x+5+2(2x+1)\ln x)/(2(1-x)^4)\,.
\end{eqnarray}
 \begingroup
 \begin{table}
 \begin{center}
 \begin{tabular*}{0.6\textwidth}{@{\extracolsep{\fill}}||c|c|c|c||}
\hline\hline 
Type of $\delta^l_{13}$ &$\tau \to e\, \gamma$ & $\tau \to e\,e\,e$ & $\tau \to e \mu
\mu$ \\[0.2pt] 
 \hline
 LL & $0.15$ & $-$ & -\\
 RR & - & - & - \\
 LR/RL & $0.04$ & $0.5$ & - \\
\hline\hline
 \end{tabular*}
 \end{center}
 \caption{Bounds on leptonic $\delta^l_{13}$ from various $\tau \to e$ 
  processes obtained using the same values of SUSY parameters as in Table
\ref{tab:mue}.}
 \label{tab:taue}
 \end{table}
 \endgroup
 \begingroup
 \begin{table}
 \begin{center}
 \begin{tabular*}{0.6\textwidth}{@{\extracolsep{\fill}}||c|c|c|c||}\hline\hline
 Type of $\delta^l_{23}$ &$\tau \to \mu\, \gamma$ & $\tau \to \mu\,\mu\,\mu$ & $\tau
\to \mu\, e\, e$ 
 \\[0.2pt] 
 \hline
 LL & $0.12$ & - & -\\
 RR & - & - & - \\
 LR/RL & $0.03$ & - & 0.5 \\
\hline\hline
 \end{tabular*}
 \end{center}
 \caption{Bounds on leptonic $\delta^l_{23}$ from various $\tau \to \mu$ 
processes  obtained using the same values of SUSY parameters as in Table
\ref{tab:mue}.}
 \label{tab:taumu}
 \end{table}
\endgroup
We note that all $ \left(\delta^l_{ij}
   \right)_{\LL}$ contributions with internal
chirality flip are $\tan\beta$-enhanced. On the other hand, the only
term proportional to $ \left(\delta^l_{ij}
   \right)_{\LR}$ arises from pure $\tilde B$
exchange and it is completely independent of $\tan\beta$, as can be
seen from Eqs.~(\ref{MIamplL}) and (\ref{MIamplR}). Therefore the
phenomenological bounds on $ \left(\delta^l_{ij}
   \right)_{\LL}$ depend on $\tan\beta$
to some extent, while those on $ \left(\delta^l_{ij}
   \right)_{\LR}$ do not.  The bounds
on LL and RL MIs are expected to approximately fulfill the relation
 $$
  \left(\delta^l_{ij}
   \right)_{\LR} \simeq\frac{m_i}{\tilde{m}}\tan\beta\,\, \left(\delta^l_{ij}
   \right)_{\LL}\,.
 $$
This is confirmed by our numerical study.

The $\delta^d_{\RR}$ sector requires some care because of the presence
of cancellations among different contributions to the amplitudes in
regions of the parameter space.  The origin of these cancellations is
the destructive interference between the dominant contributions coming
from the $\tilde B$ (with internal chirality flip and a
flavor-conserving LR mass insertion) and $\tilde B \tilde H^{0}$
exchange \cite{Paradisi:2005fk,Masina:2002mv}.  We can see this in the MI
approximation if we compare the $\tan \beta$ enhanced terms in the
second line of \eq{MIamplL} with the $\tan \beta$ enhanced terms in
\eq{MIamplR}. Here the loop function $f_3(a_1)$ corresponds to the
pure $\tilde B$ contribution while $f_{2n}(a_1,b)$ represents the
$\tilde B \tilde H^{0}$ exchange. These contributions have different
relative signs in \eq{MIamplL} and \eq{MIamplR} due to the opposite
sign in the hypercharge of $SU(2)$ doublets and singlets.  Thus, the
decay $l_i\rightarrow l_j\gamma$ does not allow to put an absolute 
bound on the RR sector.  We can still take into account other LFV 
processes such as
$l_i\rightarrow l_jl_kl_k$ and $\mu$--$e$ in nuclei.  These processes
get contributions not only from penguin diagrams (with both photon and
Z-boson exchange) but also from box diagrams.  Still the contribution
of dipole operators, being also $\tan\beta$-enhanced, is dominant.
Disregarding other contributions, one finds the relations \bea
 \label{relations}
 \frac{Br(l_i\!\rightarrow\!l_jl_kl_k)}{Br(l_i\!\rightarrow\!l_j\gamma)}
 \simeq\! 
 \frac{\alpha_{e}}{3\pi}\left(\!\log\frac{m^2_{l_i}}{m^2_{l_k}}\!-\!3\!\right)
 \,,\nn\\
 Br(\mu - e {\rm\ in \ Ti}) \simeq\,\,\alpha_{e}BR(\mu \rightarrow e \gamma)\,,
 \eea
which clearly shows that $l_i \rightarrow l_j \gamma$ is the strongest
 constraint and gives the more stringent bounds on the different 
 $\delta_{ij}$'s.  As we have mentioned above, however,
 in the case of $\delta^l_{\RR}$ 
the dominant dipole contributions interfere destructively in regions
of parameters, 
so that $Br(l_i\rightarrow l_j\gamma)$ 
is strongly suppressed while $Br(\mu - e\rm{\ in \ nuclei})$ and 
 $Br(l_i\rightarrow l_jl_kl_k)$ are dominated 
by monopole penguin (both $\gamma^*$ and Z-mediated) and box diagrams.
The formulae 
for these contributions can be found in Ref. \cite{Hisano:1995cp}.
However, given that non-dipole contributions are typically much 
smaller than dipole ones outside the cancellation region,
 it follows that the bound on $\delta^l_{\RR}$ from $\mu\to eee$ are
 expected to be less stringent than the one on $\delta^l_{\LL}$ from 
 $\mu\to e \gamma$ by a factor $\sqrt{\alpha/(8\pi)}~1/\tan \beta\simeq
 0.02/\tan\beta$, if the experimental upper bounds on the two BRs were the same.
 This is partly compensated by the fact that the present 
 experimental upper bound on the $BR(\mu\to eee)$ is one order of magnitude smaller
 than that on $BR(\mu\to e\gamma)$, as shown in Tab.~\ref{tab:exp}.  
 On the other hand, the process $BR(\mu - e\rm{\ in \ nuclei})$ suffers
 from cancellations through the interference of dipole and non-dipole 
 amplitudes as well. 
 These cancellations prevent us from getting a bound in the RR sector
 from the $\mu$--$e$ conversion in nuclei now as well as in the 
 future when their experimental sensitivity will be improved.
 However, the $\mu\rightarrow e\gamma$ and 
 $\mu$--$e$ in nuclei amplitudes exhibit cancellations in different regions
 of the parameter space so that the combined use of these two constraints
 produces a competitive or even stronger bound than the one we get 
 from $BR(\mu\rightarrow eee)$ alone \cite{Paradisi:2005fk}.

 We summarize the different leptonic bounds in tables
 \ref{tab:mue}--\ref{tab:taumu}. All these bounds are obtained making
 a scan of $m_0$ and $M_{1/2}$ over the ranges $m_{0}<$380\,GeV and
 $M_{1/2}<$160\,GeV and therefore correspond to the heaviest possible
 sfermions. As expected, the strongest bounds for $\delta^l_{\LL}$ and
 $\delta^l_{\LR}$ come always from $\mu \to e \gamma$, $\tau \to \mu
 \gamma$ and $\tau \to e \gamma$ processes.  In the case of
 $\delta^l_{\RR}$ we can only obtain a mild bound for
 $\left(\delta^l_{12}\right)_{\RR}$ from $\mu \to e e e$ and there are
 no bounds for $\left(\delta^l_{23}\right)_{\RR}$ and
 $\left(\delta^l_{13}\right)_{\RR}$. Notice, however, that does not
 mean that these LFV processes are not effective to constrain the SUSY
 parameter space in the presence of RR MIs. For most of the values of
 $m_0$ and $M_{1/2}$ there is no cancellation and the values of these
 MI are required to be of the order of the LL bounds. Only for those
 values of $m_0$ and $M_{1/2}$ satisfying the cancellation conditions
 a large value of the RR MI is allowed. Therefore, we must check
 individually these constraints for fixed values of the SUSY
 parameters.

\subsubsection{$(g-2)_{\mu}$}

One of the most stringent flavour conserving constraints in
the leptonic sector comes from the measurement of the 
anomalous magnetic moment of the muon 
[$a_\mu = (g-2)_{\mu}/2$].  This has been measured very precisely 
in the last few years \cite{Brown:2000sj,Bennett:2002jb,Bennett:2004pv}, provides a first hint of physics 
beyond the SM has been widely discussed in the recent literature.
Despite substantial progress both on the experimental 
and on the theoretical sides, the situation is not completely
clear yet (see Ref.~\cite{Passera:2004bj,Passera:2005mx} for an updated discussion).

Most recent analyses converge towards a $2\sigma$ 
discrepancy in the $10^{-9}$ range \cite{Passera:2007fk,Jegerlehner:2007xe}:
\beq
 \Delta a_{\mu} =  a_{\mu}^{\rm exp} - a_{\mu}^{\rm SM}
\approx (2 \pm 1) \times 10^{-9}~.
\label{eq:amu_exp}
\eeq

The main SUSY contribution to $a^{\rm MSSM}_\mu$ is usually 
provided by the loop exchange of charginos and sneutrinos.
The basic features of the supersymmetric contribution to $a_\mu$ are correctly
reproduced by the following approximate expression:
\beq
\frac{a^{\rm MSSM}_\mu}{ 1 \times 10^{-9}}  
\approx 1.5\left(\frac{\tan\beta }{10} \right) 
\left( \frac{300~\rm GeV}{m_{\tilde \nu}} \right)^2
\left(\frac{\mu M_2}{m^{2}_{\tilde \nu}} \right)~,
\label{eq:g_2}
\eeq
which provides a good approximation to the full one-loop result
\cite{Moroi:1995yh}

The most relevant feature of Eqs.~(\ref{eq:g_2}) is that the sign of
$a^{\rm MSSM}_\mu$ is fixed by the sign of the $\mu$ term so that the
$\mu>0$ region is strongly favored.  This is specially true for the
Standard Model prediction which uses the data from $e^+ e^-$
collisions to compute the hadronic vacuum polarization (HVP).  This
predicts a smaller value than the experimental result by about $3
~\sigma$.  In case one uses the $\tau$ data to compute the HVP, the
discrepancy with SM is reduced to about $1~ \sigma$, but it still
favors a positive correction and disfavors strongly a sizable negative
contribution. Thus, taking $\mu>0$, the region of parameter space
considered in this analysis satisfies the constraint of
\eq{eq:amu_exp}.

\subsection{Grand unification and Flavour}
In the previous section, we have seen how flavour can be used 
to constrain various supersymmetric parameters in generic MSSM. 
What happens if one has a Grand Unified theory instead of the
Standard Model ?  Well, the answer to this question crucially depends 
on the mechanism of transmission of supersymmetry breaking to the visible
sector, or more exactly on the scale of the interactions mediating SUSY
breaking. In the case, the scale of mediation is much below the scale of Grand
Unification, then the soft supersymmetry breaking parameters will not see any 
effects of Grand Unification on them. On the other hand, if indeed mediation of
supersymmetry breaking occurs above the Grand Unification scale, as
in supergravity theories then the manner of supersymmetry breaking
plays a crucial role in determining the flavour aspects of the 
theory. Lets consider two representative situations in the following :

\subsubsection{SUSY-GUTs and Flavour dependent SUSY breaking}
As we have been discussing, in a SUSY-GUT, quarks and leptons sit in 
same multiplets and are transformed ones into the others through GU 
symmetry transformations.  If the supergravity Lagrangian, and, 
in particular, its K\"ahler function 
are present at a scale larger than the GUT breaking scale, they have to 
fully respect the underlying gauge symmetry  which is the GU symmetry itself.
The subsequent SUSY breaking will give rise to  the usual soft breaking terms
in the Lagrangian. In particular, the sfermion mass matrices, originating 
from the K\"ahler potential, will have to respect the underlying GU symmetry.
Hence we expect hadron-lepton correlations among entries of the sfermion 
mass matrices. In other words, the quark-lepton unification seeps also 
into the SUSY breaking soft sector \cite{Ciuchini:2003rg}.

One of the striking aspects of this scenario is that the 
imposition of a GU symmetry on the $\mathcal{L}_{soft}$ entails relevant 
implications
at the weak scale. This is because the flavour violating (FV) mass-insertions 
do not get strongly renormalised through RG scaling from the GUT scale to the
weak scale in the absence of new sources of flavor violation. On the other 
hand, if such new sources are present, for instance due to the presence of 
new neutrino Yukawa couplings in SUSY GUTs with a seesaw mechanism for 
neutrino 
masses, then one can compute the RG-induced effects in terms of these new 
parameters. Hence,  the correlations between hadronic and leptonic flavor 
violating MIs survive at the weak scale to a good approximation. As for  
the flavor conserving (FC) mass insertions (i.e., the diagonal entries 
of the sfermion mass matrices),  they get strongly 
renormalised, but in a way which is RG computable.

To summarise, in SUSY GUTs where the soft SUSY breaking terms respect boundary 
conditions which are subject to  the GU symmetry to start with, we generally 
expect the presence of relations among the (bilinear and trilinear) 
scalar terms in the hadronic and leptonic sectors. Such relations hold 
true at the (superlarge) energy scale where the correct symmetry of the 
theory is the GU symmetry. After its breaking, the mentioned relations 
will undergo corrections which are computable through the appropriate 
RGE's which are related to the specific structure of the theory between
 the GU and the electroweak scale (for instance, new Yukawa couplings 
due to the presence of right-handed (RH) neutrinos acting down to the RH 
neutrino mass scale, presence of a symmetry breaking chain with the 
appearance of new symmetries at intermediate scales, etc.). As a result 
of such a computable running, we can infer the correlations between the 
softly SUSY breaking  hadronic and leptonic $\delta$ terms at the low scale 
where we perform our FCNC tests. 

Given that a common SUSY soft-breaking scalar term of $\mathcal{L}_{soft}$ at 
scales close to $M_{\rm Planck}$ can give rise to RG-induced  $\delta^q$'s and 
$\delta^l$'s at the weak scale, one may envisage the possibility to make use 
of the FCNC constraints on such low-energy $\delta$'s to infer bounds on the 
soft breaking parameters of the original supergravity  Lagrangian 
($\mathcal{L}_{sugra}$). Indeed, for each scalar soft parameter of 
$\mathcal{L}_{sugra}$ one can ascertain whether the hadronic or the 
leptonic corresponding bound at the weak scale yields the stronger 
constraint 
at the large scale. One can then go through an exhaustive list of the 
low-energy constraints on 
the various  $\delta^q$'s and $\delta^l$'s and, then, after RG evolving 
such $\delta$'s up to $M_{Planck}$, we will establish for each 
$\delta$ of $\mathcal{L}_{sugra}$ which one between the hadronic and 
leptonic constraints is going to win, namely which provides the strongest
 constraint on the corresponding $\delta_{sugra}$ \cite{Ciuchini:2007ha}.

Consider for example the scalar soft breaking sector of the MSSM: 
\bea 
\label{smsoft}
- {\cal L}_{soft} &=& m_{Q_{ii}}^2 \tilde{Q}_i^\dagger \tilde{Q}_i 
+ m_{u^c_{ii}}^2 \tilde{u^c}_i^\star \tilde{u^c}_i + m^2_{e^c_{ii}} 
\tilde{e^c}_i^\star \tilde{e^c}_i 
 +   m^2_{d^c_{ii}} \tilde{d^c}^\star_i \tilde{d^c}_i 
 \nonumber \\ &+&  m_{L_{ii}}^2 \tilde{L}_i^\dagger \tilde{L}_i + 
m^2_{H_1} H^\dagger_1 H_1 
+  m^2_{H_2} H_2^\dagger H_2 + A^u_{ij}~
\tilde{Q}_i \tilde{u^c}_j H_2 \nonumber \\ 
&+& A^d_{ij}~
\tilde{Q}_i \tilde{d^c}_j H_1 + A^e_{ij}~
\tilde{L}_i \tilde{e^c}_j H_1 + (\Delta^l_{ij})_{LL} \tilde{L}_i^\dagger \tilde{L}_j  + 
(\Delta^e_{ij})_{RR} \tilde{e^c}_i^\star \tilde{e^c}_j  
 \nonumber \\ 
&+& (\Delta^q_{ij})_{LL} \tilde{Q}_i^\dagger \tilde{Q}_j  + 
(\Delta^u_{ij})_{RR} \tilde{u^c}_i^\star \tilde{u^c}_j +
(\Delta^d_{ij})_{RR} \tilde{d^c}_i^\star \tilde{d^c}_j  
 \nonumber \\ 
&+&  (\Delta^e_{ij})_{LR} \tilde{e_L}_i^\star \tilde{e^c}_j +
(\Delta^u_{ij})_{LR} \tilde{u_L}_i^\star \tilde{u^c}_j
+ (\Delta^d_{ij})_{LR} \tilde{d_L}_i^\star \tilde{d^c}_j
\eea
where we have explicitly written down the various $\Delta$ parameters.

Consider now that $SU(5)$ is the relevant symmetry at the scale where the 
above soft terms firstly  show up.  Then, taking into account that matter 
is organised into the  SU(5) representations ${\bf 10}~ =~(q,u^c,e^c)$ and 
${\bf\overline 5}~ = ~(l,d^c)$, one obtains the following relations
\bea
\label{matrel1}
m^2_{Q} = m^2_{\tilde{e^c}} = m^2_{\tilde{u^c}} = m^2_{\bf 10} \\
\label{matrel2}
m^2_{\tilde{d^c}} = m^2_{L} = m^2_{\bar{\bf \overline 5}} \\
\label{trirel}
A^e_{ij} = A^d_{ji}\, .
\eea
Eqs.~(\ref{matrel1}, \ref{matrel2}, \ref{trirel}) are matrices in flavor
space.  These equations lead to relations between the slepton and squark 
flavor violating off-diagonal entries $\Delta_{ij}$. These are\footnote{The
defintion of super CKM basis becomes much more complicated in the SUSY SU(5)
models especially when one considers solutions to the fermion mass problem.
For a more detailed discussion, please see \cite{Ciuchini:2007ha}}: 
\bea
\label{cdeltas1}
(\Delta^u_{ij})_{LL} = (\Delta^u_{ij})_{RR} = (\Delta^d_{ij})_{LL} =
(\Delta^l_{ij})_{RR} \\
\label{cdeltas3}
(\Delta^d_{ij})_{RR} = (\Delta^l_{ij})_{LL} \\
\label{cdeltas4}
(\Delta^d_{ij})_{LR} = (\Delta^l_{ji})_{LR} = (\Delta^l_{ij})_{RL}^\star
\eea
These GUT correlations among hadronic and leptonic scalar soft terms that
are summarised in the second column of Table~5. Assuming that
no new sources of flavor structure are present from the $SU(5)$ scale
down to the electroweak scale, apart from the usual SM CKM one, one
infers the relations in the first column of Table~5 at low
scale. Here we have taken into account that due to their different gauge
couplings ``average'' (diagonal) squark and slepton masses acquire
different values at the electroweak scale.
\begingroup
\begin{table}
\begin{center}
\begin{tabular}{|c|c|c|}
\hline\hline
&Relations at weak-scale & Boundary conditions at $M_{GUT}$ \\[0.2pt] 
\hline
(1) & $(\delta^u_{ij})_{RR}~ \approx~ (m_{\tilde e^c}^2/ m_{\tilde u^c}^2)~ 
(\delta^l_{ij})_{RR}$ & $m^2_{\tilde u^c}(0) ~=~ m^2_{\tilde e^c}(0)$ \\
\hline
(2) & 
$(\delta^q_{ij})_{LL}~ \approx~(m_{\tilde e^c}^2/ m_{\tilde Q}^2)~ (\delta^l_{ij})_{RR}$ &
$m^2_{\tilde Q}(0) ~=~ m^2_{\tilde e^c}(0)$ \\ 
\hline 
(3) &
$(\delta^d_{ij})_{RR} ~\approx~ (m_{\tilde L}^2/ m_{\tilde d^c}^2)~ (\delta^l_{ij})_{LL}$ &
$m^2_{\tilde d^c}(0) ~=~ m^2_{\tilde L}(0)$ \\ 
\hline 
(4) &
$(\delta^d_{ij})_{LR}~\approx~ (m_{\tilde L_{\rm avg}}^2 /
m_{\tilde Q_{\rm avg}}^2) ~
(m_b/ m_\tau) ~ (\delta^l_{ij})_{LR}^\star$  & 
$A^e_{ij} = A^d_{ji}$\\
\hline\hline
\end{tabular}
\label{tb0}
\caption{Links between various transitions between up-type, down-type quarks 
and charged leptons for SU(5). $m_{\tilde f}^2$ refers to the average
mass for the sfermion $f$, $m_{\tilde Q_{\rm avg}}^2= 
\sqrt{m_{\tilde Q}^2 m_{\tilde d^c}^2}$ and $m_{\tilde L_{\rm avg}}^2= 
\sqrt{m_{\tilde L}^2 m_{\tilde e^c}^2}$}
\end{center}
\end{table}
\endgroup

Two comments are in order when looking at Table~5. First, the boundary 
conditions on the sfermion masses at the GUT scale (last column in 
Table~5) imply that the squark masses are \textit{always} going 
to be larger at the weak scale compared to the slepton masses due to the
participation of the QCD coupling in the RGEs. As a 
second remark, notice that the relations between hadronic and leptonic
$\delta$ MI in Table~5 always exhibit opposite ``chiralities'', 
i.e. LL insertions are related to RR ones and vice-versa.  This stems from 
the arrangement of the different fermion chiralities in  $SU(5)$ 
five- and ten-plets (as it clearly appears from the final column in 
Table~5). This restriction can easily be overcome 
if we move from $SU(5)$ to left-right symmetric unified models 
like  SO(10) or  the  Pati-Salam (PS) case (we exhibit the corresponding 
 GUT boundary conditions and $\delta$ MI at the electroweak scale 
in Table 6). 
\begingroup
\begin{table}
\label{tb1}
\begin{center}
\begin{tabular}{|c|c|c|}
\hline\hline
&Relations at weak-scale & Boundary conditions at $M_{GUT}$ \\[0.2pt] 
\hline
(1) & $(\delta^u_{ij})_{RR}~ \approx~ (m_{\tilde e^c}^2/ m_{\tilde u^c}^2)~ 
(\delta^l_{ij})_{RR}$ & $m^2_{\tilde u^c}(0) ~=~ m^2_{\tilde e^c}(0)$ \\
\hline
(2) & 
$(\delta^q_{ij})_{LL}~ \approx~(m_{\tilde L}^2/ m_{\tilde Q}^2)~ (\delta^l_{ij})_{LL}$ &
$m^2_{\tilde Q}(0) ~=~ m^2_{\tilde L}(0)$ 
\\
\hline\hline
\end{tabular}
\end{center}
\caption{Links between various transitions between up-type, down-type quarks 
and charged leptons for PS/SO(10) type models.}
\end{table}
\endgroup

So far we have confined our discussion within the simple $SU(5)$ model, without
the presence of any extra particles like right handed (RH) neutrinos. In the
presence of RH neutrinos, one can envisage of two scenarios 
\cite{Masiero:2002jn}: (a) with either very small neutrino Dirac Yukawa couplings 
and/or very small
mixing present in the neutrino Dirac Yukawa matrix, (b) Large Yukawa and large
mixing in the neutrino sector. In the latter case, Eqs.~(\ref{cdeltas1} -- 
\ref{cdeltas4}) are not valid at all scales in general, as large RGE effects 
can significantly modify the sleptonic flavour structure while keeping the 
squark sector essentially unmodified; thus essentially breaking the GUT 
symmetric relations. 
In the former case where the neutrino Dirac Yukawa couplings are tiny 
and do not significantly modify the sleptonic flavour structure, the 
GUT symmetric relations are expected to be valid at the weak scale. 
However, in both cases it is possible to say that there exists a upper bound 
on the hadronic $\delta$ parameters of the form \cite{Ciuchini:2003rg}:
\begin{equation}
|(\delta^d_{ij})_{\rm RR}| ~~~\geq~~~ {m_{\tilde{L}}^2 \over m_{\tilde{d}^c}^2}
|(\delta^l_{ij})_{\rm LL}|. 
\end{equation}

These powerful relations between the various soft parameters can now
be used to repeat the same excercise we have done in the previous 
section for the case of MSSM, namely, a complete analysis of hadronic
and leptonic flavour violating constraints on SUSY parameters, however
with one major difference : a given $\delta$ at high scale can now be
constrained by both leptonic as well as hadronic processes at the weak
scale. Which sector wins the \textit{match} decides the strongest 
constraint on the given $\delta$ \cite{Ciuchini:2007ha}. 

As an example of these GUT relations, let us compute the bounds on
$\left(\delta^d_{ij}\right)_{AB}$ parameters, with $A,B={\rm L,R}$,
from Lepton Flavour Violation (LFV) rare decays $l_j \to l_i, \gamma$, 
using the relations
described above.  First, we will analyse the $23$ sector, that has
been recently of much interest due to the discrepancy with SM
expectations in the measurements of the CP asymmetry $A_{CP}(B \to
\phi K_s)$, which can be attributed to the presence of large neutrino
mixing within $SO(10)$ models \cite{Hisano:1995cp,Hisano:1998fj,Hisano:2003bd,
Chang:2002mq}. Subsequently, a
detailed analysis has been presented \cite{Ciuchini:2002uv,Harnik:2002vs} 
within
the context of MSSM. It has been shown that \cite{Ciuchini:2002uv} the
presence of a large $\sim~ \mathcal{O}(1)$ $\delta^d_{23}$ of LL or RR
type could lead to significant discrepancies from the SM expectations
and in particular one could reach the present central value for the
measurement of $A_{CP}(B \to \phi K_s)$. Similar statements hold for a
relatively small $\sim~ \mathcal{O}(10^{-2})$ LR and RL type MI.

\begin{figure}
\begin{center}
\includegraphics[width=11cm]{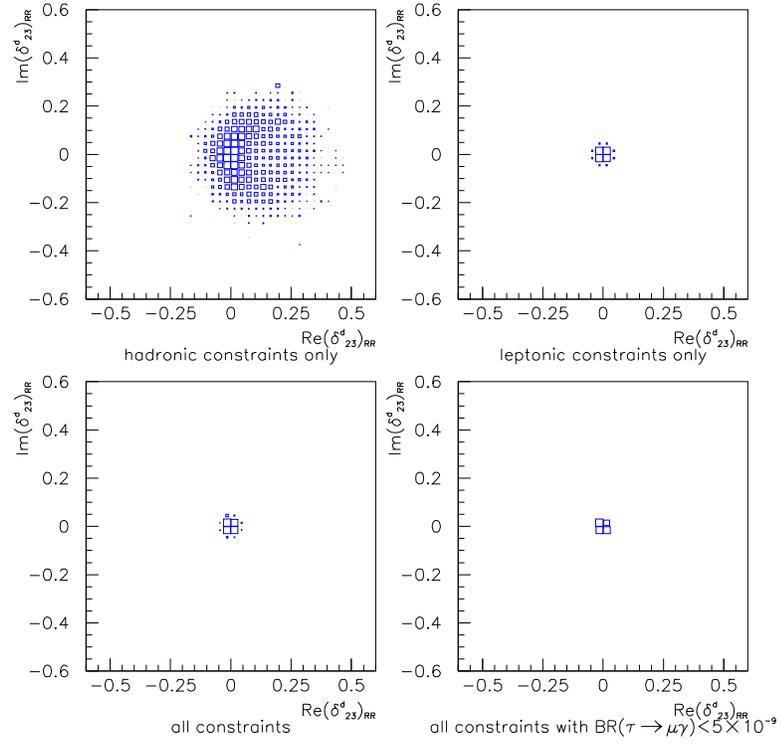}
\caption{
Allowed regions in the 
Re$(\delta^d_{23})_{RR}$--Im$(\delta^d_{23})_{RR}$ plane (top) and
in the $S_{K\phi}$--Im$(\delta^d_{23})_{RR}$ plane (bottom). Constraints
from $B \to X_s \gamma$, $BR(B \to X_s \ell^+ \ell^-)$, and 
the lower bound on $\Delta M_s$ have been used.  }
\label{fig:combi23}
\end{center}
\end{figure}

Now, we would like to analyse the impact of LFV bounds on these
hadronic $\delta$ parameters and its effect on B-physics
observables. 
\begingroup
\begin{table}[h]
\begin{center}
\label{tb3}
\begin{tabular}{|cccc|}
\hline \hline
Type &  $<~ 1.1~ 10^{-6}$ & $< ~6~ 10^{-7}$ & $<~ 1. ~10^{-7}$ \\[0.2pt]
\hline
LL & - & - & -    \\ 
RR & 0.105 & 0.075 & 0.03 \\ 
RL & 0.108 & 0.08 & 0.035\\
LR & 0.108 & 0.08 & 0.035 \\
\hline \hline
\end{tabular}
\caption{Bounds on $(\delta^d_{23})$ from $\tau \to \mu , \gamma$ for three
different values of the branching ratios for tan $\beta$ = 10.}
\end{center}
\end{table}
\endgroup
In table 7, we present upper bounds on $\left(\delta^d_{23}\right)_{\rm RR}$
with squark masses in the range 350--500 GeVs and for
three different upper bounds on Br($\tau \to \mu, \gamma$). There are
no bounds on $\left(\delta^d_{23}\right)_{\rm LL}$ because large values of
$\left(\delta^l_{23}\right)_{\rm RR}$ are still allowed due to
possible cancellations of bino and higgsino contributions for the
decay amplitudes \cite{Hisano:1995cp,Hisano:1998fj,Masina:2002mv}. In Fig~\ref{fig:combi23} we present 
the allowed ranges of 
$\left(\delta^d_{23}\right)_{\rm RR}$ and its effects on the CP asymmetry,
$A_{CP}(B \to \phi K_s)$, taking into account only hadronic constraints (left) 
or hadronic and leptonic constraints simultaneously (right).
Thus, we can see that in a $SU(5)$ GUT model where SUSY-breaking terms have 
a supergravity origin, LFV constraints are indeed very relevant for 
$(\delta^d_{23})_{RR}$  and it is not possible to generate large effects on 
$A_{CP}(B \to \phi K_s)$. Naturally we have to take into
account that the leptonic bounds and their effects on hadronic MIs
scale as $10/\tan \beta$ for different values of $\tan \beta$. However, even
for $\tan \beta \leq 5$ the leptonic bounds would be very relevant
on this MI.

Finally, we will also analyse the effects of leptonic constraints in the $12$ 
sector.
\begin{figure}
\begin{center}
\includegraphics[width=12cm]{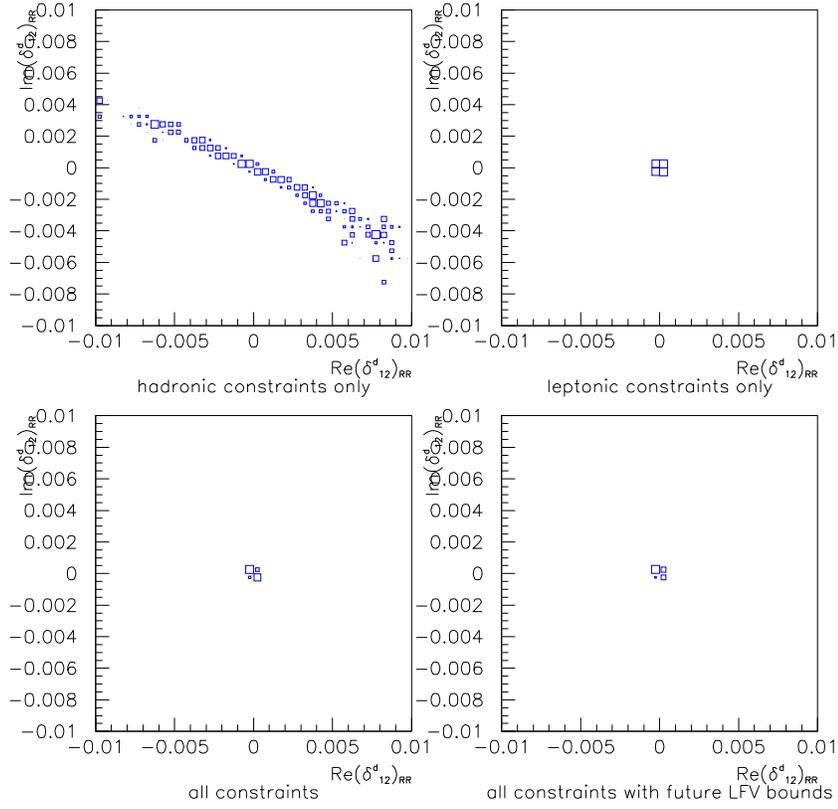}
\end{center}
\caption{Allowed regions in the ${\rm Re}
\left(\delta^d_{12} \right)_\RR$-- ${\rm Im}
\left(\delta^d_{12}\right)_\RR$ plane from hadronic and leptonic constraints.
Upper left plot takes into account only hadronic bounds, upper right plot
includes the present bound on the $\mu \to e\gamma$ decay, BR($\mu \to e,\gamma$)$<1.1~ \times~ 10^{-11}$. The second row correspond to the projected bounds 
from the proposed LFV experiments, BR($\mu \to e,\gamma$)$<10^{-13}$ and 
BR($\mu \to e,\gamma$)$<10^{-14}$ respectively. We have to take into account 
that we use $\tan \beta=10$ and leptonic bounds scale as $10/\tan \beta$.}
\label{fig:RR12}
\end{figure}
In Fig.~\ref{fig:RR12} we present the allowed values of ${\rm Re}
\left(\delta^d_{12} \right)_\RR$ and ${\rm Im}
\left(\delta^d_{12}\right)_\RR$.  The
upper left plot corresponds to the values that satisfy the hadronic
bounds, coming mainly from $\varepsilon_K = (2.284 \pm 0.014) \times 10^{-3}$. 
The upper right plot takes also into account the present
$\mu \to e\gamma$ bound, BR($\mu \to e,\gamma$)$<1.1~ \times~ 10^{-11}$, 
and the plots in the second row correspond to projected bounds from the 
proposed experiments, BR($\mu \to e,\gamma$)$<10^{-13}$ and 
BR($\mu \to e,\gamma$)$<10^{-14}$ respectively. Now 
the GUT symmetry relates $\left(\delta^d_{12} \right)_\RR$ to 
$\left(\delta^l_{12} \right)_\LL$ and in this case leptonic bounds 
(already the present bounds) are very stringent and reduce the allowed values
of $\left(\delta^d_{12} \right)_\RR$ by more than one order of magnitude to
a value $\left(\delta^d_{12} \right)_\RR \leq 4 \times 10^{-4}$ for 
$\tan \beta=10$.

In the case of $\left(\delta^d_{12} \right)_\LL$ the 
$\mu \to e \gamma$ decay does not provide a
bound to this MI due to the presence of cancellations between
different contributions. We can only obtain a relatively mild bound,
$\left(\delta_{12}^l \right)_\RR \leq 0.09$ for $\tan \beta =10$, if we
take into account $\mu \to eee$ and $\mu$--$e$ conversion in nuclei.
After rescaling this bound by
the factor ${\tilde m_{e^c}^2 \over \tilde m_{d_{\rm L}}^2}$ the
leptonic bound is still able to reduce the maximum values of 
${\rm Re} \left(\delta^d_{12} \right)_\LL$ and ${\rm Im}
\left(\delta^d_{12}\right)_\LL$ by a factor of 2, although the hadronic bound
is still more constraining in a big part of the parameter space.

In summary, Supersymmetric Grand Unification predicts links between various 
leptonic and hadronic FCNC Observables. Though such relations can
be constructed for any GUT group, we have concentrated on SU(5) and 
quantitatively studied the implications for the $23$ and $12$ sectors. 
In particular we have shown that the present limit
on $BR(\tau \to \mu, \gamma)$ is sufficient to significantly constrain the 
observability of supersymmetry in CP violating B-decays.  

\subsection{SUSY-GUTs, SUSY-Seesaw and flavour blind Supersymmetry breaking}
As discussed in the above, flavour violation can also be generated
through renormalisation group running even if one starts with 
flavour-blind soft masses at the scale where supersymmetry is mediated to the
visible sector. A classic example of this is the supersymmetric seesaw
mechanism and the generation of lepton flavour violation at the weak scale.

The seesaw mechanism  can be incorporated in the
Minimal Supersymmetric Standard Model in a manner similar to what is  done in
the Standard Model by adding right-handed neutrino superfields to the
MSSM superpotential:
\bea
\label{eq1}
W &=&  h^u_{ij} Q_i u_j^c H_2 + h^d_{ii} Q_i d_i^c H_1 + h^e_{ii} L_i e^c_i H_1
+ h^\nu_{ij} L_i \nu^c_j H_2 \nn \\
&+& M_{R_{ii}} \nu_i^c \nu_i^c + \mu H_1 H_2 ,
\eea
where we are in the basis of diagonal charged lepton, down quark and 
right-handed Majorana  mass matrices.
$M_{R}$ represents the (heavy) Majorana mass matrix
for the right-handed neutrinos.  Eq.~(\ref{eq1}) leads to the standard seesaw
formula for the (light) neutrino mass matrix
\begin{equation}
\label{seesaweq}
{\mathcal M}_\nu = - h^\nu M_{R}^{-1} h^{\nu~T} v_2^2,
\end{equation}
where $v_2$ is the vacuum expectation value (VEV) of the up-type
Higgs field, $H_2$. Under suitable conditions on $h^\nu$ and $M_R$,
the correct mass splittings and  mixing angles in $\mathcal{M}_\nu$
can be obtained. Detailed analyses deriving these conditions are
already present in the literature \cite{Altarelli:2002hx,Altarelli:2003vk,Altarelli:2004za,Masina:2001pp,Mohapatra:2002kn,Mohapatra:2003qw,King:2003jb,Smirnov:2003xe}.

Following the discussion in the
previous section, we will assume that the mechanism that breaks
supersymmetry and conveys it to the observable sector at the high scale
$\sim M_{\rm P}$ is flavour-blind, as in the CMSSM (also called mSUGRA). 
However, this flavour
blindness is not protected down to the weak scale \cite{Borzumati:1986qx}
\footnote{This is always true in a gravity mediated supersymmetry 
breaking model, but it also applies to other mechanisms under
some specific conditions \cite{Tobe:2003nx,Ibe:2004tg}.}.
The slepton mass matrices are no longer invariant under RG evolution from the
super-large scale where supersymmetry is mediated to the visible sector down to
the seesaw scale. The flavour violation present in the neutrino Dirac Yukawa
couplings $h^\nu$ is now ``felt'' by the slepton mass matrices in the 
presence of heavy right-handed neutrinos \cite{Casas:2001sr,Masiero:2004js}.

The weak-scale flavour violation so generated can be obtained
by solving the RGEs for the slepton mass matrices
from the high scale to the scale of the right-handed neutrinos. Below
this scale, the running of the FV slepton mass terms is RG-invariant
as the right-handed neutrinos decouple from the theory. For the purpose of
illustration, a leading-log estimate can easily be obtained
for these equations\footnote{Within mSUGRA, the leading-log approximation 
works very well for most of the parameter space, except for regions of 
large $M_{1/2}$ and low $m_0$. The discrepancy with the exact result 
increases with low $\tan \beta$ \cite{Petcov:2003zb}.}.
Assuming the flavour blind mSUGRA specified by
the high-scale parameters, $m_0$, the common scalar mass, $A_0$,
the common trilinear coupling, and $M_{1/2}$, the universal gaugino mass,
the flavour violating entries in these mass matrices at the weak scale
are given as:
\beq
\label{rgemi}
(\Delta^l_{ij})_{\rm LL} \approx -{3 m_0^2+A_0^2 \over 8 \pi^2} \sum_k
(h^\nu_{ik} h^{\nu *}_{jk}) \ln{M_{X} \over M_{R_k} },
\eeq
where $h^\nu$ are given in the basis of diagonal charged lepton masses
and diagonal Majorana right-handed neutrino mass matrix $M_R$, and
$M_X$ is the scale at which soft terms appear in the Lagrangian. Given
this, the branching ratios for LFV rare decays $l_j \to l_i, \gamma$
can be roughly estimated using
\beq
\label{BR}
\mbox{BR} (l_j \to l_i \gamma) \approx \Frac{ \alpha^3 ~
~| \delta^{l}_{ij}|^2 }
{G_F^2~ m^4_{\rm SUSY}} \tan^2 \beta.
\eeq
From above it is
obvious that the amount of lepton flavour violation generated by the
SUSY seesaw at the weak scale crucially depends on the flavour
structure of $h^\nu$ and $M_R$, the ``new'' sources of flavour 
violation not present in the MSSM, Eq.~(\ref{eq1}).  If either the
neutrino Yukawa couplings or the flavour mixings present in $h^\nu$
are very tiny, the strength of LFV will be significantly reduced.
Further, if the right-handed neutrino masses were heavier than the
supersymmetry breaking scale (as in GMSB models) they would decouple
from the theory before the SUSY soft breaking matrices enter into play
and hence these effects would vanish.

\subsection{Seesaw in GUTs: SO(10) and LFV}
A simple analysis of the fermion mass matrices in the $SO(10)$ model,
as detailed in the Eq.~(\ref{numats}) leads us to the following
result:  \textit{At least one
of the Yukawa couplings in $h^\nu~ =~ v_u^{-1}~M^\nu_{LR}$ has
to be as large as the top Yukawa coupling} \cite{Masiero:2002jn}. 
This result holds true in general, independently of
the choice of the Higgses responsible for the masses in
Eqs.~(\ref{upmats}), (\ref{numats}), provided that no accidental fine-tuned
cancellations of the different contributions in Eq.~(\ref{numats}) are
present. If contributions from the \textbf{10}'s solely
dominate, $h^\nu$ and $h^u$ would be equal.
If this occurs for the \textbf{126}'s, then $h^\nu =- 3~ h^u$ 
\cite{Mohapatra:1979nn}.
In case both of them have dominant entries, barring a rather precisely 
fine-tuned cancellation between $M^5_{10}$ and $M^5_{126}$ in
Eq.~(\ref{numats}), we expect at least one large entry to be present 
in $h^\nu$. A dominant antisymmetric contribution to top quark mass
due to the {\bf 120} Higgs is phenomenologically excluded, since it would
lead to at least a pair of heavy degenerate up quarks.

Apart from sharing the property that at least one eigenvalue of both $M^u$
and $M^\nu_{LR}$ has to be large, for the rest it is clear from 
Eqs.~(\ref{upmats}) and  (\ref{numats}) that these two matrices are not aligned
in general, and hence we may expect different mixing angles appearing
from their diagonalisation. This freedom is removed if one sticks to
particularly simple choices of the Higgses responsible for up quark and
neutrino masses. A couple of remarks are in order here. Firstly, note that
in general there can be an additional contribution, Eq.~(\ref{lneutmats}),
to the light neutrino mass matrix, independent of the canonical seesaw 
mechanism. Taking into consideration also this contribution leads to 
the so-called Type-II seesaw formula \cite{Lazarides:1980nt,Mohapatra:1980yp}. Secondly, the 
correlation between neutrino Dirac Yukawa coupling and the top Yukawa 
is in general independent of the type of seesaw mechanism,
and thus holds true irrespective  of the light-neutrino mass structure.

Therefore, we see that the $SO(10)$ model with only two ten-plets would 
inevitably lead to small mixing in $h^\nu$. In fact, with two Higgs fields
in symmetric representations, giving masses to the up-sector and the
down-sector separately, it would be difficult to avoid the small CKM-like
mixing in $h^\nu$. We will call this case the CKM case.
From here, the following mass 
relations hold between the quark and leptonic mass matrices at the GUT 
scale\footnote{Clearly this relation cannot hold for the first two 
generations of down quarks and charged leptons.  One expects, small 
corrections due to  non-renormalisable operators or
suppressed renormalisable operators \cite{Georgi:1979df} to be invoked.}: 
\beq
\label{massrelations}
h^u  = h^\nu \;\;\;;\;\;\; h^d  = h^e . 
\eeq
In the basis where charged lepton masses are diagonal, we have
\beq
\label{hnumg}
 h^\nu = V_{\rm CKM}^T~ h^u_{Diag}~ V_{\rm CKM}. 
\eeq 
The large couplings in $h^\nu \sim {\mathcal O}(h_t)$ induce 
significant off-diagonal entries in $m_{\tilde L}^2$ through the RG
evolution between $M_{\rm GUT}$ and the scale of the right-handed Majorana neutrinos
\footnote{Typically one has different mass scales associated with different
right-handed neutrino masses.}, $M_{R_i}$. The induced off-diagonal entries
relevant to $l_j \rightarrow l_i, \gamma$ are of the order of:  
\begin{eqnarray}
\label{wcmi1}
(m_{\tilde L}^2)_{21}&\approx& -{3 m_0^2+A_0^2 \over 8 \pi^2}~
h_t^2 V_{td} V_{ts} \ln{M_{\rm GUT} \over M_{R_3}} 
+ \mathcal{O}(h_c^2), \\
\label{wcmi2}
(m_{\tilde L}^2)_{32}&\approx& -{3 m_0^2+A_0^2 \over 8 \pi^2}~
h_t^2 V_{tb} V_{ts} \ln{M_{\rm GUT} \over M_{R_3}} 
+ \mathcal{O}(h_c^2), \\
\label{wcmi3}
(m_{\tilde L}^2)_{31}&\approx& -{3 m_0^2+A_0^2 \over 8 \pi^2}~
h_t^2 V_{tb} V_{td} \ln{M_{\rm GUT} \over M_{R_3}} 
+ \mathcal{O}(h_c^2).
\end{eqnarray}
\noindent 
In these expressions, the CKM angles are small but one would expect 
the presence of the large top Yukawa coupling to compensate such a 
suppression. The required right-handed neutrino Majorana  mass 
matrix, consistent with both the observed low energy neutrino masses 
and mixings as well as with CKM-like mixings in $h^\nu$ is easily determined 
from the seesaw formula defined at the scale of right-handed neutrinos
\footnote{The neutrino masses and mixings here are defined
at $M_{R}$. Radiative corrections can significantly modify the neutrino
spectrum from that of the weak scale \cite{Chankowski:2001mx}. 
This is more true for 
the degenerate spectrum of neutrino masses \cite{Ellis:1999my,Casas:1999tp,
Haba:1999fk} and for some 
specific forms of $h^\nu$ \cite{Antusch:2002hy}. For our present discussion, with
hierarchical neutrino masses and up-quark like neutrino Yukawa matrices, we 
expect these effects not to play a very significant role.}.  

The Br($l_i \to l_j \gamma$) are now predictable in this case. Considering
mSUGRA boundary conditions and taking $\tan \beta = 40$, we obtain that 
reaching a sensitivity of $10^{-14}$ for 
BR$(\mu \to e \gamma)$ would allow us to probe the SUSY 
spectrum  completely up to $M_{1/2} = 300$ GeV (notice that this 
corresponds to 
gluino and squark masses of order 750 GeV) and would still probe 
large regions of the parameter space up to $M_{1/2} = 700$ GeV.  
Thus, in summary, though the present limits on BR($\mu \to e, \gamma$) 
would not induce any significant constraints on the supersymmetry-breaking
parameter space, an improvement in the limit to $\sim {\mathcal O }(10^{-14})$,
as foreseen, would start imposing non-trivial constraints especially
for the large $\tan \beta$ region. 

To obtain mixing angles larger than CKM angles, asymmetric
mass matrices have to be considered. In general, it is sufficient to introduce
asymmetric textures either in the up-sector or in the down-sector. In the
present case, we assume that the down-sector couples to a combination of 
Higgs representations (symmetric and antisymmetric)\footnote{The couplings 
of the Higgs fields in the superpotential can be either renormalisable or 
non-renormalisable. See \cite{Chang:2002mq} for a non-renormalisable example.}
$\Phi$, leading to an asymmetric mass matrix in the basis where the 
up-sector is diagonal. As we will see below, this would also require that 
the right-handed Majorana mass matrix be diagonal in this basis. We have :
\beq
\label{mnsso10}
W_{SO(10)} = {1 \over 2}~  h^{u,\nu}_{ii}~ {\bf 16_i} ~{\bf 16_i} {\bf 10^u} +
{1 \over 2}~ h^{d,e}_{ij}~ {\bf 16_i} ~{\bf 16_j} \Phi \nn \\
 + {1 \over 2}~ h^R_{ii}~ {\bf 16_i}~ {\bf 16_i} {\bf 126}~,
\eeq
where the \textbf{126}, as before, generates only the right-handed neutrino
mass matrix. To study the consequences
of these assumptions, we see that at the level of $SU(5)$, we have
\beq
W_{SU(5)} = {1 \over 2}~ h^u_{ii}~ {\bf 10_i} ~{\bf 10_i} ~{\bf 5_u}
+ h^\nu_{ii} ~{\bf \bar{5}_i}~ {\bf 1_i}~ {\bf 5_u} \nn \\
+  h^d_{ij}~ {\bf 10_i} ~{\bf \bar{5}_j}~ {\bf \bar{5}_d}
+  {1 \over 2}~M^R_{ii}~ {\bf 1_i} {\bf 1_i},
\eeq
where we have decomposed the ${\bf 16}$ into ${\bf 10} + {\bf \bar{5}} +{\bf  1}$ and ${\bf 5_u}$ and
${\bf \bar{5}_d}$ are components of ${\bf 10_u}$ and $\Phi$ respectively. To have large
mixing $\sim~ U_{\rm PMNS}$  in $h^\nu$ we see that the asymmetric matrix $h^d$
should now give rise to both the CKM mixing as well as PMNS mixing.
This is possible if
\beq
V_{\rm CKM}^T~ h^d~ U_{\rm PMNS}^T = h^d_{Diag}.
\eeq

\begin{figure}[ht]
\begin{center}
\includegraphics[scale=0.35]{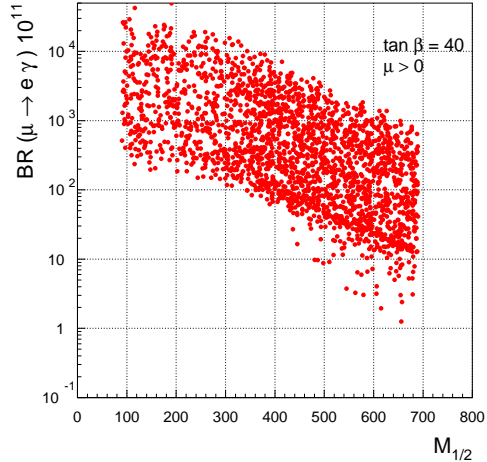}
\caption{The scatter plots of branching ratios of $\mu \to e, \gamma$
decays as a function of $M_{1/2}$ are shown for the (maximal) PMNS case for
tan $\beta$ = 40. The results do not alter significantly with the change
of sign($\mu$).}
\end{center}
\end{figure}

Therefore the ${\bf 10}$ that contains the left-handed down-quarks would
be rotated by the CKM matrix whereas the ${\bf \bar{5}}$ that contains the 
left-handed charged leptons would be rotated by the $U_{\rm PMNS}$ matrix 
to go into their respective mass bases \cite{Moroi:2000mr,Moroi:2000tk,Akama:2001em,Chang:2002mq}. Thus we have, 
in analogy with the previous subsection, the following relations in the 
basis where charged leptons and down quarks are diagonal:
\bea
h^u &=& V_{\rm CKM}~ h^u_{Diag}~ V_{\rm CKM}^T~ ,  \\
\label{hnumns}
h^\nu &=& U_{\rm PMNS}~ h^u_{Diag}.
\eea
Using the seesaw formula of Eqs.~({\ref{seesaweq}) and (\ref{hnumns}), we have
\beq
M_{R} = Diag\{ {m_u^2 \over m_{\nu_1}},~{m_c^2 \over m_{\nu_2}},
~{m_t^2 \over m_{\nu_3}} \}.
\eeq
We now turn our attention to  lepton flavour violation in this case. The
branching ratio, BR($\mu \rightarrow e, \gamma)$ would now depend on
\beq
\label{hnusqmns}
[h^\nu h^{\nu~T}]_{21} = h_t^2~ U_{\mu 3}~ U_{e 3} + h_c^2~ U_{\mu 2}~ U_{e 2} +
\mathcal{O}(h_u^2).
\eeq
It is clear from the above that in contrast to the CKM case,
the dominant contribution to the off-diagonal entries depends on the
unknown magnitude of the element $U_{e3}$ \cite{Sato:2000zh}.
If $U_{e3}$ is very close to its
present limit $\sim~0.2$ \cite{Apollonio:1999ae}, the first term on the
RHS of the Eq.~(\ref{hnusqmns})
would dominate. Moreover, this would lead to large contributions to the
off-diagonal entries in the slepton masses with $U_{\mu 3}$ of
${\mathcal O}(1)$. From \eq{rgemi} we have 
\beq
\label{bcmi1}
(m_{\tilde L}^2)_{21} \approx -{3 m_0^2+A_0^2 \over 8 \pi^2}~
h_t^2 U_{e 3} U_{\mu 3} \ln{M_{\rm GUT} \over M_{R_3}}
+ \mathcal{O}(h_c^2).
\eeq
This contribution is larger than the CKM case by a factor of
$(U_{\mu 3} U_{e3})/ (V_{td} V_{ts}) \sim 140$.
From Eq.~(\ref{BR}) we see that
it would mean about a factor $10^4$ times larger than the CKM case in
BR($\mu \rightarrow e, \gamma)$. In case $U_{e3}$ is very small, \textit{i.e}
either zero or $\ler~ (h_c^2/h_t^2)~U_{e2}~ \sim 4 \times 10^{-5}$, the
second term $\propto ~h_c^2$
in Eq.~(\ref{hnusqmns}) would dominate. However the off-diagonal 
contribution in slepton masses, now being proportional to charm Yukawa 
could be much smaller, even  smaller than the CKM contribution by a factor
\beq{h_c^2~ U_{\mu 2} ~U_{e 2} \over  h_t^2 ~V_{td} ~V_{ts}}
\sim  7 \times 10^{-2}.
\eeq
If $U_{e3}$ is close to its present limit, the current bound on
R($\mu \rightarrow e, \gamma$) would already be sufficient to
produce stringent limits on the SUSY mass spectrum. Similar
$U_{e3}$ dependence can be expected in the $\tau \to e $ transitions
where the off-diagonal entries are given by :
\beq
\label{bcmi3}
(m_{\tilde L}^2)_{31} \approx  -{3 m_0^2+A_0^2 \over 8 \pi^2}~
h_t^2 U_{e 3} U_{\tau 3} \ln{M_{\rm GUT} \over M_{R_3}}
+ \mathcal{O}(h_c^2).
\eeq
The $\tau \to \mu$ transitions are instead $U_{e3}$-independent probes of SUSY,
whose importance was first pointed out in Ref.~\cite{Blazek:2001zm}. The 
off-diagonal entry in this case is given by :
\beq
\label{bcmi2}
(m_{\tilde L}^2)_{32} \approx  -{3 m_0^2+A_0^2 \over 8 \pi^2}~
h_t^2 U_{\mu 3} U_{\tau 3} \ln{M_{\rm GUT} \over M_{R_3}}
 \mathcal{O}(h_c^2).
\eeq

In the PMNS scenario, Fig. 3 shows the plot for BR($\mu
\rightarrow e, \gamma$) for tan $\beta$ = 40. In this plot, the value of
$U_{e3}$ chosen is very close to the present experimental upper limit
\cite{Apollonio:1999ae}. As long as $U_{e3} \ger 4 \times 10^{-5}$, the plots
scale as $U_{e3}^2$, while for $U_{e3} \ler 4 \times 10^{-5}$ the term
proportional to $m_c^2$ in Eq.~(\ref{bcmi1}) starts dominating;
the result is then insensitive to the choice of $U_{e3}$.
For instance, a value of $U_{e3} =0.01$ would reduce the BR by a factor of
225 and still a significant amount of  the parameter space
for $\tan \beta = 40$ would
be excluded.
We further find that with the present limit on BR($\mu \rightarrow e, \gamma$),
all the parameter space would be completely excluded up to $M_{1/2}=
300$ GeV for $U_{e3} =0.15$, for any vale of $\tan \beta$ (not shown in the
figure).

In the $\tau \to \mu \gamma$ decay the situation is similarly
constrained.  For $\tan \beta=2$, the present bound of $3 \times
10^{-7}$ starts probing the parameter space up to $M_{1/2} \leq 150$
GeV. The main difference is that this does not depend on the value of
$U_{e3}$, and therefore it is already a very important constraint on
the parameter space of the model.  In fact, for large $\tan \beta =
40$, as shown in Fig. 4, reaching the expected limit of $1 \times
10^{-8}$ would be able to rule out completely this scenario up to
gaugino masses of $400$ GeV, and only a small portion of the parameter
space with heavier gauginos would survive. In the limit $U_{e3} =0$,
this decay mode would provide a constraint on the model stronger than
$\mu \to e, \gamma$, which would now be suppressed as it would contain
only contributions proportional to $h_c^2$, as shown in
Eq.~(\ref{bcmi1}).

\begin{figure}[ht]
\begin{center}
\includegraphics[scale=0.35]{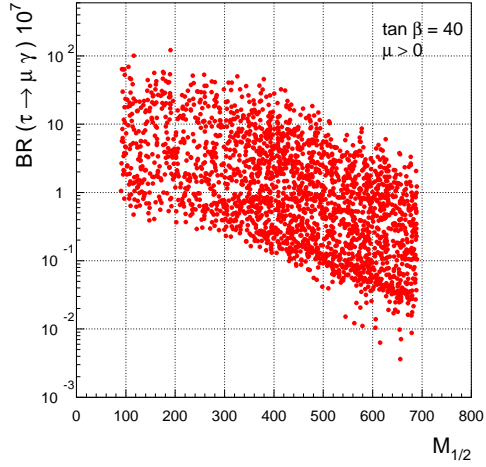}
\caption{The scatter plots of branching ratios of $\tau \to \mu, \gamma$
decays as a function of $M_{1/2}$ are shown for the (maximal) PMNS case for
n the PMNS scenario, Fig. 3 shows the plot for BR($\mu$
tan $\beta$ = 40. The results do not alter significantly with the change
of sign($\mu$).}
\end{center}
\end{figure}

In summary, in the PMNS/maximal mixing case, even the present limits
from BR($\mu \to e, \gamma$ ) can rule out large portions of the 
supersymmetry-breaking parameter space if $U_{e3}$ is either close to 
its present limit or within an order of magnitude of it (as the planned 
experiments might find out soon \cite{Goodman:2003zt}). These limits are more 
severe for large $\tan \beta$. In the extreme situation of $U_{e3}$ being 
zero or very small $\sim {\mathcal O}(10^{-4} - 10^{-5})$, 
BR($\tau \to \mu \gamma$) will start playing an important role with its 
present constraints already disallowing large regions of the parameter 
space at large $\tan \beta$.
While the above example concentrated on the hierarchical light neutrinos,
similar `benchmark' mixing scenarios have been explored in great detail, 
for degenerate spectra of light neutrinos, by Ref.~\cite{Illana:2003pj}, taking
also in to consideration running between the Planck scale and the GUT scale.

The above analysis has been restricted to one of the breaking chains of 
SO(10), namely the one which directly breaks in to MSSM. As depicted in
Fig.~\ref{fig:breakingchains}, there could be other breaking chains too. 
For example, let us consider the SU(5) breaking chain of SO(10) 
discussed in Fig.~\ref{fig:breakingchains}. The various energy scales involved
in this model can be neatly summarised as in figure \ref{scales}. In this
case, there could be addional sources of LFV other than the ones appearing
in the direct breaking chain. 
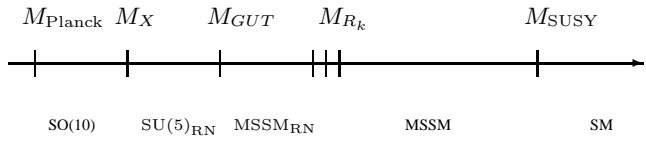
\begin{figure}[h]
{\footnotesize
\begin{center}
\begin{picture}(280,65)

\put(0,40){\vector(1,0){240}}

\put(5,55){$M_{\mathrm{Planck}}$}
\put(10,45){\line(0,-1){10}}
\put(40,55){$M_X$}
\put(45,45){\line(0,-1){10}}
\put(75,55){$M_{GUT}$}
\put(80,45){\line(0,-1){10}}

\put(117,55){$M_{R_k}$}
\put(115,45){\line(0,-1){10}}
\put(120,45){\line(0,-1){10}}
\put(125,45){\line(0,-1){10}}

\put(195,55){$M_{\mathrm{SUSY}}$}
\put(200,45){\line(0,-1){10}}

\put(15,15){\tiny SO(10)}
\put(50,15){\tiny  $\mathrm{SU(5)}_{\mathrm{RN}}$}
\put(85,15){\tiny $\mathrm{MSSM}_{\mathrm{RN}}$}
\put(150,15){\tiny MSSM}
\put(220,15){\tiny SM}
\end{picture}
\end{center}
}
\caption{\label{scales}Schematic
 picture of the energy scales involved in the model.}
\end{figure}
These sources are due to the result of $SU(5)$ running between the
scales $M_{SU(5)}$ and the $SO(10)$ scale $M_{SO(10)}$. As we have
seen several times already in SU(5) the right handed leptonic singlets sit 
in the same multiplet as the quark doublets (left handed) and up-type
singlets ($q,u,e^c$). This would have implications for the `right handed
mixing' ($\Delta_{RR}^l$) as this would get generated through RG evolution,
due to the couplings of the up-type quarks carrying the CKM information
\cite{Barbieri:1995tw}. This new contribution adds up to the $\Delta_{LL}^l$ 
contributions to LFV already present due to the 
seesaw effect. An analysis similar to the previous one could now be repeated 
for this breaking chain and it can be seen that these additional
contributions become important in some regions of the parameter
space\cite{Calibbi:2006nq}. 

Finally, let us in passing touch up on two related topics 
which are outside the realm of the present lecture series but are however
quite important in their own right. One of them relates the
\textit{visibility}  
of supersymmetry at Large Hadron Collider/ International Linear Collider with
the \textit{indirect evidence} for SUSY in flavour experiments or the 
dark matter experiments. This is an important question, in particular, in
simple models based on Grand Unification where the inter-correlations between
the various SUSY stratergies are quite interesting and could play an 
important role in \textit{"solving"} the inverse LHC problem. 
While, in general, direct searches at colliders like LHC/ILC are superior 
in comparison with indirect search strategies at flavour machines and dark 
matter experiments, there could be regions in the parameter space where the 
flavour machines can strike back. For example, in simple GUTs based on SUSY 
$SO(10)$ and in the case of the focus point region for viable dark matter, 
the sensistivity in flavour experiments is far greater than in direct 
searches at LHC\cite{Masiero:2004vk}. On an independent note, irrespective of
flavour violation, SUSY-GUTs can modify the pattern of dark matter parameter
space itselves as has been pointed out in  \cite{Calibbi:2007bk},
\cite{Calibbi:2007uw}. So, dark matter and flavour violation can both play 
an important and complementary role in unravelling the structure of SUSY-GUTs.

Finally, a second interesting aspect we have not addressed in the present
lectures  is the role of CP violation in the generation of the flavour
asymmetry of the universe in the context of SUSY-GUTs.
CP violation can play an important role for leptogenesis which can explain the
observed matter-anti-matter asymmetry. While in the simplest SO(10) models,
which we have discussed above, it might be very difficult to achieve viable
leptogenesis\cite{Branco:2002kt}, there have been simple solutions proposed 
which might make it achievable \cite{Vives:2005ra}. 

\section*{CONCLUSIONS}
\label{sec:conclu}
The ideas of the Grand Unification and Supersymmetry are closely
connected and represent the main avenue to explore in the search of
physics beyond the Standard Model. In these lectures we have presented
the reasons that make us believe in the existence of new physics
beyond the SM. We have presented the (non-supersymmetric) Grand
Unification idea and analysed its achievements and
failures. Supersymmetric grand unification was shown to cure some of
these problems and make the construction of ``realistic'' models
possible. The phenomenology of low-energy supersymmetry has been
discussed in the second part of these lectures with special emphasis
on the SUSY flavour and CP problems.  We have seen that, quite
generally, SUSY extensions of the SM lead to the presence of a host of
new flavour and CP violation parameters.  The solution of the ``SUSY
flavour problem'' and the ``SUSY CP problem'' are intimately
linked. However, there is an ``intrinsic'' CP problem in SUSY which
goes beyond the flavour issue and requires a deeper comprehension of
the link between CP violation and breaking of SUSY. We tried to
emphasise that the these two problems have not only a dark and
worrying side, but also they provide promising tools to obtain
indirect SUSY hints. We have also seen that the presence of a grand
unified symmetry and/or new particles, like right-handed neutrinos,  
at super-large scales has observable consequences in the structure of
soft masses at the electroweak scale. Thus the discovery of low energy
SUSY at the LHC or low energy FCNC experiments and the measurement of the
SUSY spectrum may provide a fundamental clue for the assessment of SUSY GUTs 
and SUSY seesaw in nature.

\section*{ACKNOWLEDGEMENTS}
Although some part of the material of these lectures reflects the
personal views of the authors, much of it results from works done
in collaboration with several friends, in particular M. Ciuchini,
P. Paradisi and L. Silvestrini. We thank them very much.
A.M. is very grateful to the organisers and participants of this school
for the pleasant and stimulating environment they succeeded to create
all along the school itself.

A.M. acknowledges partial support from the MIUR PRIN "Fisica
Astroparticellare" 2004-2006. O.V. acknowledges partial support 
from the Spanish  MCYT FPA2005-01678. All the figures have been plotted 
with Jaxodraw \cite{Binosi:2003yf}.

\bibliography{MSSM}
\end{document}